\documentclass[twocolumn]{aastex7}
\usepackage{times,amsmath}
\usepackage[T1]{fontenc}

\hypersetup{pdfauthor={Kamlesh Rajpurohit \& Ewan O'Sullivan},
            pdftitle={A combined radio and X-ray view of Stephan's Quintet},
            pdfkeywords={Galaxy collisions; Galaxy interactions; Galaxy groups; Hickson compact group; Intergalactic medium; Intergalactic filaments; Radio continuum emission; Shocks},
            bookmarksnumbered=true}
\pdfoutput=1

\usepackage[utf8]{inputenc}
\usepackage{graphicx}
\usepackage{amssymb}
\usepackage{amsmath}
\usepackage{float}
\usepackage{multirow}
\usepackage{textcomp}
\usepackage{gensymb}
\usepackage{enumitem}  
\usepackage{hyperref}
\usepackage{natbib}
\usepackage{comment}
\usepackage{systeme}
\usepackage[Symbol]{upgreek}
\usepackage{xcolor}
\usepackage{xspace}
\definecolor{xlinkcolor}{cmyk}{1,1,0,0}

\newcommand{\arcm}{\hbox{$^\prime$}}

\newcommand{\chandra}{\emph{Chandra}}

\newcommand{\hst}{\emph{HST}}
\newcommand{\jwst}{\emph{JWST}}
\newcommand{\arcs}{\mbox{\arcm\arcm}}

\newcommand{\Msol}{\ensuremath{\mathrm{~M_{\odot}}}}
\newcommand{\Msolpyr}{\ensuremath{\mathrm{~M_{\odot}~yr^{-1}}}}

\newcommand{\s}{\ensuremath{\mbox{~s}}}
\newcommand{\ps}{\ensuremath{\s^{-1}}}

\newcommand{\km}{\ensuremath{\mbox{~km}}}

\newcommand{\kmps}{\ensuremath{\km \ps}}

\newcommand{\gtsim}{\,\rlap{\raise 0.5ex\hbox{$>$}}{\lower 1.0ex\hbox{$\sim$}}\,} 

\newcommand{\Hi}{H\textsc{i}}

\definecolor{green}{RGB}{44,160,44}

\newcommand{\SQ}{SQ}

\shorttitle{Radio Continuum View of a Stephan's Quintet}

\begin{document}

\title{A radio continuum view of Stephan's Quintet: age, dynamics and origin of the shock}

\begin{NoHyper}
\correspondingauthor{Kamlesh Rajpurohit, Ewan O'Sullivan}
\end{NoHyper}
\email{kamlesh.rajpurohit@cfa.harvard.edu, eosullivan@cfa.harvard.edu }
\author[0000-0001-7509-2972]{K. Rajpurohit} 
\altaffiliation{These authors contributed equally to this work}
\affil{Center for Astrophysics $|$ Harvard \& Smithsonian, 60 Garden Street, Cambridge, MA 02138, USA}
\email{kamlesh.rajpurohit@cfa.harvard.edu}

\author[0000-0002-5671-6900]{E. O'Sullivan}
\altaffiliation{These authors contributed equally to this work}
\affil{Center for Astrophysics $|$ Harvard \& Smithsonian, 60 Garden Street, Cambridge, MA 02138, USA}
\email{eosullivan@cfa.harvard.edu}

\author[0000-0002-4962-0740]{G. Schellenberger}
\affil{Center for Astrophysics $|$ Harvard \& Smithsonian, 60 Garden Street, Cambridge, MA 02138, USA}
\email{gerrit.schellenberger@cfa.harvard.edu}
\author[0000-0002-9325-1567]{A. Botteon}
\affil{INAF-IRA, via Gobetti 101, 40129 Bologna, Italy} 
\email{andrea.botteon@inaf.it}
\author[0000-0002-0587-1660]{R. J. van Weeren }
\affil{Leiden Observatory, Leiden University, PO Box 9513, 2300 RA Leiden, The Netherlands} 
\email{rvweeren@strw.leidenuniv.nl}
\author[0000-0002-9471-5423]{U. Lisenfeld}
\affil{Departamento de F\'{i}sica Te\'{o}rica y del Cosmos, Universidad de Granada, 18071 Granada, Spain}
\affil{Instituto Carlos I de F\'{i}sica Te\'{o}rica y Computacional, Facultad de Ciencias, 18071 Granada, Spain}
\email{ute@ugr.es}
\author[0009-0007-0318-2814]{J. M. Vrtilek}
\affil{Center for Astrophysics $|$ Harvard \& Smithsonian, 60 Garden Street, Cambridge, MA 02138, USA}
\email{jvrtilek@cfa.harvard.edu} 
\author{L. P. David}
\affil{Center for Astrophysics $|$ Harvard \& Smithsonian, 60 Garden Street, Cambridge, MA 02138, USA}
\email{ldavid@head.cfa.harvard.edu}
\author[0000-0002-1634-9886]{S. Giacintucci}
\affil{Naval Research Laboratory, 4555 Overlook Avenue Southwest, Code 7213, Washington, DC 20375, USA}
\email{simona.giacintucci.civ@us.navy.mil}

\author[0000-0002-9478-1682]{W. Forman}
\affil{Center for Astrophysics $|$ Harvard \& Smithsonian, 60 Garden Street, Cambridge, MA 02138, USA}
\email{wforman@cfa.harvard.edu}
\author[0000-0003-2206-4243]{C. Jones}
\affil{Center for Astrophysics $|$ Harvard \& Smithsonian, 60 Garden Street, Cambridge, MA 02138, USA}
\email{cjones@cfa.harvard.edu}
\author[0000-0002-1588-6700]{C. K. Xu}
\affil{Chinese Academy of Sciences South America Center for astronomy, National Astronomical Observatories, CAS, Beijing 100101, People's Republic of China}
\affil{National Astronomical Observatories, Chinese Academy of Sciences, 20A Datun Road, Chaoyang District, Beijing 100101, People's Republic of China}
\email{coxu@ipac.caltech.edu}
\author[0000-0001-5042-3421]{A. Togi}
\affil{Department of Physics, 601 University Drive, Texas State University San Marcos, TX 78666, USA}
\email{aditya.togi@txstate.edu}

\begin{abstract}
We present a detailed, multi-band radio continuum study of the compact galaxy group Stephan's Quintet (HCG~92). We use a combination of new (MeerKAT, uGMRT) and archival (LOFAR, VLA) observations covering the $\rm 120\,MHz{-}8\,GHz$ frequency range to examine the radio properties of the group, focusing on the famous radio ridge and surrounding diffuse emission. We find filamentary substructure and branching in the southern half of the ridge, confirm an extension of the ridge to the northwest, and identify for the first time a radio counterpart to the gas bridge linking the ridge and NGC~7319. The northern ridge, northwest extension and diffuse emission have relatively steep, curved spectra, with a high-frequency spectral index gradient running north-south along the ridge.  We show that the ridge emission primarily arises from a single physical mechanism, probably strong ($\mathcal{M}\simeq 40-100$) shocks in cold gas, caused by the $\sim$850-1000\kmps\ collision between NGC~7318B and tidal gas filaments produced by past galaxy interactions. Synchrotron spectral age estimates suggest the collision began at the north end of the ridge $\sim$20~Myr ago, and finished only $\sim$5-6~Myr ago in the south, with the southern end of the shocked ridge likely still within or close to the disk of NGC~7318B. Based on this age gradient, we find that the angle between the intruder galaxy's motion and the tidal filaments was probably only $\sim$15\degree, and combining this with the lack of a spectral index gradient in the diffuse radio emission suggests that NGC~7318B's direction of motion is probably close to the line of sight.
\end{abstract}

\keywords{Galaxy collisions (585); Galaxy interactions (600); Galaxy groups (597); Hickson compact group (729); Intergalactic medium (813); Intergalactic filaments (811); Radio continuum emission (1340); Shocks (2086)}

\section{Introduction}

\begin{figure*}[!thbp]
\begin{center}
    \includegraphics[width=0.90\textwidth]{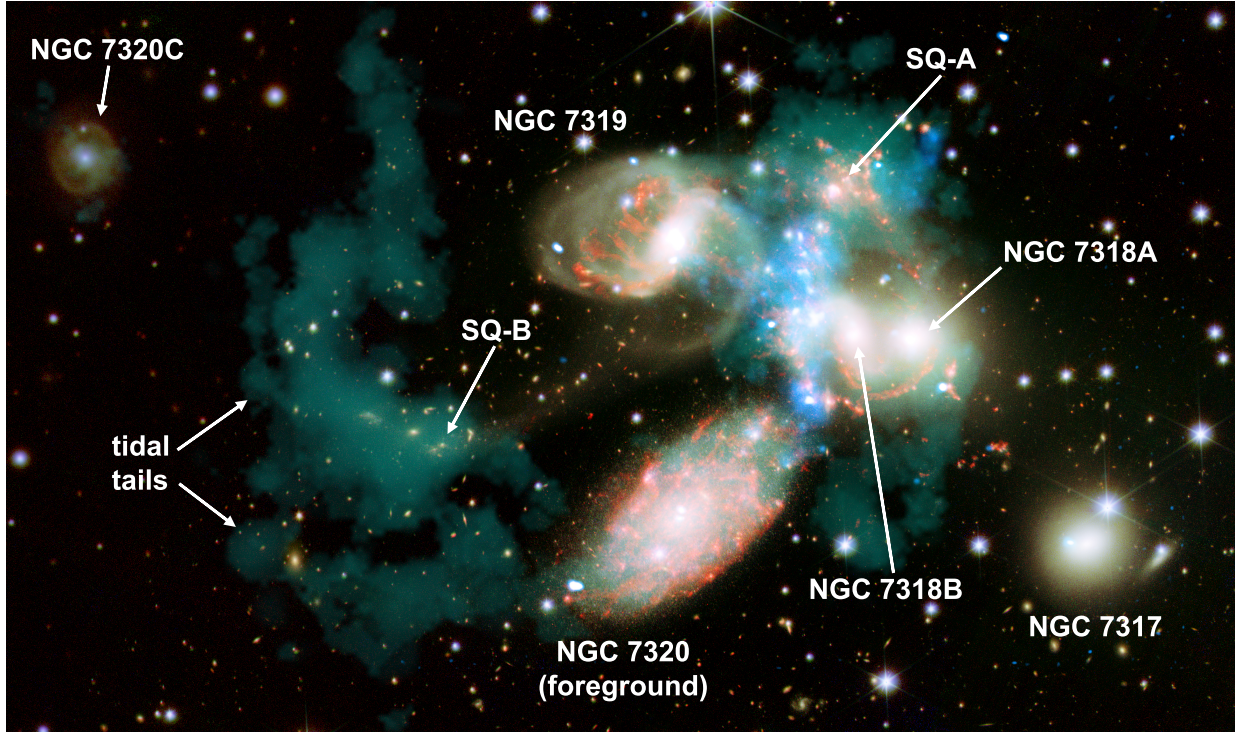}
\end{center}
\vspace{-0.5cm}
\caption{Overview of Stephan's Quintet (HCG~92). The galaxies and stellar structures are shown using a combination of PanSTARRS optical and \jwst/NIRCAM near-infrared imaging, with \jwst/MIRI 7.7 and 10~$\mu$m imaging overlaid in red to highlight dust and molecular hydrogen emission. X-ray emission (\chandra\ 0.3-2~keV) from the shock ridge is shown in blue. \Hi\ emission from cold neutral hydrogen (MeerKAT 5400-7400\kmps) is shown in pale green. Galaxies, the tidal tails, and the star formation regions SQ-A and SQ-B are labeled.}
\label{fig::overview}
\end{figure*} 

Galaxy groups, and particularly compact groups, bring galaxies into close physical proximity at low relative velocities, promoting tidal interactions and mergers. This makes galaxy groups an ideal laboratory in which to study the processes by which cold gas rich, star-forming spirals evolve into gas poor, quiescent elliptical and lenticular galaxies. These processes can include tidal stripping of cold gas \citep[e.g.,][]{Verdes-Montenegro2001,Borthakur2010,Konstantopoulos2010,Ianjamasimananaetal25}, viscous and ram-pressure stripping \citep[e.g.,][]{Rasmussenetal06a,Rasmussenetal08}, ejection of gas from galaxies by starburst winds \citep[e.g.,][]{OSullivanetal14b,OSullivanetal14c}, and collisional shocks \citep[e.g.,][]{Appletonetal15,Joshietal19,OSullivanetal25}. Perhaps the best-known example of the latter process is Stephan's Quintet \citep{Stephan1877}, also known as HCG~92 \citep{Hickson82}, which contains a $\sim$50~kpc long S-shaped ridge of radio and X-ray emission which appears to be the result of a $\sim$900\kmps\ collision between a galaxy and a filament of tidally-stripped cold gas.

In its modern definition, the group consists of five principal galaxies, four of which (NGC~7317, NGC~7318A, NGC~7319 and NGC~7320C) have recession velocities in the range $\sim$6550-6750\kmps. The fifth galaxy, NGC~7318B, is an intruder with a (blueshifted) relative velocity of $\sim$900\kmps\ compared to the other members, despite overlapping (in projection) its neighbor NGC~7318A. The original definition of the group included NGC~7320, whose redshift shows it to be an unrelated foreground spiral. Several group members show signs of past or ongoing interactions, including tidal tails, arms, and optical filaments (see Figure~\ref{fig::overview}).

Early radio interferometric observations of the group \citep{Allen1972} revealed a ridge of continuum emission running roughly north-south along the east side of the intruder \citep[see also][]{vanderHulst1981} and it was fairly quickly recognized that this emission likely arose from collisional shocks \citep{Shostak1984}. Later observations confirmed this picture, showing that the radio continuum ridge links \Hi\ structures at the group velocity \citep[e.g.,][]{Williams2002} and has an integrated spectral index\footnote{We define the radio spectral index $\alpha$ as $S_{\nu}\propto \nu^{\alpha}$ where $S_\nu$ is the flux at frequency $\nu$.} indicative of non-thermal emission \citep[$\rm \alpha^{4.8~GHz}_{1.4~GHz}=-0.93$;][]{Xu2003}. Deeper radio imaging revealed even steeper-spectrum low surface-brightness emission extending up to $\sim$1\arcm\ ($\sim$27~kpc) around the ridge \citep{NikielWroczynski2013,Arnaudova2024}. Both the ridge and extended emission are only weakly polarized \citep{NikielWroczynski2013} though, uniquely, there is evidence of a large-scale ordered magnetic field associated with the group \citep{NikielWroczynski2020}. The radio picture of the group is complicated by background sources \citep[e.g., SQ-R,][]{Xu2003}, contributions from the Active Galactic Nuclei (AGNs) of the member galaxies \citep[notably the Seyfert nucleus and jets of NGC~7319,][]{Aoki1999,Xanthopoulos2004}, and star formation (SF). As well as emission from the foreground galaxy NGC~7320, the group contains SF regions associated with tidal structures, most notably the collision-triggered starburst SQ-A \citep{Xuetal99,Xuetal25} and the tidal dwarf galaxy candidate SQ-B \citep{Xuetal99,Lisenfeldetal04} in the younger, northern optical tidal tail.

\begin{deluxetable*}{lclccccc}
\tablewidth{0pt}
\tablecaption{\label{tab:obs}Summary of the radio observations}
\tablehead{
\colhead{Telescope} & \colhead{Project} & \colhead{Observation} & \colhead{Frequency Band} & \colhead{Array} & \colhead{On-source time} & \colhead{Channel width} & \colhead{Total channels} \\
 & & \colhead{Date} & \colhead{(GHz)} &  & \colhead{(hr)} &  & 
}
\startdata
\multirow{2}{*}{MeerKAT} & \multirow{2}{*}{SCI-20241101} & 2024 December 19 & \multirow{2}{*}{$0.90{-}1.70$} & - &3.0 &\multirow{2}{*}{21.6\,kHz} & \multirow{2}{*}{32768}\\
 & & 2024 December 28 & (L-band) & - &3.0 & & \\
\hline
\multirow{2}{*}{LOFAR} & \multirow{2}{*}{LC8\_014} & 2017 August 21 & \multirow{2}{*}{$0.120{-}0.168$} & - &3.8 &\multirow{2}{*}{781.2\,kHz} & \multirow{2}{*}{64}\\
 & & 2017 November 11 & (HBA) & - &3.8 & & \\
\hline
\multirow{2}{*}{uGMRT}   & \multirow{2}{*}{47$\_$057}  & 2024 December 14     & $0.30{-}0.50$  & - &7.5 & \multirow{2}{*}{97.7\,kHz}& \multirow{2}{*}{4096}\\
 & & 2024 December 15     & $0.55{-}0.85$  & - & 7.5& & \\
\hline
\multirow{10}{*}{VLA} & \multirow{10}{*}{21A$-$334} & 2021 September 15 & \multirow{4}{*}{$2{-}4$} & CnB &0.33 & \multirow{4}{*}{2\,MHz}&\multirow{4}{*}{64} \\
        &  & 2022 August 31 &  & D & 0.33& & \\
         & & 2022 September 03 & (S-band)  & D & 0.33& & \\
        & & 2022 September 05 & & D & 0.33& & \\
        \cline{3-8}
        &  & 2021 August 10 & \multirow{6}{*}{$4{-}8$} & C &0.33 & \multirow{6}{*}{27\,MHz}&\multirow{6}{*}{64} \\
        & & 2022 August 04 &  & C & 0.33&  & \\
        & & 2022 July 22 & & C & 0.33& & \\
        & & 2022 August 31 & (C-band)  & D &0.33 & & \\        
        & & 2022 August 28 &  & D &0.33 & & \\
        & & 2022 August 24 &  & D &0.33& & \\
\enddata
\tablecomments{ VLA S-band (2-4~GHz) observations have a total of 1024 channels per array, while the C-band (4-8~GHz) has 2048 channels per array. The VLA observations have 16 spectral windows per array at S-band and 34 at C-band}
\end{deluxetable*}

Stephan's Quintet has an exceptionally rich multi-wavelength dataset, which reveals an extraordinary range of emission from the shock region. This includes hot ($\sim$0.6~keV) X-ray emitting plasma \citep{Trinchieri2003,Trinchierietal05,OSullivan2009}; ionized gas emitting in, among other lines, H$\alpha$ \citep[e.g.,][]{DuartePuertas2019,Arnaudova2024} [C\textsc{ii}] \citep{Appleton2013}, H$\beta$ and Ly$\alpha$ \citep{Guillardetal22}; neutral hydrogen \citep{Xu2022,Cheng2023,paperI}; and molecular gas emitting via CO \citep{Maedaetal2025,Emonts2025}, H$_2$ \citep{Appleton2023} and other lines including H$_2$O \citep{Appleton2013}. Spatial correlation of gas across these different phases, decreasing shock velocities in cooler components, and a remarkable similarity in emission luminosity over four orders of magnitude in temperature led to the suggestion of a turbulent cascade helping to dissipate the energy injected by the collision \citep{Guillard2009,Guillardetal22}.

\begin{deluxetable*}{c c c r c c c r}[!thbp]
\tablecaption{Imaging properties of radio maps used in the analysis}
\tablehead{& Name & Restoring Beam & Robust  & \textit{uv}-cut & \textit{uv}-taper & RMS noise\\ 
&&& parameter &&&$\upmu\rm Jy\,beam^{-1}$}
\startdata
\multirow{2}{3cm}{LOFAR HBA (120--169\,MHz)} &IM1&$8\arcsec \times 8\arcsec$&$-0.5$&$ \geq\rm0.2\,k\uplambda$&$-$&110\\
  &IM2&$15\arcsec \times 15\arcsec$&$-0.5$&$ \geq\rm0.2\,k\uplambda$&$5\arcsec$&180\\
\hline
 \multirow{3}{3cm}{uGMRT Band\,3 (300--500\,MHz)} &IM3&$8\arcsec \times 8\arcsec$& $-0.5$&$\geq\rm0.2\,k\uplambda$&$-$&30\\
  &IM4&$15\arcsec \times 15\arcsec$&$0.0$&$ -$&7\arcsec&40\\
  &IM5&$15\arcsec \times 15\arcsec$&$-0.5$&$ \geq\rm0.2\,k\uplambda$&7\arcsec&46\\ 
 \hline   
\multirow{2}{3cm}{uGMRT Band\,4 (550--850\,MHz)} &IM6&$8\arcsec \times 8\arcsec$&$-0.5$&$\geq\rm0.2\,k\uplambda$&$-$&20\\
  &IM7&$15\arcsec \times 15\arcsec$&$0.0$&$ -$&7\arcsec&40 \\
\hline     
\multirow{4}{3cm}{MeerKAT L-band (0.9--1.7\,GHz)} &IM8&$8\arcsec \times 8\arcsec$&$-0.5$ &$\geq\rm0.2\,k\uplambda$&$-$&12\\
  &IM9&$15.5\arcsec \times 14\arcsec$&$0.0$&$-$&7\arcsec& 10 \\
  &IM10&$15\arcsec \times 15\arcsec$&$-0.5$&$ \geq\rm0.2\,k\uplambda$&7\arcsec& 15\\
  \hline 
\multirow{3}{3cm}{VLA S-band (2--4\,GHz)}   &IM11&$8\arcsec \times 8\arcsec$&$0$ &$-$&5\arcsec&8\\
&IM12&$8\arcsec \times 8\arcsec$&$-0.5$ &$\geq\rm0.2\,k\uplambda$&$-$&10\\
  &IM13&$15\arcsec \times 15\arcsec$&$0$&$-$&10\arcsec&10\\
  &IM14&$15\arcsec \times 15\arcsec$&$-0.5$&$ \geq\rm0.2\,k\uplambda$&10\arcsec& 12\\  
  \hline 
\multirow{3}{3cm}{VLA C-band (4--8\,GHz)} &IM15&$8\arcsec \times 8\arcsec$&$0$ &$-$&5\arcsec&7\\
&IM16&$8\arcsec \times 8\arcsec$&$-0.5$ &$\geq\rm0.2\,k\uplambda$&$-$&8\\
  &IM17&$15\arcsec \times 15\arcsec$&$-0.5$&$ \geq\rm0.2\,k\uplambda$&10\arcsec& 10\\
\enddata
\tablecomments{Final imaging was performed in \texttt{WSCLEAN} using {\tt multiscale} and with the {\tt Briggs} weighting scheme.}
\label{tab:imaging}
\end{deluxetable*}

Despite this wealth of data available for the group, questions remain about its interaction history, the direction of motion and timescale of the collision between NGC~7318B and the tidal filament, and the mechanisms responsible for the radio emission in the shock ridge. In \citet[hereafter paper~I]{paperI} we used MeerKAT observations to map \Hi\ in the group, detecting neutral hydrogen in NGC~7319 and NGC~7320C for the first time and providing a detailed picture of the tidal structures (see also Figure~\ref{fig::overview}). In this work, we attempt to address some of the issues noted above, using new wideband radio observations including the MeerKAT observations used in paper~I and data from the upgraded Giant Metrewave Radio Telescope (uGMRT). In addition, we make use of archival data from the Karl G. Jansky Very Large Array (VLA) and public Low-Frequency Array (LOFAR) observations.

Throughout the paper, we assume a redshift $z$=0.0215 and a distance to Stephan's Quintet (hereafter \SQ) of 94~Mpc, to facilitate comparison with other recent studies \citep{Appleton2023,Emonts2025,Aromaletal25,Xuetal25}. At this distance, 1\arcs\ corresponds to 456~pc. All radio images are in the J2000 coordinate system and corrected for primary beam attenuation.

The paper is structured as follows. In Section~\ref{sec::observtaions}, we present our new MeerKAT and uGMRT observations and data reduction steps. The new images are discussed in Section~\ref{sec::results} followed by analysis in Section~\ref{sec::analysis}. We discuss potential mechanisms responsible for the radio ridge in Section~\ref{sec::origin} and the implications of our results for the dynamics of the group in Section~\ref{sec::dynamics}, summarizing our findings in Section~\ref{sec::summary}.

\section{Observations and Data Reduction}
\label{sec::observtaions}
We analyzed new MeerKAT (0.9-1.7~GHz) and uGMRT observations of \SQ, along with the archival LOFAR (120-168~MHz) and VLA S (2-4~GHz) and C-bands (4-8~GHz) data. Details of these observations are shown in Table~\ref{tab:obs}.

\subsection{MeerKAT}
We observed \SQ\ during two separate runs with MeerKAT at L-band covering a frequency range 0.9-1.7~GHz, for a total on-source time of 6~hr. Each Meerkat observing run was calibrated using the Containerized Automated Radio Astronomy Calibration pipeline \citep[{\tt CARACal};][]{caracal2020}\footnote{\url{https://ascl.net/2006.014}}. The initial step in {\tt CARACal} involved flagging shadowed antennas, autocorrelations, and known radio frequency interference (RFI) channels using the {\tt tfcrop} algorithm. Subsequently, {\tt AOflagger} \citep{Offringa2010} was employed to identify and remove additional RFI-contaminated data. The primary calibrator (J0408-6545) was modeled using the MeerKAT Local Sky Models within {\tt CARACal}. Following this, cross-calibration was conducted to solve for time-dependent delays, complex antenna gains, and bandpass corrections.

After calibration, we created an initial image of the target field by combining both the data sets. The imaging was performed in {\tt WSClean} \citep{Offringa2014} within {\tt CARACal}. Four rounds of phase only self-calibration were performed using {\tt CubiCal} \citep{Kenyon2018}, followed by a final round of amplitude-phase calibration. The calibrated data were imaged using the Briggs weighting scheme \citep{Briggs1995} with a robust parameter of $0$, and multiscale cleaning.  The properties of the MeerKAT continuum images are summarized in Table~\ref{tab:imaging}.

\begin{figure*}[!thbp]
    \centering
    \includegraphics[width=1.0\textwidth]{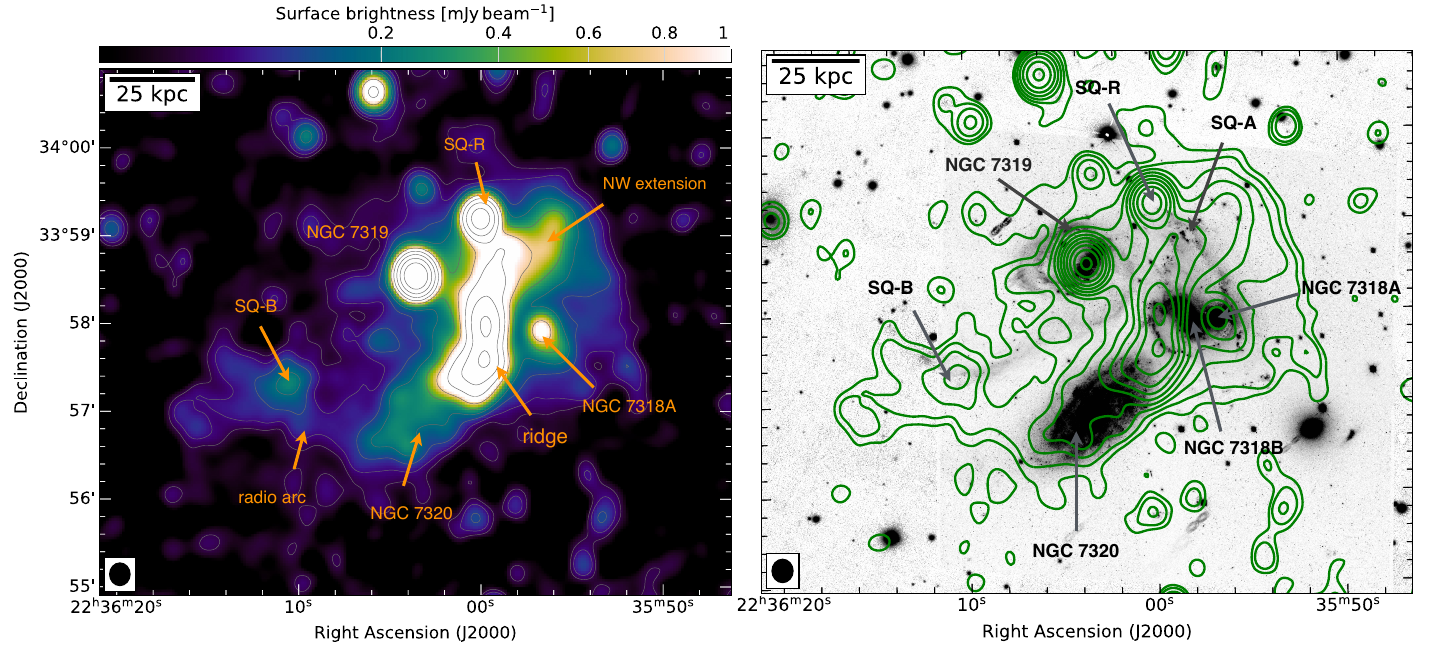}
 \caption{\textit{Left:} MeerKAT L-band (0.9-1.7\,GHz) image of \SQ\ with an angular resolution 15.5\arcs\ $\times$ 14.0\arcs, highlighting the newly detected low surface brightness emission around the ridge, radio arc and NW extension. The image is displayed using a square-root intensity scale. \textit{Right}: The combined Hubble Space Telescope (HST) and Digitized Sky Survey (DSS) optical image is overlaid with a MeerKAT L-band (0.9-1.7~GHz) continuum contours, showing the overview of the galaxies in \SQ. In both the maps, the radio contour levels are drawn at  $[1, 2, 4, 8 ...]\times 3.0\sigma_{\rm rms}$. The radio beam size is indicated in the bottom left corner of each image. For the radio image properties, see Table\,\ref{tab:imaging} IM9.}
      \label{fig::radio-optical}
\end{figure*}

\subsection{uGMRT}
We also observed \SQ\ with uGMRT in Band\,3 and Band\,4 covering the frequency ranges 300-500~ MHz and 550-850~MHz, respectively. The data reduction was carried out using the SPAM (Source Peeling and Atmospheric Modeling) pipeline \citep{Intema2009}. Each of the two wideband datasets was first divided into six sub-bands. We adopted the flux scale of the primary calibrator 3C\,48 \citep{Scaife2012}. Standard processing steps including averaging, flagging, and bandpass correction were applied. Phase calibration was performed using a global sky model derived from GMRT narrow-band data. Each sub-band was subsequently imaged with the {\tt WSClean} software using multiscale cleaning to generate deep continuum images. Imaging parameters are again shown in Table~\ref{tab:imaging}.

\subsection{VLA}
We analyzed archival, previously unpublished VLA S-band (2-4\,GHz) and C-band (4-8\,GHz) observations of Stephan's Quintet. The S-band data were obtained in the D and CnB configurations, while the C-band observations were carried out in the C and D arrays. A summary of the observational details is provided in Table\,\ref{tab:obs}. Primary calibrators for the S-band observations were either 3C147 or 3C48, while 3C147 or 3C138 were used for the C-band. The phase calibrator for all observations was J2236+2828.

The VLA C and S band data were calibrated with the Common Astronomy Software Applications \citep[\texttt{CASA};][]{McMullin2007,casa2022} package. Data obtained from different observing runs were calibrated separately but in the same manner. The initial data reduction steps included Hanning smoothing and RFI flagging using the \texttt{tfcrop} mode within the \texttt{flagdata} task. Additionally, low-amplitude RFI was flagged using {\tt AOFlagger}. Following flagging, we determined and applied elevation-dependent gain  and antenna offset corrections. To prevent flagging of good data due to bandpass roll-off at the edges of the spectral windows, we corrected for bandpass using the primary calibrator. 

We used \texttt{CASA} 3C48, 3C147, and 3C138 models for the S- and C-bands and set the flux density scale according to \cite{Perley2013}.  An initial phase calibration was performed using the primary calibrator over a few channels per spectral window. After this we corrected the antenna delays and determined the bandpass response. Applying the bandpass and delay solutions, we proceeded with the gain calibration. All solutions were applied to the target field.  

To create an initial image of the target field for each individual dataset, we used \texttt{WSClean} with Briggs weighting and a robust parameter of 0. Following the initial imaging, several rounds of self-calibration (phase and amp-phase) were performed to refine the calibration. After self-calibration, the uv-data from all configurations in each band were imaged together to produce deeper S and C band continuum images. We also produced a combined S plus C band (2-8~GHz) continuum image. The combined SC-band image improved the sensitivity provided by the larger bandwidth and provides higher angular resolution and recovers more extended emission than the C-band image alone.

\begin{figure*}[!thbp]
    \centering
    \includegraphics[width=1.00\textwidth]{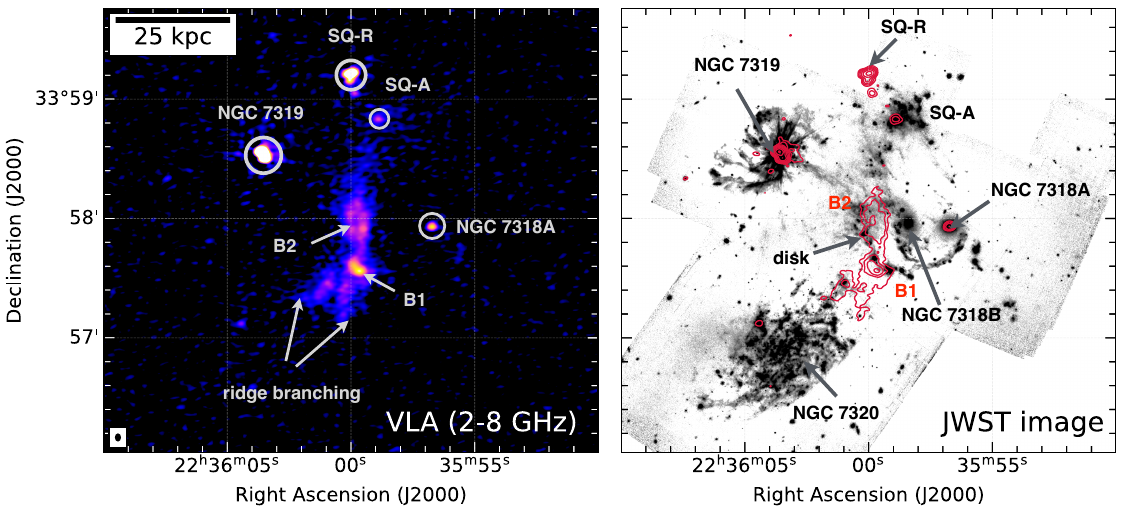}
 \caption{\textit{Left}: High resolution ($3.4\arcsec\times2.3\arcsec$) VLA combined SC-bands (2-8~GHz) image, revealing substructures in the ridge region. The image was created using Briggs weighting with {$\tt \rm robust=0$ and without any \textit{uv}-cut or tapering. \textit{Right}: \jwst\ MIRI 10$~\mu$m image with VLA 5~GHz radio contours overlaid. The image reveals that the brighter part of the ridge overlaps the southeast spiral arm of NGC 7318B. Radio contour levels are drawn at  $[1, 2, 4, 8 ...]\times 3.5\sigma_{\rm rms}$ ($\rm rms=4~\mu Jy\,beam^{-1}$) and are from the VLA SC-band combined image.}}
      \label{fig::radio-optical_high_res}
      \vspace{0.5cm}
\end{figure*}

\subsection{LOFAR}
\SQ\ was observed by the LOFAR Two-metre Sky Survey \citep[LoTSS;][]{Shimwell2019, Shimwell2022, Shimwell2026} at 144~MHz. The group is located at $\sim$47.5\arcm\ from the pointing center P339+33, which we adopt for the subsequent analysis. This pointing was observed in August and November 2017 under project code LC$8\_014$. Each observation lasted 3.8~hours and used 3C48 as primary calibrator. We reprocessed the survey data using the ``extract+self-calibration'' method (i.e., also called \texttt{facetselfcal}) described in \cite{vanWeeren2021}, aimed at improving the calibration in the direction of the target and providing greater flexibility in the re-imaging of a small portion of the field containing only the target of interest. In the case of \SQ, all sources outside a region of 0.29 degree were removed from the visibilities by adopting the models derived with the {\tt ddf-pipeline}, used by the LOFAR Surveys Key Science Project team to process LoTSS observations \cite[see, e.g.,][]{Tasse2021}. We found that the observation conducted in August 2017 was of bad quality and therefore we opted to analyze only the observation performed in November of the same year.

\subsection{Flux density scale}

The flux density scale for all observations (LOFAR, uGMRT, and VLA) was validated by comparing the spectra of compact sources across the 144~MHz to 6~GHz frequency range. The uncertainty in the flux density measurements was estimated as follows:
\begin{equation}
\Delta S =  \sqrt {(f \cdot S)^{2}+{N}_{{\rm{ beams}}}\ (\sigma_{{\rm{rms}}})^{2}},
\end{equation}
where $f$ is the absolute flux density calibration uncertainty, $S$ is the flux density, $\sigma_{{\rm{ rms}}}$ is the noise level, and $N_{{\rm{beams}}}$ is the number of beams. We assumed absolute flux density uncertainties of 10\% for LOFAR \citep{Shimwell2026} and uGMRT \citep{Chandra2017},  5\% for MeerKAT L-band, and 3\% for the VLA C and S-bands data \citep{Perley2013}.

\section{Results: Radio continuum morphology}
\label{sec::results}
Our high-sensitivity wideband radio data allow us to study the continuum emission from \SQ\ in great detail across a broad frequency range, spanning from 120~MHz to 8~GHz. Figure~\ref{fig::radio-optical} presents our new MeerKAT L-band (0.9-1.7~GHz) image at $15.5\arcsec\times14\arcsec$ resolution, with the various components labeled. The group hosts the main shock ridge as well as several other distinct continuum sources associated with star formation and AGNs, surrounded by diffuse radio emission. We discuss the main structures below. 

\subsection{Shock Ridge}
As shown in Figure~\ref{fig::radio-optical}, the most prominent extended source is the ridge (shock). It has the largest linear size (LLS) of 45\,kpc and is surrounded by diffuse radio emission.  In Figure\,\ref{fig::radio-optical_high_res} left panel, we present the highest resolution combined C- and S-bands VLA image ($3.4\arcsec\times2.4\arcsec$) at a central frequency of 5~GHz. The image reveals that the ridge contains significant substructures. In particular, the southern end of the ridge forks into two branches. Just north of these is the brightest clump of radio emission in the ridge (labeled B1) and above this a larger region of bright emission (B2) apparently dominated by two parallel filaments. The northern part of the ridge is significantly fainter at high frequencies. 

Figure\,\ref{fig::radio-optical_high_res} right panel provides a comparison of the high-frequency radio and infrared (IR) emission. Clump B1 has no clear IR counterpart but is located adjacent to bright IR emission in the southeast spiral arm of NGC\,7319B. Region B2 overlaps this spiral arm, though much of the emission is located west of the arm, inside its curve.

\begin{figure*}[!thbp]
    \centering
    \includegraphics[width=1.0\textwidth]{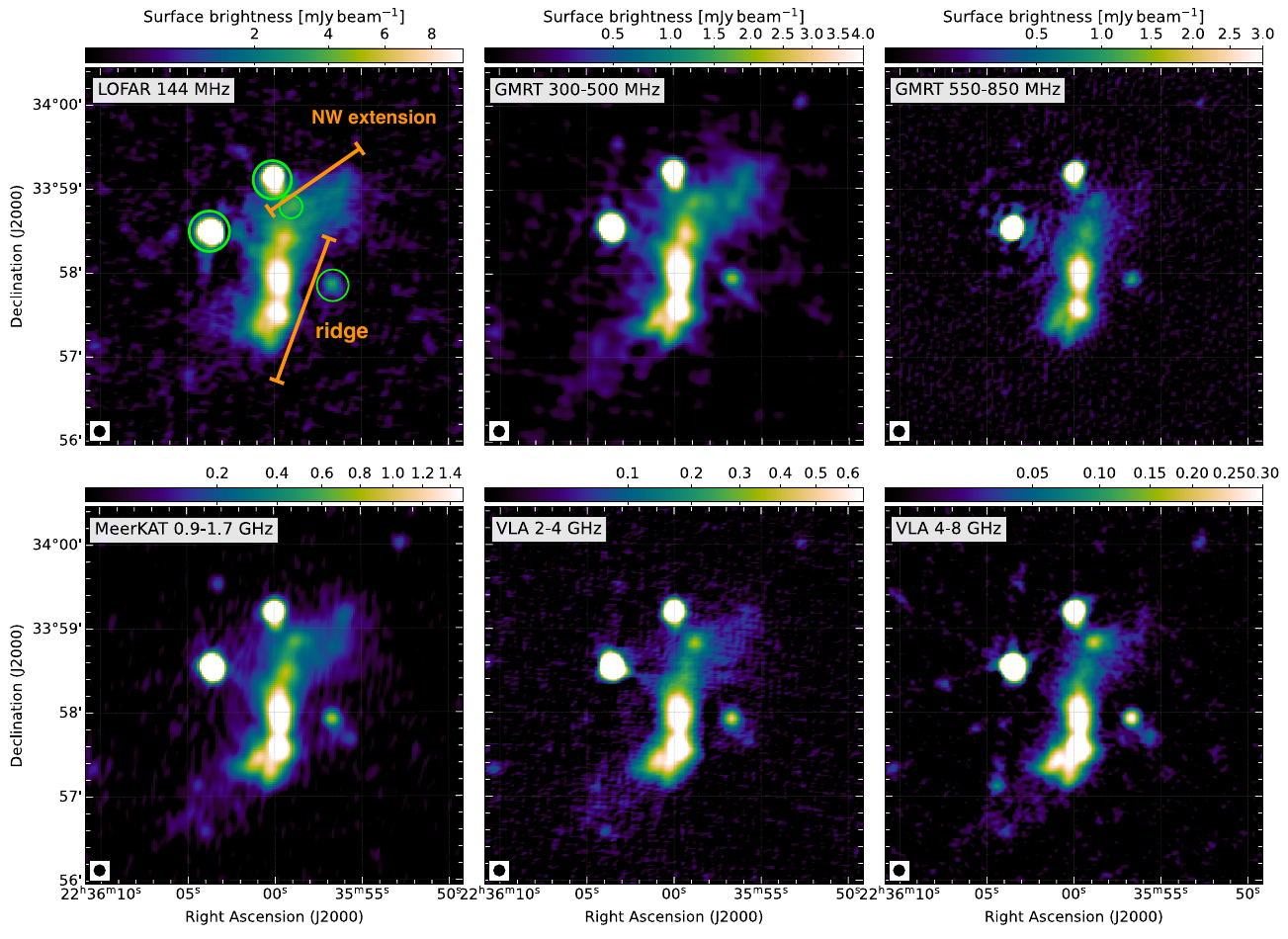}
 \caption{High resolution images of \SQ\ seen from LOFAR (144~MHz), uGMRT (350-850~MHz), MeerKAT (0.9-1.7~GHz)  and VLA (2-8~GHz). All images have a common resolution of 8\arcsec. The images are in a square-root intensity scale. Embedded discrete sources are marked with green circles in the top left image. The radio beam size is indicated in the bottom left corner of each image. For the images properties, see Table\,\ref{tab:imaging} IM1, IM3, IM6, IM8, IM11, and IM15.}
      \label{fig::continumm_high}
\end{figure*}

\begin{figure*}[!thbp]
    \centering
    \includegraphics[width=1.0\textwidth]{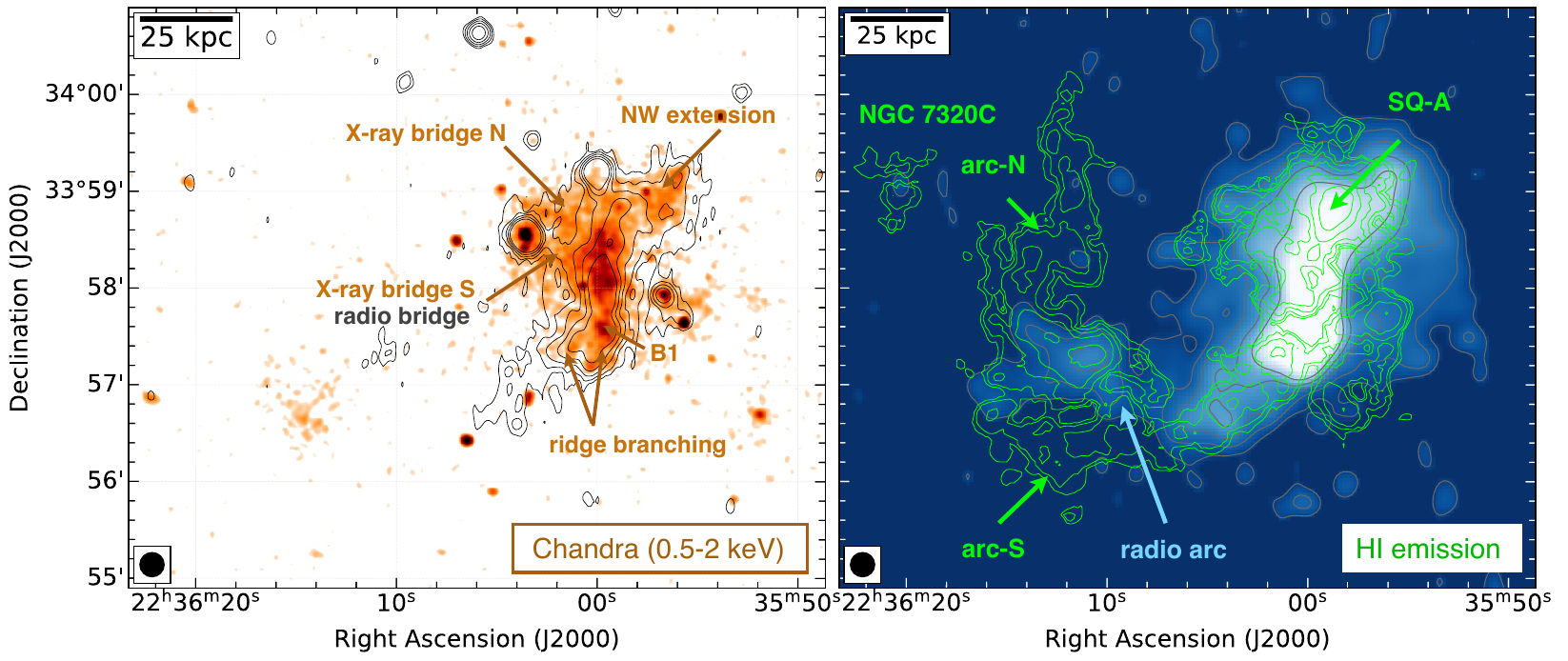}
\caption{\textit{Left}: \chandra\ X-ray image Gaussian smoothed with a FWHM of 3\arcsec\ and overlaid with MeerKAT 0.9-1.7~GHz (8\arcs\ resolution) radio contours. The X-ray and radio morphologies are remarkably similar across the shock and NW extension. The MeerKAT contours reveal a radio bridge connecting the radio/X-ray ridge to NGC~7319. The continuum contour levels are drawn at  $[1, 2, 4, 8 ...]\times 3.5\sigma_{\rm rms}$. \textit{Right}: MeerKAT 0.9-1.7~GHz point source subtracted continuum image at 15\arcsec\ resolution overlaid with MeerKAT \Hi\ emission contours. The \Hi\ image beam size is $17\arcsec\times14\arcsec$ and the contours are drawn at $5.6, 11.1, 22.3, 44.5, 89.1, 178 \times 10^{19}\rm \,atoms\,cm^{-2}$. For the left panel continuum image properties, see Table\,\ref{tab:imaging} IM8.}
      \label{fig::Xray-radio-HI}
\end{figure*}

\begin{figure*}[!thbp]
    \centering
    \includegraphics[width=0.95\textwidth]{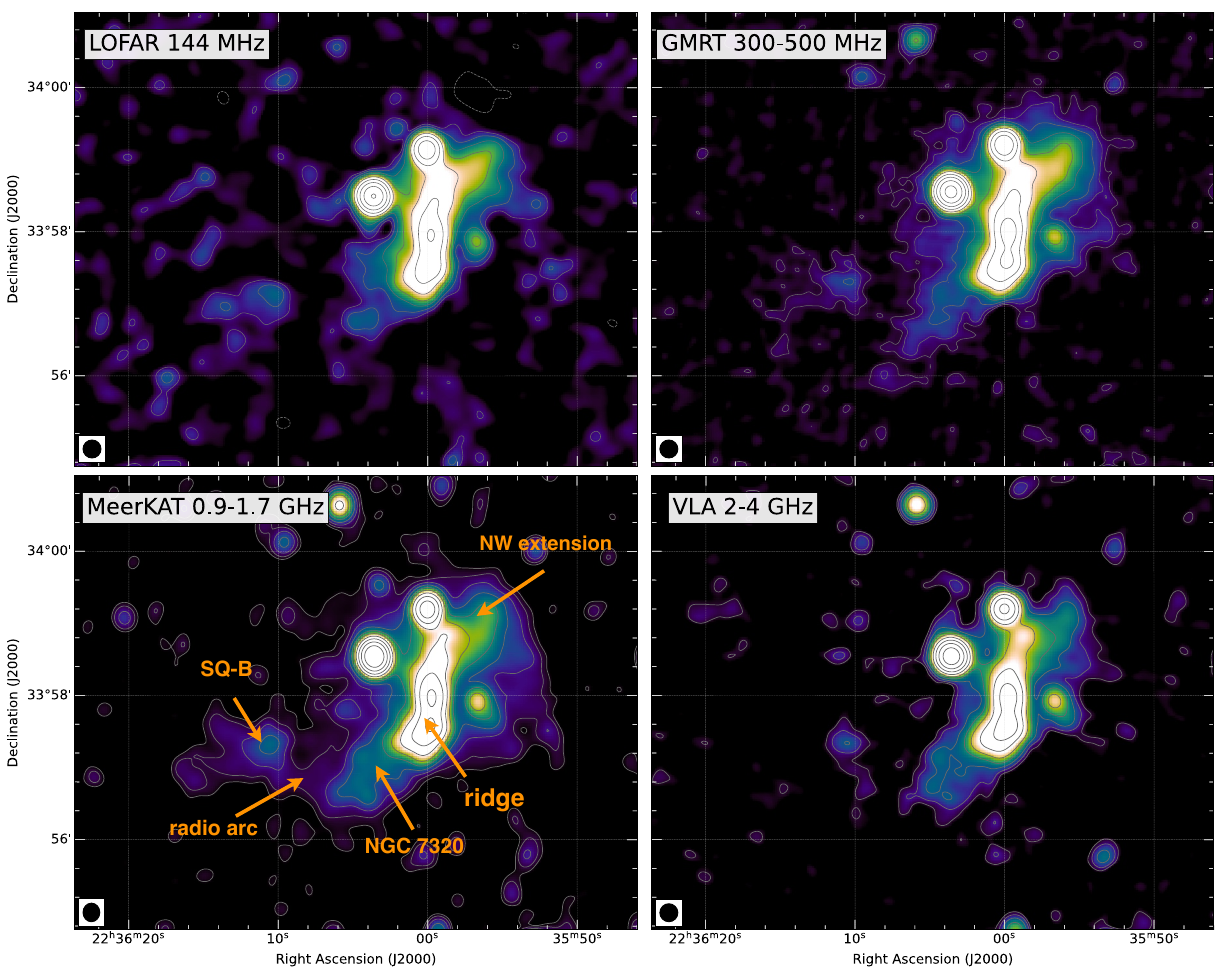}
 \caption{Medium resolution (15\arcsec) images of \SQ\ seen from LOFAR, uGMRT Band3, MeerKAT, and VLA S-band. The radio surface brightness is in units of $\rm mJy\ beam^{-1}$. Radio contour levels are drawn at  $[1, 2, 4, 8 ...]\times 3.0\sigma_{\rm rms}$. Dashed contours depict the $-3.0\sigma_{\rm rms}$ contours. The radio beam size is indicated in the bottom left corner of each image. For the images properties, see Table\,\ref{tab:imaging} IM2, IM4, IM9, and IM13.}
      \label{fig::low_res}
\end{figure*} 

High resolution images of the ridge are shown in Figure\,\ref{fig::continumm_high}, covering the frequency range from 144\,MHz to 8\,GHz with a common resolution of 8\arcsec. The radio emission from the ridge dominates the group at both GHz and MHz frequencies. The overall morphology of the ridge is similar across all the observed frequencies and is consistent with earlier studies \citep[e.g.,][]{Allen1972, Williams2002,Xu2003, NikielWroczynski2013, Arnaudova2024}. While the size of structures such as B2 remains similar with frequency, the overall ridge width changes significantly, from 8\,kpc at 6\,GHz to 18~kpc at 144~MHz.

Figure~\ref{fig::Xray-radio-HI} left panel shows the \textit{Chandra} image overlaid with the MeerKAT 8\arcs\ image radio contours. While the southern part of the ridge is brighter than the northern part in the radio continuum, the X-ray emission shows the opposite trend. Remarkably, the X-ray image also reveals a branching structure at the southern end of the ridge, closely following the radio morphology. As reported by \citet{OSullivan2009}, clump B1 shows the brightest X-ray emission within the ridge.

In Figure\,\ref{fig::Xray-radio-HI} right panel, we present a point source subtracted 15\arcs\ resolution L-band continuum image, overlaid with MeerKAT neutral atomic hydrogen contours. While we do not observe a one-to-one correspondence between the \Hi~ and radio continuum emission morphologies, there are some interesting correlations, particularly around SQ-A, the northwest extension, and in the tidal tail near SQ-B. The comparison is complicated given the multiple velocity structures in the \Hi\ (paper I). However, it is clear that although there is a link between the cold gas and star formation regions, the diffuse radio emission extends well beyond the limits of the known \Hi\ structures.

\subsection{NW extension}

The radio emission from the ridge extends toward the northwest of SQ-A, labeled as the ``NW extension'' in Figure~\ref{fig::radio-optical}. This structure \citep[identified in previous VLA observations,][]{Xu2003} is detected at all the observed frequencies (except that it is not visible in the high resolution uGMRT Band\,4 550-850~MHz image). The radio surface brightness of the NW extension is lower than that of the ridge by a factor of six. From the morphology alone, it is unclear whether this emission represents an extension of the shock ridge region or a separate structure.

Interestingly, the \textit{Chandra} image (Figure\,\ref{fig::Xray-radio-HI} left panel) also reveals X-ray emission at the same location, with a broadly similar morphology, indicating a possible physical connection. However, in the X-ray a band of fainter emission separates the NW extension from the main ridge. Part of the emission from the NW extension and the ridge is also detected in \Hi, see Figure\,\ref{fig::Xray-radio-HI} right panel. Overall, compared to the X-ray and \Hi~emission, the radio continuum emission is more extended toward the west and east.

\subsection{Radio bridge}

Our MeerKAT L-band (0.9-1.7~GHz) images reveal a radio bridge linking the nucleus of NGC~7319 to the ridge, running roughly southwest-northeast (see Figure\,\ref{fig::Xray-radio-HI} left panel). The radio bridge is also visible in the 8\arcsec~ resolution uGMRT Band3 (300-500~MHz) and S-band maps (see Figure\,\ref{fig::continumm_high}). As shown in Figure\,\ref{fig::Xray-radio-HI}, a counterpart X-ray filament (labeled as X-ray bridge S) is seen in the \chandra\ data. Gas in this southern bridge has previously been detected via a number of emission lines, including H$\alpha$, H$_2$ and CO \citep[e.g.,][]{Sulentic2001,Cluveretal10,Guillard2012, Emonts2025}. MeerKAT revealed that the southern bridge contains \Hi\ (paper~I) and confirmed a strong velocity gradient  previously seen in CO observations \citep{Emonts2025}. 

A second X-ray bridge connects the northern edge of the ridge to NGC~7319 (labeled "X-ray bridge N"). We do observe faint radio continuum emission filling the region between NGC~7319 and the northern end of the ridge (see Figure\,\ref{fig::Xray-radio-HI} left panel), but no clear morphological correspondence between the diffuse radio emission and filamentary X-ray bridge.

\begin{figure*}[!thbp]
    \centering
    \includegraphics[width=0.49\textwidth]{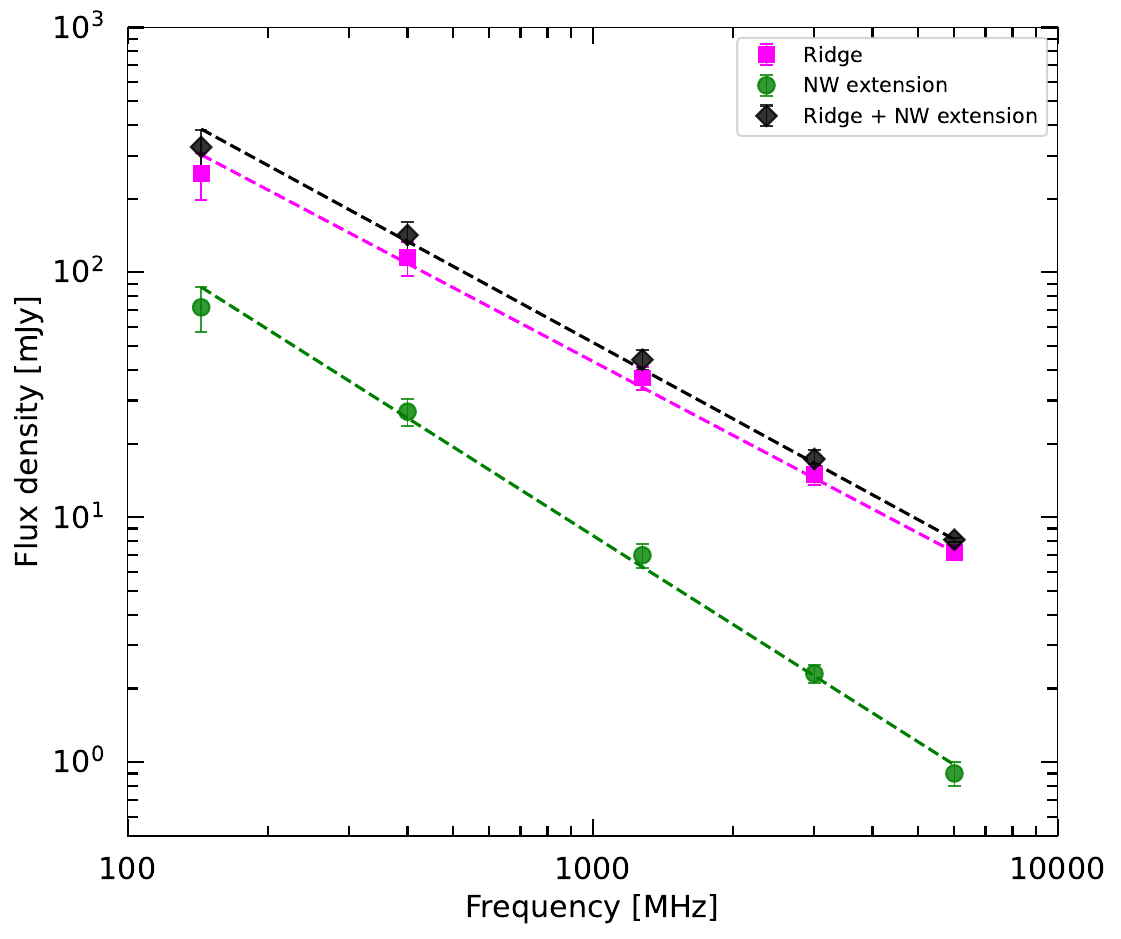}
     \includegraphics[width=0.46\textwidth]{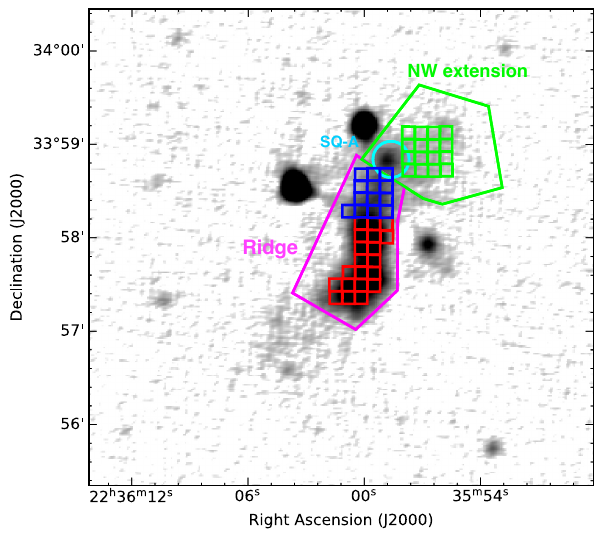}
 \caption{\textit{Left:} Total integrated spectra of the ridge and the NW extension (between 144~MHz and 6~ GHz). Both regions spectra can be explained by a power law fit (dashed lines). \textit{Right:} Regions, overlaid on S-band continuum image, used to extract flux density from radio continuum maps for the integrated spectra (cyan polygons) and radio color-color plots (red, blue and green squares).}
      \label{fig::spectrum}
\end{figure*}

\subsection{Faint diffuse radio emission}
In Figure\,\ref{fig::low_res}, we present 15\arcsec\ resolution images obtained with MeerKAT at 1.28~GHz, uGMRT at 400~MHz, VLA at 3~GHz, and LOFAR at 144\,MHz. The low resolution images reveal large scale diffuse emission ($150\,kpc \times 100$\,kpc) surrounding the ridge and NW extension in all directions \citep[as shown, at lower spatial resolution, by][]{NikielWroczynski2013}. This extended emission does not appear to be associated with any identified structures in the system (see Figure\,\ref{fig::radio-optical} right panel).  The faint diffuse emission has a more or less similar extent between 1.28~GHz and 400~MHz. The LOFAR 144~MHz image does not show the same extent, likely owing to the relatively poor quality of the data. Moreover, the southwestern outer edge of the diffuse emission overlaps with the stellar structures of the Southern Debris Region \citep[SDR]{Fedotovetal11} $\sim$45\arcs\ ($\sim$20~kpc) south and southwest of NGC~7318A. As with the shock ridge, such large scale diffuse emission is exceptionally rare among galaxy groups, and its origin is unknown. 

In our MeerKAT L-band image, we clearly detect for the first time an arc-like emission structure extending eastward from the group, connecting NGC~7320 with SQ-B and continuing beyond. This feature is labeled as the ``radio arc" in Figure\,\ref{fig::low_res}. As shown in Figure\,\ref{fig::Xray-radio-HI} right panel, this radio arc closely follows the southern section of the \Hi\ arc-N feature, with the continuum emission tracing much of its extent until it turns to the north. By contrast, the \Hi\ arc-S feature does not show any associated continuum counterpart in our L-band continuum image.  At low frequencies (in the LOFAR and uGMRT images), there is a hint of some diffuse radio emission surrounding SQ-B, however, it does not form the continuous radio arc seen in the MeerKAT L-band image (see Figures\,\ref{fig::radio-optical} and \ref{fig::low_res}). This suggests that the radio arc likely has a flat spectral index. The clear detection of the radio arc only at L-band is probably attributable to the superior sensitivity and dense inner \textit{uv}-coverage of MeerKAT.

\subsection{AGN and Star Formation regions}
The brightest compact radio source in the group is the Seyfert 2 galaxy NGC~7319 which exhibits an FR~II-like morphology at high resolution \citep{Xanthopoulos2004}, with lobes separated by $\sim$5\arcs\ ($\sim$2.5~kpc), see Figure\,\ref{fig::continumm_high} left panel. Although the source is not resolved in our highest resolution image, it is consistent with the previously reported morphology. In the low resolution radio maps, the nuclear source in NGC 7319 is surrounded by diffuse emission (see Figure\,\ref{fig::low_res}). 

We do not detect any significant radio emission from the nucleus of the intruder galaxy NGC~7318B. As shown in Figure\,\ref{fig::continumm_high}, the core of NGC~7318A is detected at all observed frequencies. Additionally, we detect a compact source to the south-west of the NGC~7318A core, most clearly at high frequencies.

The SQ-A SF region is detected as a compact radio source (aperture of 2.8\arcsec) embedded within the radio emission from the ridge (Figure\,\ref{fig::radio-optical_high_res} left panel). It has a counterpart in the X-ray and coincides with a bright star-forming region seen in the optical and IR (Figure\,\ref{fig::radio-optical_high_res} right panel). To the southeast of the ridge lies SQ-B, another relatively compact SF region embedded in the younger inner tidal tail (see Figure\,\ref{fig::radio-optical}) right panel). In our high resolution images, SQ-B is detected as a compact source, see Figure\,\ref{fig::continumm_high}. The MeerKAT and VLA observations show a pronounced \Hi\ emission enhancement around SQ-B (\citealt{Williams2002}, paper~I).

The foreground galaxy NGC~7320 is detected across all observed radio frequencies, see Figures~\ref{fig::continumm_high} and \ref{fig::low_res}. It is prominent at GHz frequencies suggesting a flat spectral index, as expected for an actively star forming system.

SQ-R, the bright source north of the ridge, is believed to be unrelated to the group. Our comments on its properties can be found in Appendix~\ref{app:SQR}.

\setlength{\tabcolsep}{20pt}
\begin{table*}[htp]
\caption{Flux densities of the different sources/regions.}
\begin{center} 
\begin{tabular}{*{10}{c}}
\hline \hline
\multirow{1}{*}{Source} &\multirow{1}{*}{LOFAR} &\multirow{1}{*}{uGMRT} &\multirow{1}{*}{MeerKAT}& \multicolumn{2}{c}{VLA} \\
  \cline{5-6} 

& $S_{\rm144\,MHz}$ &${S_{\rm400\,MHz}}$& ${S_{\rm1.28\,GHz}}$&${S_{\rm3.0\,GHz}}$&${S_{\rm6.0\,MHz}}$\\
  & (mJy) & (mJy) & (mJy)  & (mJy) &(mJy) \\

  \hline 
Total emission$^\star $& $365\pm40$ &$175\pm20$ & $57\pm6$ &$24\pm1$&$10\pm1$\\
Ridge & $250\pm55$ & $115\pm20$  & $37\pm4$ & $15.0\pm1.5$& $7.2\pm0.1$\\
NW extension & $72\pm15$& $27\pm4$&$7.0\pm0.8$& $2.3\pm0.2$& $0.9\pm0.1$\\
radio bridge & $-$ &$2.7\pm0.3$&$0.74\pm0.04$&$0.20\pm0.01$&$-$\\
\hline 
\end{tabular}
\end{center} 
{Notes. All reported flux densities were extracted from  8\arcsec\ or 15\arcsec\ images created with ${\tt robust}=-0.5$ and a \textit{uv}-cut of 0.2k$\lambda$. The image properties are given in Table\,\ref{tab:imaging}. The regions where the flux densities were extracted for the ridge and NW extension are indicated in Figure~\ref{fig::spectrum} right panel (shown in cyan). Absolute flux density scale uncertainties are assumed to be 10\% for LOFAR, uGMRT Band\,3, and uGMRT Band\,4, 5\% for MeerKAT L-band data, and for the VLA S-band and C-band data 2.5\%. $^\star$Total flux density excludes contribution from the nuclei of NGC~7319, NGC~7318A, and SQ-R. }
\label{Table:flux}   
\end{table*} 

\section{Analysis: Spectral index and curvature}
\label{sec::analysis}
The radio observations used in this work were acquired using MeerKAT, uGMRT, VLA and LOFAR, each with different \textit{uv}-coverages. As a result, comparing flux densities and deriving spectral indices, particularly for extended radio emission, requires careful analysis. We created images using Briggs weighting with a robust parameter of $-0.5$ and applied a common inner \textit{uv}-cut at $200\lambda$, corresponding to the shortest well-sampled baseline in the uGMRT observations. This implies that the radio emission is recovered at the same angular scales across all the observed frequencies from 144~MHz to 1.28~GHz. We note that this inner \textit{uv}-cut has no effect on the VLA S- and C-band data. However, comparison with single-dish flux density measurements, together with the observed power law radio spectrum, indicates that no significant flux is missing at these frequencies. For the spectral analysis, we use two sets of radio maps at different angular resolutions: the high (synthesized beam 8\arcs) and low (15\arcs) resolution images.

\begin{figure*}[!thbp]
    \centering
    \includegraphics[width=0.45\textwidth]{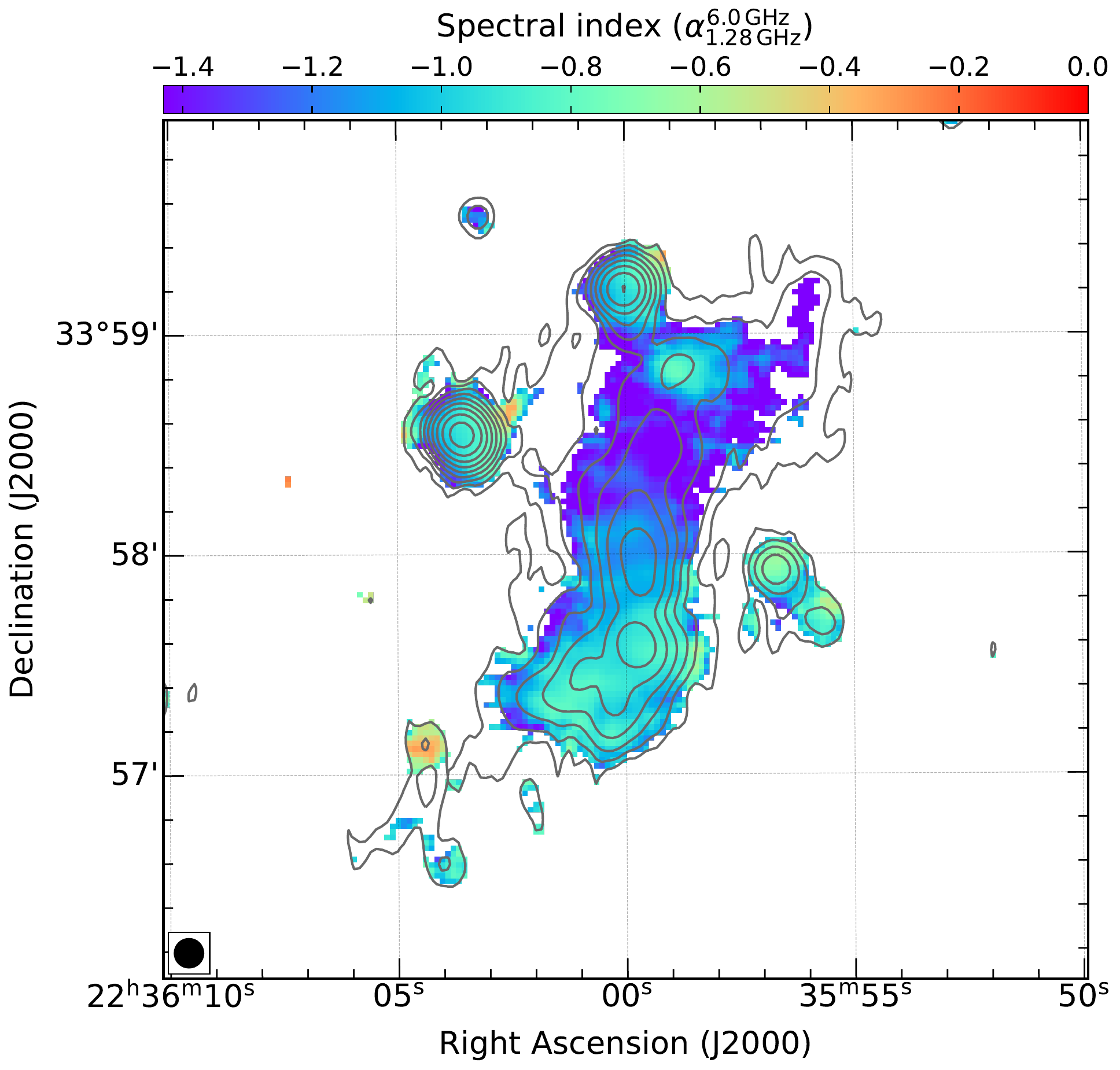}
    \includegraphics[width=0.45\textwidth]{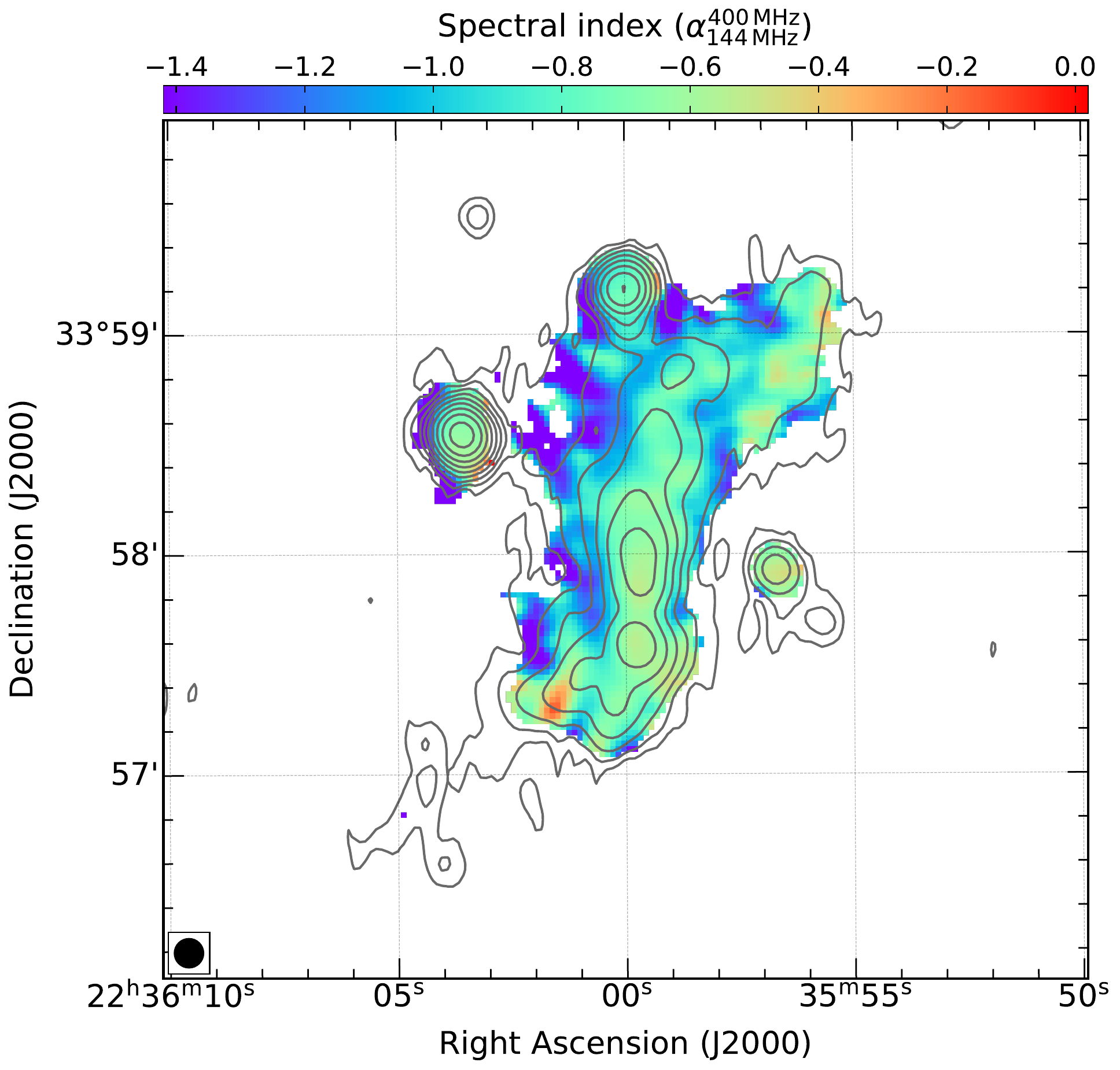}
       \includegraphics[width=0.45\textwidth]{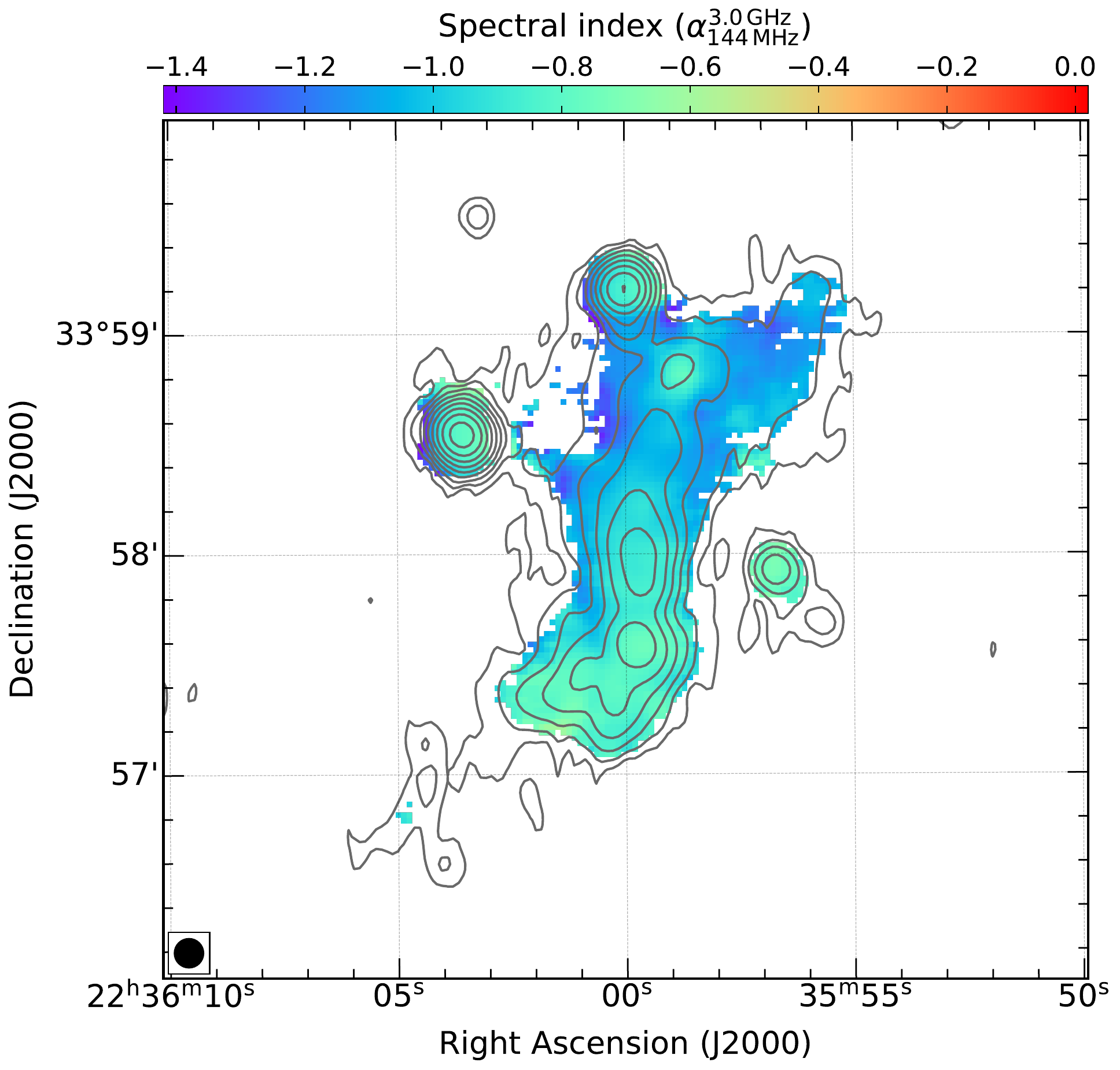} 
      \includegraphics[width=0.47\textwidth]{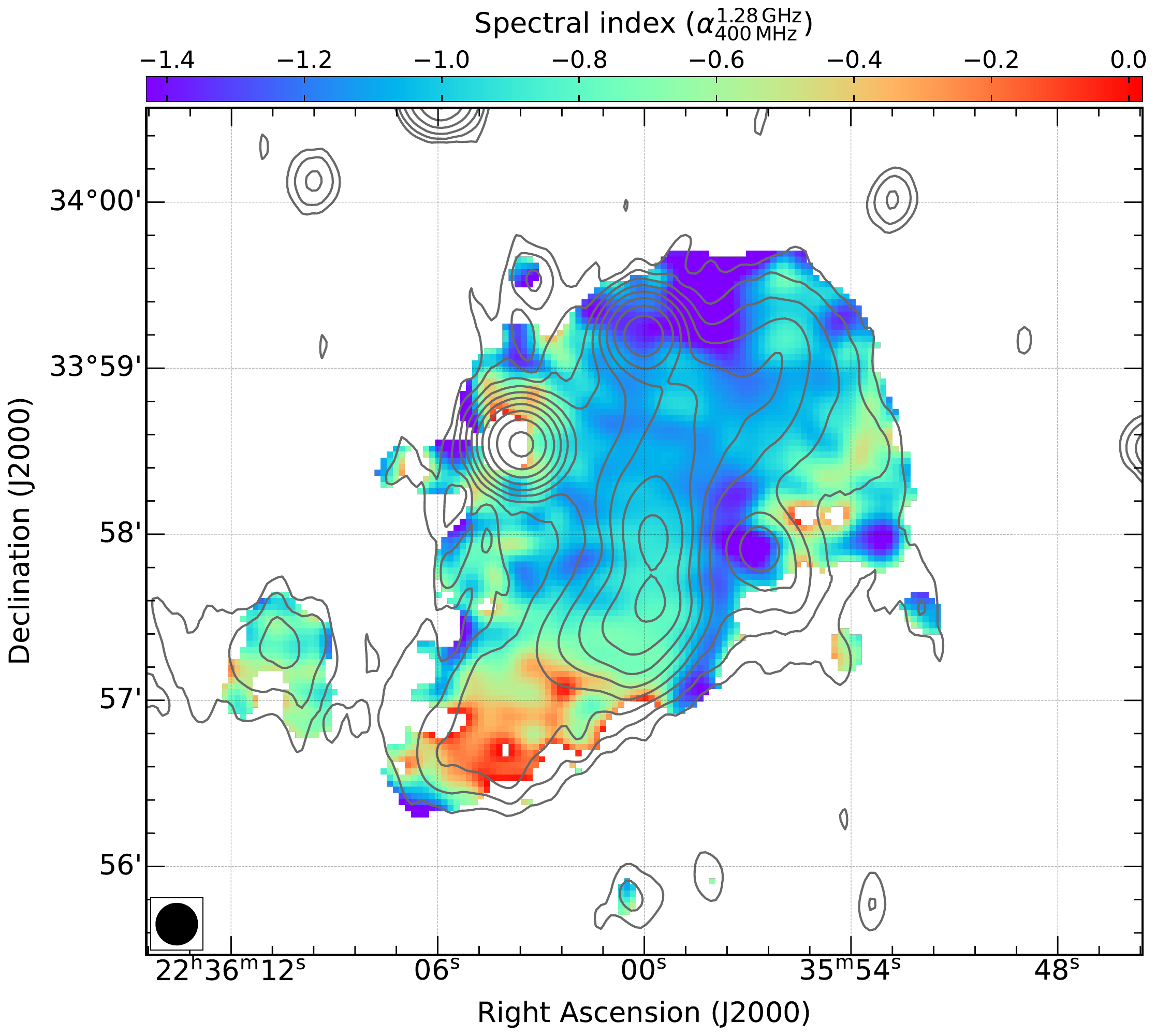}  
 \caption{Spectral index maps of \SQ. The high resolution (8\arcs) maps were created between 1.28~GHz and 6~GHz (top-left) and 144~MHz and 400~MHz (top-right). The wideband 8\arcsec spectral index map was created using 144~MHz, 400~MHz, 1.28~GHz and 3~GHz (bottom left). The low resolution spectral index map (bottom-right) was created after subtracting point sources from the \textit{uv}-data. The map has a beam size of 15\arcsec\ and is created between 400~MHz and 1.28~GHz. Discrete sources are marked with circles. In all maps, MeerKAT L-band radio contour levels are drawn at  $[1, 2, 4, 8 ...]\times 3.5\sigma_{\rm rms}$. The radio beam size is indicated in the bottom left corner of each image.}
      \label{fig::index_maps}
\end{figure*} 

\subsection{Integrated spectra}

The total integrated spectra of the ridge and the NW extension are shown in Figure\,\ref{fig::spectrum} left panel with the regions used to extract the flux densities indicated in the right panel (shown in cyan). The extraction regions were defined using the 144~MHz map, where the radio emission is most extended and detected above $3\sigma_{\rm rms}$. In Table\,\ref{Table:flux}, we report the measured flux density of the ridge at the observed frequencies. We note that the flux density contribution from SQ-A is excluded. The total emission from the ridge can be explained by a power law spectrum from 144~MHz to 6~GHz with no clear evidence of spectral steepening toward high frequencies. The ridge has a relatively flat spectral index of $\alpha_{144\,\rm MHz}^{6\,\rm GHz} = -1.00 \pm 0.03$, consistent with that reported by \cite{Xu2003} between 1.4 and 4.86~GHz. The integrated spectrum of the NW extension also broadly follows a power-law. However, it is significantly steeper than the ridge, $\alpha_{\rm 144\,\rm MHz}^{6.0\,\rm GHz} = -1.20 \pm 0.04$, with a hint of spectral curvature. Considering the ridge and NW extension as a single structure, we obtain an integrated spectral index of $-1.03\pm0.04$. We also measure spectral indices for the radio bridge, finding a low frequency index of $\alpha_{400\,{\rm MHz}}^{1.28\,{\rm GHz}}=-1.11\pm0.10$ which steepens to $\alpha_{1.28\,{\rm GHz}}^{3\,{\rm GHz}}=-1.54\pm0.08$ at higher frequencies.

Recently, \cite{Arnaudova2024} reported spectral indices for the ridge as a whole of $\alpha_{144\,\rm MHz}^{1.7\,\rm GHz}=-0.87 \pm 0.16$ and $\alpha_{1.7\,\rm GHz}^{4.86\,\rm GHz}=-1.22 \pm 0.08$, indicating spectral steeping at higher frequencies. We do not find clear evidence of high frequency spectral steepening between 144~MHz and 6~GHz in the total integrated spectra of the ridge and NW extension. We note that our data are more sensitive than those used by \citet{Arnaudova2024} and to obtain the total integrated spectral index, we used higher resolution radio maps. These allowed us to better identify and subtract flux density contributions from unrelated sources. In addition, we created maps using a common \textit{uv}-cut. As will be shown in  Sections\,\ref{sec:SPIX} and \ref{sec::curvature} we do see evidence of spectral steepening when we examine the spectra of smaller regions. The combination of spectra with differing brightness and curvature from different regions can produce the total integrated spectra which are well-described by a simple power law \citep[e.g., ][]{Rajpurohit2021a, Rajpurohit2022b, Wittor2021}.

\begin{figure*}[!thbp]
    \centering
    \includegraphics[width=1.0\textwidth]{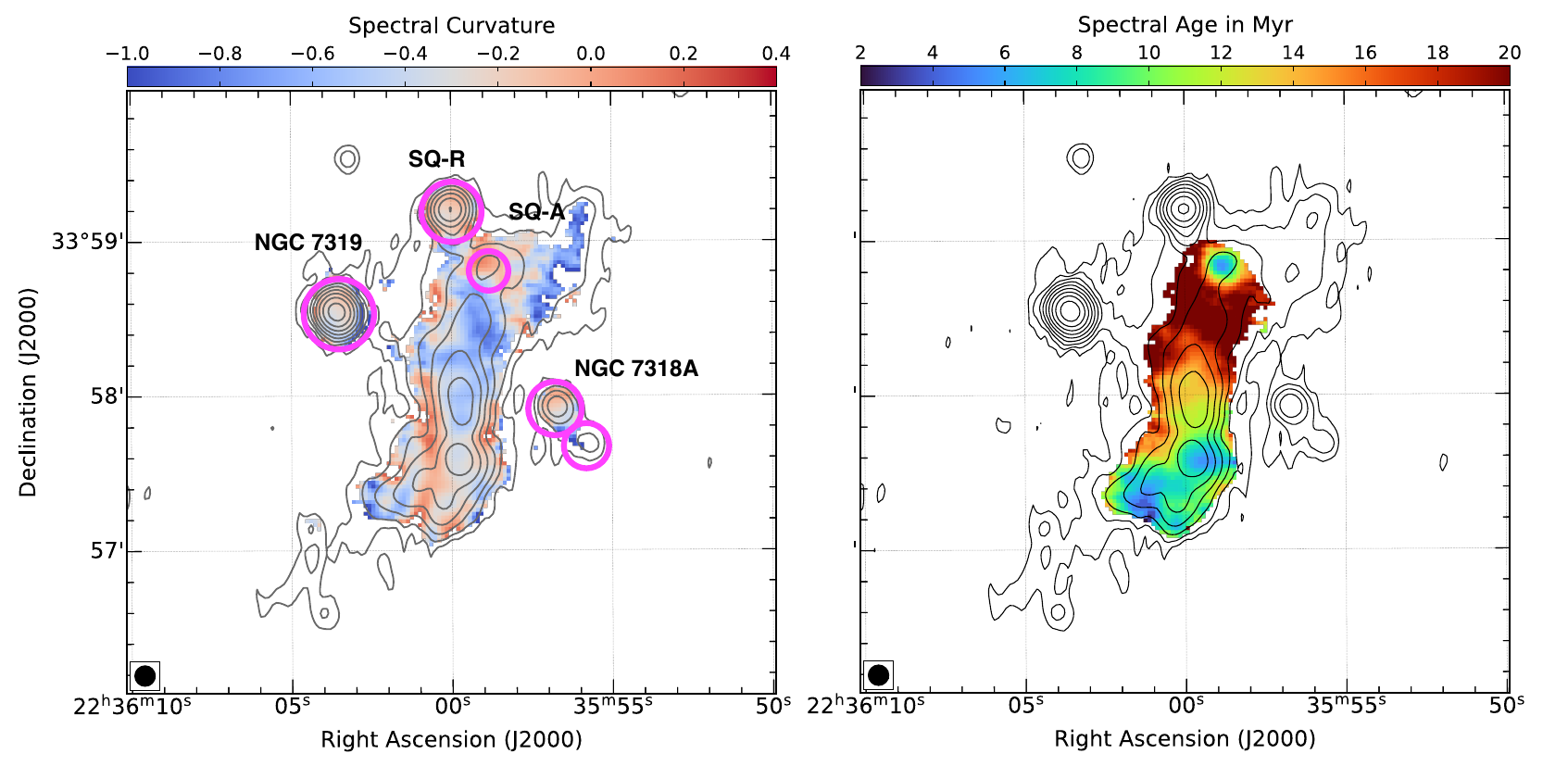}
 \caption{\textit{Left:} High resolution (8\arcsec) spectral curvature maps of \SQ\ obtained using $144-400$~MHz and $1.28-3.0$~GHz images. \textit{Right}: Radio spectral age map of the shock ridge derived from the 8\arcs\ resolution 144~MHz, 400~MHz, 1.28~GHZ, 3~GHz and 6~GHz data. The spectral age is obtained adopting the JP model, an injection index of $-0.70$ and a magnetic field of $\rm 10\mu G$. In both maps, radio contour levels are drawn at  $[1, 2, 4, 8 ...]\times 3.5\sigma_{\rm rms}$ and are from the MeerKAT L-band. The radio beam size is indicated in the bottom left corner of each image. }
      \label{fig::curvature_age}
\end{figure*}

\begin{figure}[!thbp]
    \centering
    \includegraphics[width=0.438\textwidth]{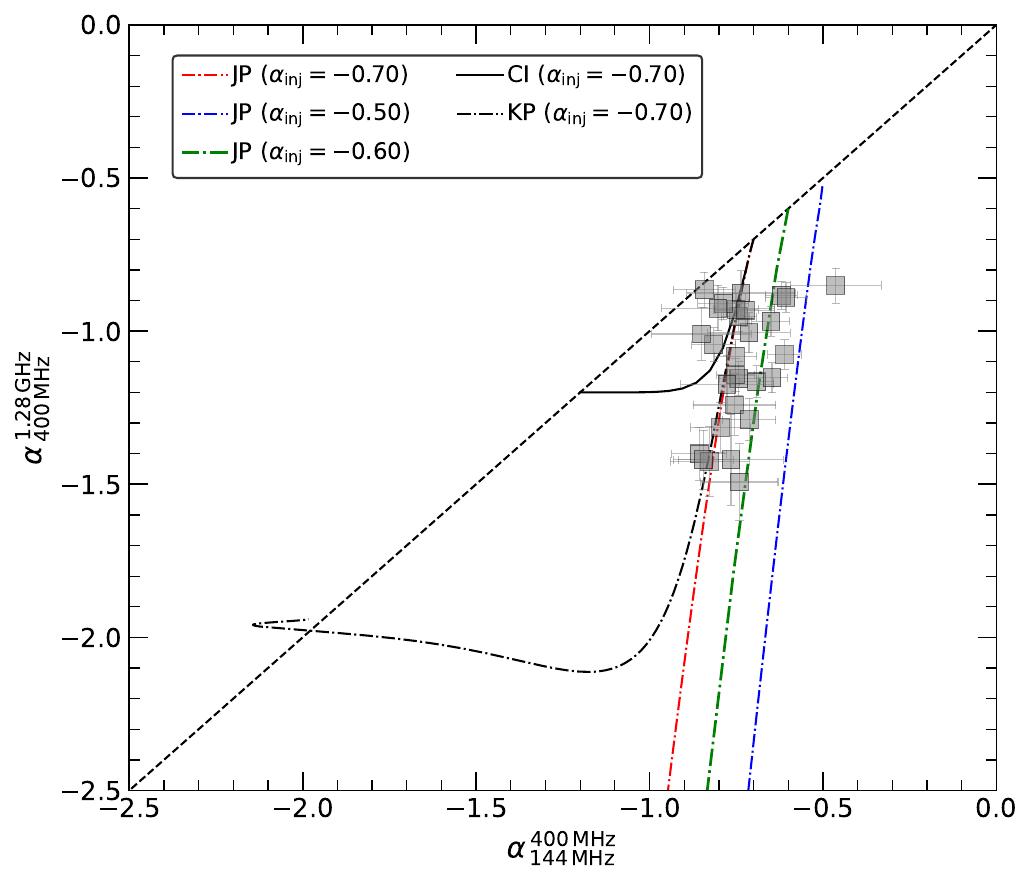}
     \includegraphics[width=0.438\textwidth]{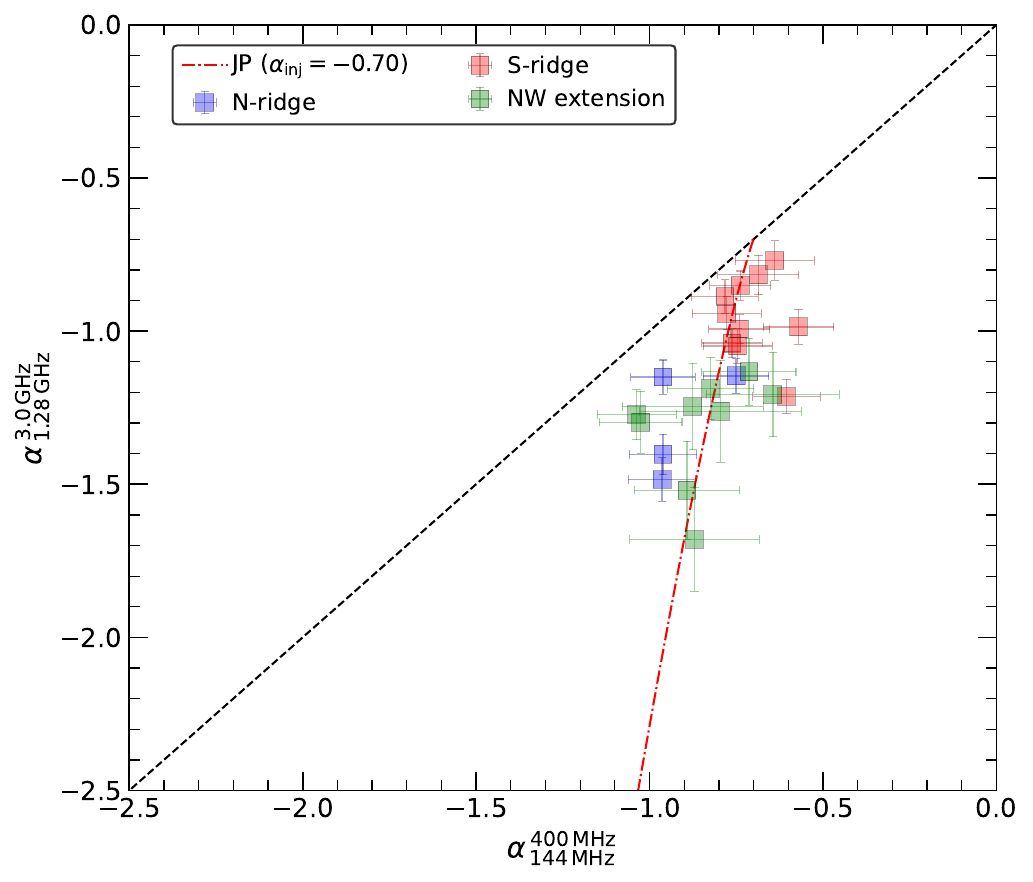}
  \vspace{-0.3cm} 
 \caption{\textit{Upper}: Radio color-color plot of the ridge (shock) between 144~MHz and 6.0~GHz, superimposed with the JP, KP, and CI model curves. Spectral index values were measured in the grid of 8\arcs\ square boxes shown in Figure~\ref{fig::spectrum}. In the shock region, the majority of the observed data points are consistent with the JP model with an $\alpha_{\rm inj}=-0.70$. \textit{Lower}: Radio color-color plot of the subregions between 144~MHz and 3.0~GHz using 15\arcsec\ maps. The southern part of the ridge is less curved while the NW extension is the most curved. Spectral index values were measured using a grid of 15\arcs\ square boxes.}
      \label{fig::cc_plots}
\end{figure}

\subsection{Spectral index mapping}
\label{sec:SPIX}
In Figure\,\ref{fig::index_maps}, we show high resolution (8\arcsec) spectral index maps of the ridge in high frequency (1.28 and 6~GHz), low frequency (144 and 400~MHz), and wide bands, the latter (bottom-left) measured from the 144~MHz, 400~MHz, 1.28~GHz, and 3~GHz maps using BRATS. For all spectral index maps, we considered only pixels with a flux density above $3.5\sigma_{\rm rms}$. The spectral index maps reveal a clear gradient along the ridge from south to north, with the index steepening from $-0.7$ in the south to roughly $-1.4$ in the north. The spectral index gradient is most notable in the high frequency map, and less pronounced between 144~MHz and 400~MHz. The gradient suggests the presence of a younger electron population in the S-Ridge and older electrons in N-Ridge. We note that radio relics, produced by merger driven shocks in the intracluster medium, generally show a clear spectral index gradient across their minor axis \citep{Rajpurohit2018, Gennaro2018, Rajpurohit2024a} due to aging of electrons in the downstream regions; the gradient traces the motion of the shock front. We do not observe any significant, coherent spectral index gradient along the ridge minor axis. 

The flattest spectral indices ($-0.53$) are observed around clump B2 and the branches at the southern end of the ridge, between 144~MHz and 400~MHz. We find that N-Ridge and the NW extension exhibit comparable spectral indices, with a mean spectral index of about $-1.2$ between 144~MHz and 3~GHz. This is consistent with the value obtained from the integrated spectral index between 144~MHz and 6~GHz for the NW extension. However, N-Ridge and the NW extension show clear spectral steepening: the spectral index steepens from about $-0.9$ at low frequencies (144-400~MHz) to $-1.4$ at high frequencies (1.28-6~GHz). In contrast, the S-Ridge shows little spectral steepening.

NGC~7319, SQ-A, and NGC~7318A are all characterized by flat spectral indices between 144~MHz and 6~GHz (Figure\,\ref{fig::index_maps}). We measured a spectral index of $-0.80\pm0.03$ for NGC~7319, consistent with those reported by \citet{NikielWroczynski2013} and \citet{Xu2003}. The radio emission around the core of NGC~7318A exhibits a flat spectral index of roughly $-0.6$, consistent with the values reported by \cite{Xu2003} but differing significantly from the steep index reported by \cite{NikielWroczynski2013}. 

In Figure\,\ref{fig::index_maps}, bottom-right panel, we show the low resolution (15\arcsec) 400-1280~MHz spectral index map. Point-like  sources were subtracted from the \textit{uv}-data using an inner-uv cut of 3k$\lambda$. This map reveals the spatial distribution of the spectral index of the low surface brightness diffuse emission. We do not observe any distinct spectral index trends (for example spectral index gradient) within the diffuse emission region; the spectral index distribution is broadly similar to that of the ridge region. Overall, the spectral index varies between $-0.6$ and $-1.4$ (excluding NGC~7320). To the north of the NW extension, the spectral index steepens to values as low as $-1.8$. 

The foreground galaxy NGC~7320 has the flattest spectral index of about $-0.50$ (see Figure\,\ref{fig::index_maps} bottom-right) between 400~MHz and 1.28 GHz. Our measured spectral index is significantly flatter than that reported by \citet{NikielWroczynski2013}. NGC~7320 is a late-type, rotating, disk galaxy whose optical colors clearly indicate ongoing star formation, for which a flat spectral index is expected. The galaxy lies along the line of sight to the head of the \Hi\ tidal tail. Our measured spectral index of $-0.5$ indicates that the radio emission in that region is dominated by NGC~7320.

The spectral index of SQ-B is $-0.65\pm0.04$ between 144~MHz and 3~GHz. The measured value is in agreement (within errors) with \citet{Xu2003} and \citet{NikielWroczynski2013}.  It is important to note that SQ-B is embedded in low surface brightness arc-like emission, suggesting its true spectral index may be even flatter than the measured value. In any case, the flat spectral index provides strong evidence that the emission in this region is associated with star formation rather than predominantly non-thermal emission as suggested by \cite{NikielWroczynski2013}.

\subsection{Spectral curvature}
\label{sec::curvature}

In Figure\,\ref{fig::curvature_age}, we present the (8\arcs\ resolution) spectral curvature map. Spectral curvature (SC) was defined as:
\begin{equation}
\mathrm{SC} = -\alpha_{\rm low} + \alpha_{\rm high},
\end{equation}
where $\alpha_{\rm low}$ and $\alpha_{\rm high}$ are the low ($144-400$~MHz) and high ($1.28-6$~GHz) frequency spectral index maps, respectively. By convention, curvature is negative for a concave spectrum, positive for a convex spectrum, and a value of SC=0 indicates no curvature. Consistent with the resolved spectral index trends, a distinct curvature gradient is observed along the ridge extending from south to north ranging from 0 to $-0.8$ (Figure\,\ref{fig::curvature_age} left-panel). The southern section of the ridge and SQ-A exhibit little to no curvature, while the northern part of the ridge is the most curved. We do not observe any clear curvature trend from east to west in the ridge region. 

Similar curvature trends are observed in a low resolution curvature map (not shown) obtained from the 400~MHz, 1.28~GHz and 3~GHz maps. The NW extension and the northern section of the ridge display a comparable curvature, with an average value of about $-0.3$. 

\cite{Arnaudova2024} presented a 14\arcs\ resolution spectral curvature map of \SQ\ using LOFAR 144~MHz, and archival VLA 1.7 GHz and 4.6 GHz radio images. They reported that the spectral curvature in the ridge is uniform (about $-0.3$) and suggested that this indicates a uniform age for the plasma throughout the shock region. They further noted that the ridge is surrounded by regions with higher spectral curvature (about $-0.5$), implying that the shock is energizing or compressing pre-existing radio plasma. However, our high-resolution curvature map reveals that the curvature instead exhibits a gradient from south to north in the ridge. 

Radio color-color plots are commonly used to investigate the spectral properties of radio sources \citep{Katz1993}. By comparing spectral indices across different frequency intervals (e.g., low versus high frequencies), these plots can be used to identify spectral curvature, multiple electron populations, and projection effects, etc. \citep[e.g.,][]{Rajpurohit2020a, Rajpurohit2024}.  An advantage of color-color plot is that their spectral shapes (trajectories) are conserved for changes in the magnetic field, adiabatic expansion or compression, and the radiation losses, for standard spectral models. 

We used our high and low resolution maps to create color-color plots, extracting indices in a grid of square boxes with a width 8\arcs\ (see Figure\,\ref{fig::spectrum} right panel) or 15\arcsec\ (not shown) covering the ridge and NW extension. We included regions with flux density $\geq3.5\sigma_{\rm rms}$. The 8\arcsec\ maps and regions are also used for making the global spectrum (see Section\,\ref{sec::global_spectrum}).

In Figure~\ref{fig::cc_plots}, we present the resulting plots for the ridge and sub-regions. The color-color plot of the ridge, derived from 144~MHz, 400~MHz, 1.28~GHz, and 6~GHz data, shows a clear spectral curvature, with the observed data points following a single spectral trajectory.  

We overlay the observed trends with the Jaffe-Perola \citep[JP;][]{Jaffe1973}, Kardashev-Pacholczyk \citep[KP;][]{Kardashev1962} and continuous injection \citep[CI:][]{Pacholczyk1970} spectral aging models. The JP model assumes efficient pitch angle scattering while in the KP model the pitch angle remains constant. The CI model assumes a continuous injection of particles throughout the lifetime of the source. As shown in Figure~\ref{fig::cc_plots} top panel, the CI model is inconsistent with the observed radio colors. We utilized three injection index values, namely $-0.50$, $-0.60$ and $-0.70$ for the JP model. The majority of the ridge data points are broadly consistent (within errors) with the JP model with an injection index of $-0.70$.

In the bottom panel of Figure\,\ref{fig::cc_plots}, we present the color-color plots for the ridge and NW extension, using the 144~MHz, 400~MHz, 1.28~GHz  and 3 GHz data. We divide the ridge into northern and southern subregions. The S-ridge exhibits the least curvature and the NW extension the strongest curvature. We do not find evidence for distinct electron populations or significantly different curvature or spectral-index trends between the N-ridge and the NW extension. This suggests that the NW extension is likely a part of  the shock ridge.

\subsection{Spectral age}
\label{sec::age}
The spectral curvature of a source is governed by the radiative losses experienced by the particle population and is therefore directly related to the plasma age. The spectral age can be determined as:
\begin{equation}
t_{\rm age}=1.59\,\frac{B^{1/2}}{B^2+B_{\rm CMB}^2}\left[\nu_{\rm br}(1+z)\right]^{-1/2}\,{\rm Gyr}
\label{equ::age}
\end{equation}
where $t_{\rm age}$ is the synchrotron spectral age in Gyr, $B$ is the magnetic field strength in $\mu\mathrm{G}$, $B_{\rm CMB}$ is the equivalent  magnetic field of the Cosmic Microwave Background (CMB), given by $B_{\rm CMB}=3.25(1+z)^2~\mu\mathrm{G}$, $\nu_{\rm br}$ is the spectral break frequency in GHz, and z is the source redshift. The spectral age of a source can be estimated either from the measured spectral break frequency using Equation\,\ref{equ::age} or by fitting the observed radio spectrum with model spectra generated through numerical solutions of the radiative loss equations, including synchrotron and inverse Compton (IC) losses. Here, we adopt the latter approach and model the radio spectra using the Broadband Radio Astronomy Tools ({\tt BRATS)} package \citep{Harwood2013}. We note that IC losses from scattering of non-CMB photons are likely negligible, based on estimates of the radiation energy density of IR and radio emission within the ridge itself.

In order to estimate the age of the ridge on a pixel-by-pixel basis, we used the 8\arcsec resolution radio maps at 144~MHz, 400~MHz, 1.28~GHz, 3~GHz, and 6~GHz. Since our goal is to determine the age of the ridge and NW extension, all discrete sources (except SQ-A) and other structures were masked. Pixels with a flux density below $5\sigma_{\rm rms}$ were blanked. We first searched for the best-fit injection index by performing a series of JP fitting iterations in the range $\alpha_{\rm inj}=-0.5$ to $-1.0$. We adopt the JP model as it is more physically realistic than the KP model. The best fit injection index is found to be in the range $-0.70$ to $-0.75$. Since the radio color-color analysis also suggests a typical injection index of $-0.70$ for the shock ridge, we adopt this value for the final spectral age estimate. 

The equipartition magnetic field strength was estimated with the {\tt pysynch} package\footnote{https://github.com/mhardcastle/pysynch}. We assumed a cylindrical geometry for the ridge, with a radius of 15\arcsec\ and a length of 90\arcsec. For equipartition, the flattest observed injection spectral index should be used, rather than the typical value for the emission region as a whole; we therefore adopted $\alpha_{\rm inj}=-0.55$, based on the low-frequency spectral index map. The flux density within the emitting region was measured to be $\rm S_{144\,MHz}=198\pm25 mJy$. We adopted a proton-to-electron energy density ratio of $\kappa=40\pm20$, as suggested for strong shocks \citep{Beck2005} and $\gamma_{\rm min}=20$ \citep{Brunetti1997}. Using these values, we obtain an equipartition magnetic field strength of $B_{\rm eq}=10\pm2\mu{\rm G}$. The uncertainty is estimated by propagating the errors in the integrated flux density, injection index and $\kappa$. This value is consistent with previous equipartition magnetic field estimates for \SQ\ \citep{Arnaudova2024,NikielWroczynski2013}, albeit these assumed different input parameters.
 
We then created a spectral age map by adopting a constant magnetic field strength of $B=10~\mu{\rm G}$ in {\tt BRATS} (Figure~\ref{fig::curvature_age} right panel). We find a range of ages in the ridge spanning $\sim$2-20 Myrs with a mean value of $14\pm2$~Myr. The mean value is comparable to that reported by \citet[11~Myr, with $B=9.5~\mu{\rm G}$]{Arnaudova2024}. However, while they suggested that the shock ridge contains a homogeneous relativistic particle population (i.e., with a single age) our spectral age map reveals significant age variations along the ridge. The southern part of the ridge is the youngest, with typical ages 5-6~Myr, the northern region shows ages as high as 20~Myr, with intermediate ages $\sim$10-12~Myr in region B2.  

Radio spectral age estimates are subject to a variety of uncertainties, with the magnetic field strength being the most crucial parameter. The inferred age of the ridge depends significantly on the assumed magnetic field strength: for $B=5~\mu{\rm G}$, the mean age increases by 100\% ($\rm 28\,Myr$), while for $B=15~\mu{\rm G}$, it decreases by 35\% ($\rm 8~Myr$). The age also depends on the spectral coverage of the radio data, as it is constrained by the observed spectral curvature and  therefore sensitive to the highest frequency measurements available. In Appendix~\ref{app:agemap2}, we present an alternative age map made without the C-band data. The impact on the fitted ages is minimal, with the same north-south age gradient visible, and the map extends further into the NW extension, where the age is as high as $\sim$25~Myr. The age estimates from the {\tt BRATS} modeling imply quite high break frequencies, extending beyond the range of our observed data for the younger emission regions. We note that the older emission at the northern end of the shock has an inferred break frequency of 5-6~GHz, well within the frequency range of our data, and spectra extracted from small regions exhibit clear spectral curvature. Finally, we note that age estimates are dependent on the model assumed. We have adopted the JP model \citep{Jaffe1973} consistent with the results of the color-color plots and global spectrum. We also tested the Tribble model \citep{tribble1991}, which accounts for an inhomogeneous magnetic field. However, the best fit provided ages and reduced $\chi^2$ distributions comparable with the JP model.

\begin{figure}
\centering
\includegraphics[width=\columnwidth]{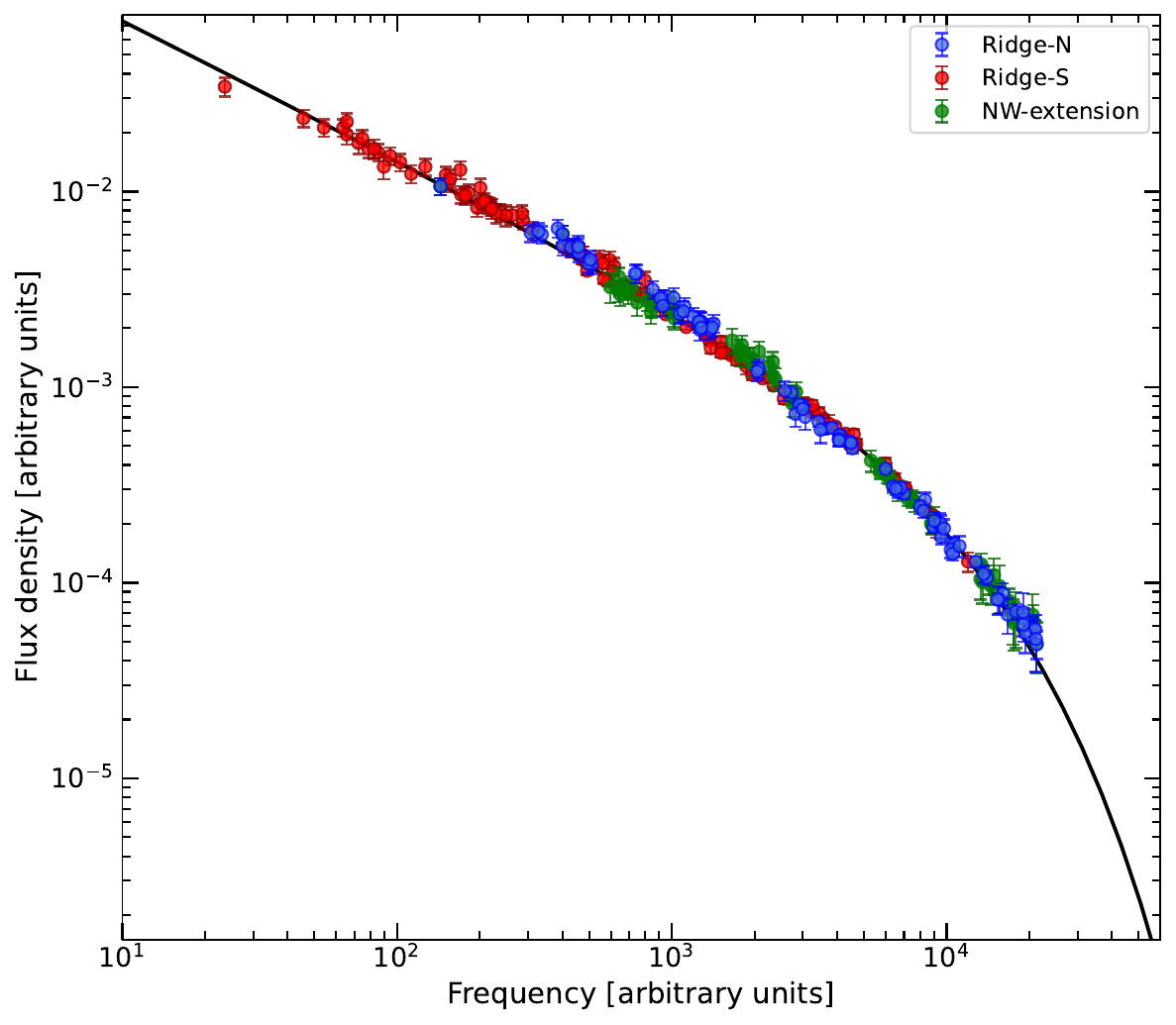}
\caption{Global spectrum of different regions of the shock ridge (Ridge-N, Ridge-S, and NW-extension) created using 8\arcsec~ resolution maps and five frequency data points (144~MHz-6.0~GHz). The spectrum of each region has been shifted in $\log(\nu)$ and $\log(I)$ space to create the ``global'' spectrum. The JP model line for an injection index of $-0.70$ is shown in black. The plot reveals that there is a single global spectrum for the entire ridge+NW extension region.}
\label{fig:global_spectrum}
\end{figure}

\subsection{Global spectrum}
\label{sec::global_spectrum}
To investigate whether a single global spectrum exists in the ridge and NW extension, we used the spectral-shift technique \citep{Katz1993, Rudnick1994, vanWeeren2012a, Rajpurohit2024}. This technique allows the identification of regions with different properties (magnetic fields, electron density, electron energy), if present. 

In Figure\,\ref{fig:global_spectrum}, we show the resulting global spectrum extracted in different subregions of the ridge and NW extension. In this plot, the spectra from individual regions were shifted in intensity, $\log(I)$ and frequency, $\log(\nu)$ to align them with the reference JP model with the $\alpha_{\rm inj}=-0.70$. Remarkably the S-ridge, N-ridge and NW extension regions are all consistent with a single global spectrum indicating a single underlying electron energy distribution. We note that a small amount of scatter is present at 144~MHz, particularly for the Ridge-S regions. The presence of a single global spectrum implies that these three subregions are consistent with the same underlying physical parameters and points to a connected origin for the radio emission. This provides additional evidence that the NW extension is also associated with the shock ridge. 

\section{Origin of the ridge and diffuse emission}
\label{sec::origin}
Our new radio images reveal complex structures in \SQ. The most remarkable is the shock ridge, in which we identify substructures including clumps B1 and B2, and the branches at its southeast tip. The ridge is linked to NGC~7319 by a radio bridge, has an extension to its NW, and is surrounded by large scale diffuse radio emission.  

While we observe flat spectral indices in known star-forming regions (the foreground galaxy NGC~7320, SQ-A, SQ-B) and from AGN (the nuclei of NGC~7319 and NGC~7318A), the northern ridge and NW extension are steep-spectrum diffuse structures and are thus non-thermal in nature. The nature of the emission in the southern ridge is somewhat less clear. The ridge exhibits clear spectral index and curvature gradients, with the flattest, least curved spectra observed in the south. Below we discuss possible explanations for the diffuse radio emission in \SQ.

\subsection{Star formation} 
Star formation (SF) cannot produce the diffuse emission we observe in \SQ\ but may contribute in specific regions. The steep spectra of the northern half of the ridge and NW extension are incompatible with SF. The large-scale diffuse radio emission extends well beyond the bright stellar components of the galaxies (see Figure\,\ref{fig::radio-optical} right panel). Even in the southern ridge, where flatter spectral indices are seen, the spatial correlation is poor, with the radio clearly not following the southeast spiral arm. Clump B2 overlaps the SE arm, but most of the radio emission is found to the west of the stars (see Figure\,\ref{fig::radio-optical_high_res}); this agrees with the behavior of the X-ray, H$\alpha$ and warm H$_2$, all of which suggest a separation between the shocked material and spiral arm in this region (paper~I). Clump B1 is located next to two bright stellar regions in the SE arm, which are known to be CO-rich \citep{Emonts2025}. However, south of B1, the radio emission completely diverges from the spiral arm and instead turns east. SF emission may be a confusing factor for the branches at the southeast tip of the ridge, as they overlap what appear to be stellar features in NGC~7320, but the fact that both branches are also seen in soft X-ray and 10~$\mu$m warm H$_2$ emission (paper~I) strongly suggests that the radio emission from the branches is not SF-dominated.

It is also notable that known SF regions (SQ-A, SQ-B, NGC~7320) are relatively faint at radio frequencies compared to the shock ridge. \citet{Xuetal05} find a (UV-based) star formation rate (SFR) of 1.45\Msolpyr\ for the ridge excluding SQ-A. We estimate the expected 144~MHz luminosity from the SF using the SFR:L$_{\rm 144~MHz}$ relation for ram-pressure stripped galaxies \citep{Edleretal24}, finding a predicted luminosity L$_{\rm 144~MHz}$=3.2$\times$10$^{22}$~W~Hz$^{-1}$, equivalent to a flux density of $\sim$30~mJy. This is only $\sim$12\% of the total measured 144~MHz flux density of the ridge, $\sim$250~mJy, and should be considered a conservative estimate, since relations for undisturbed spirals would predict lower values. Using a relation for more typical spirals \citep{Heesenetal22} we would predict a flux density of only $\sim$6~mJy. We therefore conclude that while SF may contribute to the radio emission from the ridge, its true origin must lie elsewhere.

For diffuse radio emission outside the shock ridge and northwest extension, we can consider the SFRs estimated for regions III, IV and VII of \citet{Xuetal05}, essentially the inner disks of NGC~7318B and NGC~7319, and part of the young, inner tidal tail. These total to $\sim$3.2\Msolpyr, giving an expected flux density of $\sim$13-66~mJy depending on the SFR:L$_{\rm 144~MHz}$ relation adopted. Table~\ref{Table:flux} shows that the diffuse emission totals to $\sim$43~mJy, within the predicted range. However, much of this emission has spectral indices $\sim -1$ (Figure~\ref{fig::index_maps}) suggesting a non-thermal origin, and it is not clearly correlated with SF regions. SF is likely the dominant factor in regions of flat indices (e.g., north of the nucleus of NGC~7318A) but we again conclude that it is not the primary source of the diffuse, extended radio emission.

\subsection{Shock driven radio emission} \citet{Shostak1984} first suggested that the radio ridge was the product of shocks driven by the collision of NGC~7318B with a tidal filament of gas stripped from one or more of the group members, and this is the most widely adopted explanation for the continuum emission. The shock ridge connects the \Hi\ tails to the high-velocity ($\sim$6600 and $\sim$6900\kmps) \Hi\ components around SQ-A, and it is thought that it marks a section of a pre-existing tidal filament which incorporated both structures.

Previous works have noted the strong spatial correlation between the radio continuum and emission at other wavelengths. Our own comparisons (including in paper~I) confirm the close correlations between the radio continuum, X-ray, and 10~$\mu$m warm H$_2$ emission in the main ridge, branches, northwest extension, and in the southern bridge linking the ridge to NGC~7319. We also note that both X-ray and H$_2$ follow the morphology of H$\alpha$ maps of high velocity-dispersion ionized gas at the group redshift \citep{Sulentic2001,Arnaudova2024}. By contrast, lower velocity-dispersion H$\alpha$ emission at the velocity of NGC~7318B follows the southeast spiral arm of that galaxy. These correlations strongly imply that the ridge emission is colocated with the shock-heated and shock-excited gas components, not with the galaxies. However, as previous radio studies and our new maps show, the continuum emission extends well beyond the radio ridge. Excluding AGN, SQ-A and SQ-B, diffuse continuum emission extends across much of the northern and central parts of NGC~7318A/B, into NGC~7319, and possibly out into the tidal tails (the radio arc). Much of this extended component has spectral indices comparable to that of the northern shock ridge, suggesting a common origin.

Our expectations for shock-driven emission depend on the Mach number of the shock, and therefore on the medium in which it propagates, as well as the velocity of the collision. We expect that tidal filaments stripped from spiral galaxies would consist largely of \Hi\ and cold molecular gas, but it would also contain other elements of the interstellar medium (ISM) and it would be embedded within the hot X-ray emitting intragroup medium (IGrM). We therefore need to consider two possibilities: weak shocks propagating through a hot medium, and strong shocks propagating primarily in \Hi.

Simple estimates of the line-of-sight velocity of the collision can be found by comparing the recession velocity of the intruder galaxy, NGC~7319 (5774\kmps) with the peak velocity of the most extended \Hi\ component ($\sim$6600\kmps) or the mean velocity of the three core group members (NGC~7317, NGC~7318A, NGC~7319, $\sim$6670\kmps). Assuming the group halo to be centered on the core members, this would suggest a collision velocity of $\sim$825-895\kmps\ between the intruder and IGrM. Higher velocity differences, $\sim$1000\kmps, are seen between the different \Hi\ components along the line of sight through the SQ-A SF region, or between the base of the \Hi\ tidal arc at the southern tip of the shock ridge, and \Hi\ in the southern disk of NGC~7318B. These may be more applicable to shocks in the cold phase.

\subsubsection{Strong shocks in cold gas}
As discussed by \citet{Trinchieri2003}, a $\sim$1000\kmps\ collision between cold gas clouds in NGC~7318B and a tidally-stripped gas filament would be expected to drive strong shocks. Sound speeds in the cold ISM are of order 10\kmps, so such a shock would have a very high Mach number and would produce maximal (factor 4) compression and strong heating. The X-ray ridge would thus consist largely of shock-heated \Hi\ \citep{OSullivan2009}.

\citet{LisenfeldVolk10} presented a model of radio emission from a similar fast ($\sim$500\kmps) collision between the \Hi\ disks of two spiral galaxies. They argued that high Mach number shocks are comparable to those of middle-aged supernova remnants in the Sedov phase. Such shocks are highly efficient accelerators, capable of boosting particles from the \Hi\ component up to the relativistic energies necessary to produce synchrotron emission. Radio spectral indices for such supernova remnants are typically flat, $\alpha$$\simeq$0.5 \citep[e.g.,][]{Reynolds10}. For the range of collision velocities we see in \SQ, even considering the possibility that obliquity might reduce shock strength, the Mach number in the \Hi\ will always be in this hypersonic regime, so no variation in spectral index with shock strength is expected. The shock front would also compress and enhance the magnetic field. However, for a line-of-sight collision we would be looking through the shock front, so strong polarization would not be expected.

Such a model would suggest that the shock ridge, northwest extension, and southern bridge represent regions in which the impact of NGC~7318B has driven fast shocks into tidally-stripped \Hi\ filaments. The range of spectral indices and curvature along the ridge would indicate variations in the age of the shock, with the youngest regions at the south end of the ridge having had essentially all \Hi\ at the group velocity destroyed. This age gradient is consistent with the findings of \citet{Appleton2017} who show a N-S gradient in temperature and shock strength in the molecular gas phase, and interpret the lower shock velocities and cooler H$_2$ temperatures in the northern ridge as indicating that the energy injected by the collision has had longer to dissipate in this region. 

The variation in age suggests an angle between the tidal \Hi\ filaments and the disk of NGC~7319, with the collision beginning in the northern ridge and northwest extension. Either the filaments were not in the plane of the sky, or the galaxy disk is slightly tilted toward us in the north, as well as being more significantly inclined east-west. Moving at $\sim$1000\kmps\, the $\sim$15~Myr age difference along the ridge would imply a line-of-sight distance of $\sim$15~kpc, so the ridge would only need to be tilted by $\sim$15\degree\ to the plane of the sky. The clumpy radio emission observed in the ridge would be expected to be a product of the clumpiness of the cold gas in the filament and intruder galaxy. 

In the strong shock approximation \citep{ShullDraine87,Trinchieri2003} for an ideal gas with a ratio of specific heats $\gamma$=5/3, we would expect the post-shock gas temperature to be:

\begin{equation}
    kT = \frac{3}{16}\mu mv^2  \sin^2\phi,
\end{equation}

where $\mu$ is the mean molecular weight (1 for pure \Hi, fractionally greater for more realistic neutral gas), $m$ is the mean mass per particle, $v$ is the shock velocity, and $\phi$ is the angle between the flow direction and shock surface. In a perpendicular shock, $\phi$=90\degree, giving the maximum post-shock temperature. A 1000\kmps\ collision would produce temperatures as high as 2~keV, far higher than observed in \SQ. The temperature is affected by shock obliquity, potentially reducing it significantly. NGC~7318B has an optical inclination of 58.4\degree\ to the line of sight \citep{Makarovetal14}, with the eastern or northeastern side of its disk tilted toward us \citep{Sulentic2001}, so this could reduce the expected temperature to $\sim$0.6~keV, very similar to what is observed in the shock ridge. There is also evidence that the highly multi-phase gas in the shock zone is efficiently cooled via a turbulent cascade, in which energy from the hotter phases is transferred to cooler gas before being radiated away via H$_2$ line emission; the H$_2$ luminosity exceeds its X-ray luminosity by at least a factor of 3 \citep{Guillard2009,Cluveretal10,Appleton2017}. This may also explain why the ridge shows cooler X-ray temperatures than the surrounding diffuse IGrM \citep{OSullivan2009}.

As this model requires the collision of cold gas, it is not applicable in regions outside the NGC~7318B disk, or in which no tidal \Hi\ structures existed in the IGrM. If strong shocks are the origin of the diffuse emission west of the ridge, this would imply that stripped \Hi\ was scattered through much of that region, perhaps as the result of the tidal influence of NGC~7318A. Continuum emission in the radio arc, around SQ-B in the tidal tails southeast of the group, seems less likely to arise from this mechanism.

\subsubsection{Weak shocks in hot plasma}
An alternative mechanism is acceleration by the weaker shocks expected from the collision of NGC~7318B with hot thermal plasma in the IGrM or stripped ISM. Shocks with Mach number 2-4 are believed to be responsible for radio relics in galaxy clusters \citep[e.g.,][]{vanWeeren2009,Rajpurohit2018, Rajpurohit2022b, Gennaro2018}, either through hadronic processes or via a first-order Fermi acceleration process referred to as diffusive shock acceleration \citep[DSA;][]{Drury1983, Hoeft2007}. In the DSA test particle approximation, a shock of Mach number $\mathcal{M}$ generates a population of relativistic electrons with a power-law distribution in momentum. The integrated spectral index ($\alpha_{\rm int}$) is related to the Mach number as:

\begin{equation}
\mathcal{M}=\sqrt{\frac{\alpha_{\rm int}-1}{\alpha_{\rm int}+1}}
\label{eq::DSA}
\end{equation}

In this DSA framework, the integrated spectral index of such a shock is $0.5$ steeper than the injection index ($\alpha_{\mathrm{int}} = \alpha_{\mathrm{inj}} - 0.5$). 
For \SQ, where we are viewing the shock close to face-on, this raises a problem; DSA cannot produce emission with the flat integrated spectral indices ($\alpha>-1.0$) observed at the south end of the shock ridge.

In addition, it should be noted that radio relics have never been observed in galaxy groups, even though examples of merger shocks with suitable Mach numbers are known \citep[e.g.,][]{OSullivanetal19}. We can compare the radio luminosity of the shock ridge with the established relationship between cluster mass and radio luminosity for relics \citep[e.g.,][]{Jones_relics2023}. The monochromatic radio power of the ridge at a frequency $\nu$ is computed as
\begin{equation}
P_{\nu} = 4\pi D_{L}^{2}\, S_{\nu}\,(1+z)^{-(1+\alpha)},
\end{equation}
where $S_{\nu}$ is the observed flux density at frequency $\nu$, $D_{L}$ is the luminosity distance, $z$ is the redshift, and $\alpha$ is the spectral index. The factor $(1+z)^{-(1+\alpha)}$ accounts for the $k$-correction. For the ridge, this yields a radio power of $P_{1.4\,{\rm GHz}} = (3.9 \pm 0.4)\times 10^{22}~{\rm W~Hz^{-1}}$ and $P_{144\,{\rm MHz}} = (2.6 \pm 0.4)\times 10^{23}~{\rm W~Hz^{-1}}$. If the NW extension is a part of the ridge, we obtain a total power  of $P_{1.4\,{\rm GHz}} = (4.6 \pm 0.5)\times 10^{22}~{\rm W~Hz^{-1}}$ and $P_{144\,{\rm MHz}} = (3.4 \pm 0.7)\times 10^{23}~{\rm W~Hz^{-1}}$. \SQ's mass is not well established, but assuming a mass of order 10$^{13}$\Msol, we would expect a 150~MHz radio power several orders of magnitude below what is observed.
 
In short, even though NGC~7318B seems likely to be driving a weak shock into the hot IGrM, particle acceleration by such a shock seems unlikely to be the source of the diffuse radio emission in \SQ.

\subsection{Adiabatic compression} 
\citet{Arnaudova2024} suggest that the radio emission from the shock ridge is best explained by adiabatic compression of a pre-existing cosmic ray population associated with the hot phase of the IGrM. This would boost the radio luminosity by compressing the relativistic plasma, without reaccelerating the electrons, and thus without changing the shape of the emission spectrum \citep[e.g.,][]{Colafrancesco2017}. They argue that the compression associated with a $\mathcal{M}$$\sim$3.8 shock could produce the enhancement in radio flux seen between the ridge and its surroundings, if a population of mildly-relativistic electrons was already present.

There are a number of issues with the application of this model to the shock ridge. Firstly, the high Mach number implies a large plane-of-sky velocity component for the intruder. \citet{OSullivan2009} argue that the likely line-of-sight shock Mach number in the X-ray emitting component is $\mathcal{M}$=2.3-2.9. A plane-of-sky velocity of $\sim$1450\kmps\ would be required to raise this to $\mathcal{M}$=3.8. The direction of motion of the shock would thus be largely in the plane of the sky. If this were the case, we would expect to see a very significant X-ray temperature difference between the pre- and post-shock regions on either side of the ridge, and an X-ray surface brightness discontinuity associated with the shock front; no such features are observed.

The spectral indices observed also imply a young age, particularly at the south end of the ridge. Adiabatic compression would increase the magnetic field in the ridge and mildly boost the energy of the emitting electrons, increasing the break frequency of the spectrum \citep{Colafrancesco2017,EnsslinGopalKrishna2001}. Both of these factors would increase our estimate of the age of the relativistic seed population. However, this age would represent the time since the seed population was first accelerated, not since the compression occurred. To produce the observed ridge, there would have had to be seed electrons contained within the tidal filament. The lack of radio relics in galaxy groups suggest a general lack of a pool of seed electrons in systems on this mass scale, and the lack of cosmic rays injected from AGN with large-scale radio jets in \SQ\ rules out one mechanism of producing such a pool. We note that the jets of the Seyfert NGC~7319 are confined within the galaxy \citep{Xanthopoulos2004}. Any relativistic seed electrons in the filament would therefore have to have been produced by supernovae in the galaxy from which the filament was stripped. The timescale for the formation of the filament is far too long to be consistent with the observed spectral indices in the southern ridge, even allowing for the impact of the strong compression advocated by \citet{Arnaudova2024}; the timescale of formation of the tidal tails is estimated to be at least 200~Myr \citep{Sulentic2001}. Adiabatic compression therefore seems implausible as the source of the radio emission.

\section{Implications for the dynamical history of the group}
\label{sec::dynamics}
Our radio continuum observations, in combination with the \Hi\ results presented in paper~I, suggest that the intruder galaxy, NGC~7318B, is moving toward us on a path close to the line of sight, has collided with a collection of gas filaments and clouds tidally stripped from the other members of the group in previous tidal interactions. The line-of-sight velocity of the collision, $\sim$850-1000\kmps, was sufficient to shock heat significant amounts of cold \Hi\ into the X-ray and relativistic regimes, producing the bright ridge of emission we observe in our radio maps. The collision began at the northwest end of the shock ridge $\sim$20-25~Myr ago, and probably finished at the branched south end of the shock only 5-6~Myr ago. While these timescales are dependent on a number of assumptions (notably the magnetic field strength) they are comparable to the typical age of young star clusters in the system, $\sim$5-10~Myr \citep{Aromaletal25}, and consistent with the gradients in H$_2$ properties found by \citet{Appleton2017}. As the line-of-sight depth of the \Hi\ structure was probably comparable to its width ($\sim$15~kpc), this implies that much of the shock ridge is likely still within (or close to) the disk of NGC~7318B, particularly at its southern end.

Structures observed in multiple wavebands support a collision along a path close to the line of sight. As noted in paper~I, the alignment of \Hi\ structures in SQ-A and the northern ridge over a velocity range $\sim$1000\kmps, and their close correlation with the H$\alpha$ and optical/NIR structures in the ridge, suggests that NGC~7318B cannot have a large plane-of-sky velocity, at least on an east-west axis. The low radio polarization fraction found by \citep{NikielWroczynski2013} is suggestive of a shock moving primarily along the line of sight. The morphology of the main shock ridge is similar in radio, H$\alpha$, 10~$\mu$m and X-ray emission, and is not consistent with a bow shock at the edge of the NGC~7318B disk. The bridge structures linking the shock to NGC~7319 are difficult to explain in any scenario where NGC~7318B is moving eastward.

We have estimated an angle of $\sim$15\degree\ between the disk of NGC~7318B and the pre-existing \Hi\ filament. The \Hi\ filament is likely to have been aligned close to the plane of the sky, since the galaxies from which it was probably stripped have very similar recession velocities and there is only a small velocity gradient between the remaining sections of the filament (i.e., the tidal tails and high velocity \Hi\ component around SQ-A). A similar argument can be made for the fainter diffuse emission surrounding the radio ridge which, setting aside emission from NGC~7320 and SQ-B, shows no clear spectral index gradient. The dynamical timescale for the intruder galaxy to cross the E-W extent of the diffuse radio emission would be $>$150~Myr if it were moving in the plane of the sky at the sound speed of the hot IGrM (400-500\kmps). This would produce an E-W age gradient much stronger than the N-S gradient we observe along the ridge. If we conservatively consider that there is unlikely to be an age difference $\gtrsim$30~Myr between the east and west edges of the radio emission, this implies an angle to the line of sight of $\lesssim$20\degree. We note however that this is compatible with the $\sim$250\kmps\ transverse velocity suggested by \citet{Xuetal25} based on the morphology of \Hi\ clouds in the SQ-A region; small plane-of-sky velocities are not ruled out. 

One remaining mystery is the origin of the 700\kmps\ velocity gradient along the bridge structure that links the core of NGC~7319 and the shock ridge. The similarity of its multi-wavelength emission (CO, H$_2$, H$\alpha$, X-ray and radio continuum) to that seen in the main shock ridge, suggests that the bridge may be another tidal gas filament shock heated by collision with NGC~7318B. However, this is complicated by the relatively low radio luminosity and steep spectral index compared to the main ridge, which could indicate an older (and perhaps weaker) shock. Optical estimates of the extent of the disks of NGC~7319 and NGC~7318B suggest they overlap, in which case our limits on the direction of motion of the shock suggest that the outer disks have passed through each other. One possibility is that tidal forces associated with this encounter produced the gradient, with the impact on the gas increasing southwestward as the gravitational influence of NGC~7319 decreases and that of NGC~7318B increases. However, it seems questionable that tidal forces alone could accelerate the gas from the group's mean velocity to that of the intruder galaxy. More detailed observations of the bridge to determine its age and internal dynamics may be needed to resolve this question.

\section{Summary and Conclusions}
\label{sec::summary}

This paper follows on from our previous work on the \Hi\ structures in \SQ\ (paper~I), presenting the radio continuum properties of this unique group. Our analysis is based on new and previously unpublished wideband observations with MeerKAT, uGMRT, and the VLA, as well as reprocessed LOFAR HBA (LoTSS) observations, with a combined frequency coverage from 144~MHz to 6~GHz. These data provide crucial insight into the spectral properties of the shock-driven diffuse emission in the group, as well as higher resolution ($\sim$3\arcs) mapping of the main shock ridge. The combination of broad spectral coverage and high spatial resolution allowed us to constrain the age and dynamics of the interaction responsible for the famous radio/X-ray ridge, and the mechanism responsible for the radio emission. Our principal results are summarized below:

\begin{enumerate}
\item The new observations provide a detailed view of the ridge, its extension to the northwest, and surrounding diffuse emission which extends out to $\sim$30~kpc. Our view of the group is complicated by emission from the intergalactic SF regions SQ-A and SQ-B, SF in the foreground galaxy NGC~7320, and AGNs in some member galaxies. MeerKAT traces the faintest emission, including a northwest extension to the shock ridge, and an arc running along the line of the younger, northern tidal tail, connecting the shock ridge to SQ-B and extending somewhat beyond it. The VLA observations reveal substructures within the ridge, notably bright knots and filaments in its southern half, and branching at the southern tip. Overall, the radio morphology of the ridge is correlated with the X-ray and H$_2$ line emission, and with the high velocity component of the H$\alpha$ emission, but poorly correlated with star forming structures in the southern disk of NGC~7318B.

\item The ridge shows a distinct north-south spectral index gradient, with flatter indices ($\alpha\simeq$-0.53) in the south and steeper indices ($\alpha\simeq$-1.4) in the north. A stronger curvature in is also observed in the north ridge and NW extension. The spectral index of the diffuse emission around the ridge is relatively steep ($\alpha_{\rm 400MHz}^{\rm 1.28 GHz}\simeq$-1.2). Radio color-color plots show that the emission from the ridge and NW extension is well-described by the JP spectral aging model with $\alpha_{\rm inj}$=-0.7. We find that the emission is consistent with a single global spectrum, indicating that ridge and NW extension are a single structure with a dominant relativistic particle population originating via a single physical mechanism.

\item We detect a radio counterpart to the gaseous bridge which links the core of NGC~7319 to the shock ridge. While much fainter than the main ridge, the bridge emission also appears to have a non-thermal origin, with a steep spectral index and strong curvature ($\alpha_{400\,{\rm MHz}}^{1.28\,{\rm GHz}}=-1.11$, $\alpha_{1.28\,{\rm GHz}}^{3\,{\rm GHz}}=-1.54$). As with the main ridge, the bridge has previously been found to be highly multi-phase, and it therefore seems likely to have the same origin as the main ridge.

\item We estimate the equipartition magnetic field in the ridge to be 10~$\mu$G, in good agreement with past estimates. The spectral index gradient along the ridge translates to an age gradient, with ages of $\sim$20~Myr in the northern ridge, 10-12~Myr in the central clump B2, and 5-6~Myr in the southern tip and branch structures. This finding is consistent with previously identified trends in H$_2$ temperature and shock strength along the ridge \citep{Appleton2017} and our estimated ages are comparable to the typical ages of young star clusters identified in \hst\ and \jwst\ observations of \SQ, $\sim$5-10~Myr \citep{Aromaletal25}.

\item We consider different physical mechanisms by which the shock ridge may have formed, finding that the most likely is strong ($\mathcal{M}$=40-100) shocks in cold gas (primarily neutral atomic hydrogen) caused by the $\sim$1000\kmps\ collision between the intruder galaxy NGC~7318B and gas filaments drawn out of the other member galaxies during past tidal interactions. While star formation is clearly important in particular regions, we rule it out as a significant source of diffuse emission in the ridge. We find that weak ($\mathcal{M}\simeq$2-4) shocks in the hot IGrM are implausible as a formation mechanism for the ridge, as they would not produce the flat spectral indices observed in the southern ridge, and their expected radio luminosity is far too low. We also consider adiabatic compression of a pre-existing relativistic electron population in the IGrM, but find that implausible, as there is no clear source for such a seed population, and the required Mach numbers conflict with the X-ray observations. Strong shocks in the cold gas phase provide a highly efficient acceleration mechanism and naturally explain the correlated X-ray and radio emission; both relativistic and thermal plasmas would be produced by the same shock, with differences in structure likely arising from differences in pre-shock gas density, dust content, relative velocities, and post-shock aging.    

\item Based on the north-south age gradient along the shock ridge, we estimate the angle between the shock front and pre-existing tidal gas filaments was probably $\sim$15\degree. The lack of clear spectral index gradients in the diffuse radio emission surrounding the ridge suggests that there is unlikely to be a strong east-west velocity component, again probably $<$15-20\degree. We therefore conclude that the direction of motion of the shock is close to the line of sight, with the intruder galaxy moving toward us. This is consistent with the low radio polarization fraction found in prior studies \citep{NikielWroczynski2013,NikielWroczynski2020}, and with the close alignment between structures at different velocities within the group, e.g., \Hi, H$\alpha$, optical, and NIR structures in the ridge and around the SQ-A SF region. 

\end{enumerate}

\SQ\ provides one of the clearest, and certainly the most spectacular example of collisional shock heating on galaxy scales. Such shocks have been shown to be effective in heating cold, low-density gas from the interstellar medium into the hot, X-ray emitting phase \citep[as in, e.g., the Taffy galaxies or HCG~57A/D][]{Appletonetal15,OSullivanetal25} and the radio ridge in \SQ\ demonstrates their effectiveness as particle accelerators. While the collision in \SQ\ may be especially energetic, such interactions are inevitable as galaxies evolve and merge, and will drive shocks even at the more modest velocities typical among bound members of galaxy groups. Radio observations provide a crucial window on the physical processes involved and, as we have demonstrated, offer opportunities to constrain the dynamics and timescales of the interactions. The observations described in this paper provide an exceptionally deep and detailed view of the relativistic particle component of \SQ, limited mainly by the spatial resolution of our data. Higher resolution multi-band observations would complement the rich datasets being collected in other wavebands for this unique object (e.g., from \jwst\ and ALMA) shedding light on the physical properties of the radio bridge, branches, and filamentary ridge substructures.

\section*{acknowledgments}
We thank the anonymous referee for their careful reading and constructive comments on the manuscript. The authors thank the staff of the MeerKAT observatory for their help with the observations presented in this work.  The MeerKAT telescope is operated by the South African Radio Astronomy Observatory, which is a facility of the National Research Foundation, an agency of the Department of Science and Innovation. We thank the staff of the GMRT that made these observations possible. GMRT is run by the National Centre for Radio Astrophysics of the Tata Institute of Fundamental Research. Some of the computations in this paper were conducted on the Smithsonian High Performance Cluster (SI/HPC), Smithsonian Institution (\url{https://doi.org/10.25572/SIHPC}). This research has made use of the NASA/IPAC Extragalactic Database, which is funded by the National Aeronautics and Space Administration and operated by the California Institute of Technology. We acknowledge the usage of the HyperLeda database (http://leda.univ-lyon1.fr). Basic research in radio astronomy at the Naval Research Laboratory is supported by 6.1 Base funding.
UL acknowledges support from the project PID2023-150178NB-100, as well as from the  Consejer\'{i}a de Universidad, Investigaci\'{o}n e Innovaci\'{o}n and Gobierno de Espa\~{n}a and Uni\'{o}n Europea - NextGenerationEU through grant AST22$\_$4.4, financed by MCIN/AEI/10.13039/501100011033, and FQM108, financed by the Junta de Andaluc\'{i}a (Spain).

LOFAR \citep{Haarlem2013} is the Low Frequency Array designed and constructed by ASTRON. It has observing, data processing, and data storage facilities in several countries, which are owned by various parties (each with their own funding sources), and which are collectively operated by the LOFAR ERIC under a joint scientific policy. The LOFAR resources have benefited from the following recent major funding sources: CNRS-INSU, Observatoire de Paris and Universit\'e d'Orl\'eans, France; BMFTR, MKW-NRW, MPG, Germany; Science Foundation Ireland (SFI), Department of Business, Enterprise and Innovation (DBEI), Ireland; NWO, The Netherlands; The Science and Technology Facilities Council, UK; Ministry of Science and Higher Education, Poland; The Istituto Nazionale di Astrofisica (INAF), Italy.
This research made use of the Dutch national e-infrastructure with support of the SURF Cooperative (e-infra 180169) and the LOFAR e-infra group. The J\"ulich LOFAR Long Term Archive and the German LOFAR network are both coordinated and operated by the J\"ulich Supercomputing Centre (JSC), and computing resources on the supercomputer JUWELS at JSC were provided by the Gauss Centre for Supercomputing e.V. (grant CHTB00) through the John von Neumann Institute for Computing (NIC).
This research made use of the University of Hertfordshire high-performance computing facility and the LOFAR-UK computing facility located at the University of Hertfordshire and supported by STFC [ST/P000096/1], and of the Italian LOFAR-IT computing infrastructure supported and operated by INAF, including the resources within the PLEIADI special ``LOFAR'' project by USC-C of INAF, and by the Physics Department of Turin university (under an agreement with Consorzio Interuniversitario per la Fisica Spaziale) at the C3S Supercomputing Centre, Italy.
This research is part of the project LOFAR Data Valorization (LDV) [project numbers 2020.031, 2022.033, and 2024.047] of the research programme Computing Time on National Computer Facilities using SPIDER that is (co-)funded by the Dutch Research Council (NWO), hosted by SURF through the call for proposals of Computing Time on National Computer Facilities.

\facilities{MeerKAT}, {uGMRT}, {LOFAR}, {VLA}

\software{CARACal \citep{caracal2020}, AOflagger \citep{Offringa2010}, WSClean \citep{Offringa2014}, SPAM \citep{Intema2009}, Astropy \citep{astropy2013, astropy2018}, APLpy \citep{aplpy}, Matplotlib \citep{matplotlib}}

\bibliographystyle{aasjournal}
\bibliography{ref}

@ARTICLE{Chandra2017,
       author = {{Chandra}, Poonam and {Kanekar}, Nissim},
        title = "{Giant Metrewave Radio Telescope Monitoring of the Black Hole X-Ray Binary, V404 Cygni during Its 2015 June Outburst}",
      journal = {\apj},
         year = 2017,
        month = sep,
       volume = {846},
       number = {2},
          eid = {111},
        pages = {111},
          doi = {10.3847/1538-4357/aa85a2},
archivePrefix = {arXiv},
       eprint = {1708.02739},
 primaryClass = {astro-ph.HE},
       adsurl = {https://ui.adsabs.harvard.edu/abs/2017ApJ...846..111C}
}

@ARTICLE{Beck2005,
       author = {{Beck}, R. and {Krause}, M.},
        title = "{Revised equipartition and minimum energy formula for magnetic field strength estimates from radio synchrotron observations}",
      journal = {Astronomische Nachrichten},
         year = 2005,
        month = jul,
       volume = {326},
       number = {6},
        pages = {414-427},
          doi = {10.1002/asna.200510366},
archivePrefix = {arXiv},
       eprint = {astro-ph/0507367},
 primaryClass = {astro-ph},
       adsurl = {https://ui.adsabs.harvard.edu/abs/2005AN....326..414B}
}

@ARTICLE{Rajpurohit2024a,
       author = {{Rajpurohit}, K. and {Lovisari}, L. and {Botteon}, A. and {Jones}, C. and {Forman}, W. and {O'Sullivan}, E. and {van Weeren}, R.~J. and {HyeongHan}, K. and {Bonafede}, A. and {Jee}, M.~J. and {Vazza}, F. and {Brunetti}, G. and {Cho}, H. and {Dom{\'\i}nguez-Fern{\'a}ndez}, P. and {Stroe}, A. and {Finner}, K. and {Br{\"u}ggen}, M. and {Vrtilek}, J.~M. and {David}, L.~P. and {Schellenberger}, G. and {Wittman}, D. and {Lusetti}, G. and {Kraft}, R. and {De Gasperin}, F.},
        title = "{Abell 746: A Highly Disturbed Cluster Undergoing Multiple Mergers}",
      journal = {\apj},
         year = 2024,
        month = may,
       volume = {966},
       number = {1},
          eid = {38},
        pages = {38},
          doi = {10.3847/1538-4357/ad29fa},
archivePrefix = {arXiv},
       eprint = {2309.01716},
 primaryClass = {astro-ph.CO},
       adsurl = {https://ui.adsabs.harvard.edu/abs/2024ApJ...966...38R}
}

@ARTICLE{Harwood2013,
       author = {{Harwood}, Jeremy J. and {Hardcastle}, Martin J. and {Croston}, Judith H. and {Goodger}, Joanna L.},
        title = "{Spectral ageing in the lobes of FR-II radio galaxies: new methods of analysis for broad-band radio data}",
      journal = {\mnras},
         year = 2013,
        month = nov,
       volume = {435},
       number = {4},
        pages = {3353-3375},
          doi = {10.1093/mnras/stt1526},
archivePrefix = {arXiv},
       eprint = {1308.4137},
 primaryClass = {astro-ph.CO},
       adsurl = {https://ui.adsabs.harvard.edu/abs/2013MNRAS.435.3353H}
}

@ARTICLE{Tasse2021,
       author = {{Tasse}, C. and {Shimwell}, T. and {Hardcastle}, M.~J. and {O'Sullivan}, S.~P. and {van Weeren}, R. and {Best}, P.~N. and {Bester}, L. and {Hugo}, B. and {Smirnov}, O. and {Sabater}, J. and {Calistro-Rivera}, G. and {de Gasperin}, F. and {Morabito}, L.~K. and {R{\"o}ttgering}, H. and {Williams}, W.~L. and {Bonato}, M. and {Bondi}, M. and {Botteon}, A. and {Br{\"u}ggen}, M. and {Brunetti}, G. and {Chy{\.z}y}, K.~T. and {Garrett}, M.~A. and {G{\"u}rkan}, G. and {Jarvis}, M.~J. and {Kondapally}, R. and {Mandal}, S. and {Prandoni}, I. and {Repetti}, A. and {Retana-Montenegro}, E. and {Schwarz}, D.~J. and {Shulevski}, A. and {Wiaux}, Y.},
        title = "{The LOFAR Two-meter Sky Survey: Deep Fields Data Release 1. I. Direction-dependent calibration and imaging}",
      journal = {\aap},
         year = 2021,
        month = apr,
       volume = {648},
          eid = {A1},
        pages = {A1},
          doi = {10.1051/0004-6361/202038804},
archivePrefix = {arXiv},
       eprint = {2011.08328},
 primaryClass = {astro-ph.IM},
       adsurl = {https://ui.adsabs.harvard.edu/abs/2021A&A...648A...1T}
}

@ARTICLE{vanWeeren2021,
       author = {{van Weeren}, R.~J. and {Shimwell}, T.~W. and {Botteon}, A. and {Brunetti}, G. and {Br{\"u}ggen}, M. and {Boxelaar}, J.~M. and {Cassano}, R. and {Di Gennaro}, G. and {Andrade-Santos}, F. and {Bonnassieux}, E. and {Bonafede}, A. and {Cuciti}, V. and {Dallacasa}, D. and {de Gasperin}, F. and {Gastaldello}, F. and {Hardcastle}, M.~J. and {Hoeft}, M. and {Kraft}, R.~P. and {Mandal}, S. and {Rossetti}, M. and {R{\"o}ttgering}, H.~J.~A. and {Tasse}, C. and {Wilber}, A.~G.},
        title = "{LOFAR observations of galaxy clusters in HETDEX. Extraction and self-calibration of individual LOFAR targets}",
      journal = {\aap},
         year = 2021,
        month = jul,
       volume = {651},
          eid = {A115},
        pages = {A115},
          doi = {10.1051/0004-6361/202039826},
archivePrefix = {arXiv},
       eprint = {2011.02387},
 primaryClass = {astro-ph.CO},
       adsurl = {https://ui.adsabs.harvard.edu/abs/2021A&A...651A.115V}
}

@ARTICLE{Shimwell2026,
       author = {{Shimwell}, T.~W. and {Hardcastle}, M.~J. and {Tasse}, C. and {Drabent}, A. and {Botteon}, A. and {Williams}, W.~L. and {Best}, P.~N. and {R{\"o}ttgering}, H.~J.~A. and {Br{\"u}ggen}, M. and {Brunetti}, G. and {Callingham}, J.~R. and {Chy{\.z}y}, K.~T. and {Conway}, J.~E. and {De Gasperin}, F. and {Haverkorn}, M. and {Horellou}, C. and {Jackson}, N. and {Miley}, G.~K. and {Morabito}, L.~K. and {Morganti}, R. and {O'Sullivan}, S.~P. and {Schwarz}, D.~J. and {Smith}, D.~J.~B. and {van Weeren}, R.~J. and {Vedantham}, H.~K. and {White}, G.~J. and {Ahmadi}, A. and {Alegre}, L. and {Arias}, M. and {Asabere}, B. and {Bahr-Kalus}, B. and {Barkus}, B. and {Bilicki}, M. and {B{\"o}hme}, L. and {Brentjens}, M. and {Brienza}, M. and {Bomans}, D.~J. and {Bonafede}, A. and {Bonato}, M. and {Bonnassieux}, E. and {Boxelaar}, J.~M. and {Camera}, S. and {Cassano}, R. and {Chilufya}, J. and {Cianfaglione}, M. and {Croston}, J.~H. and {Cuciti}, V. and {Dabhade}, P. and {De Rubeis}, E. and {de Jong}, J.~M.~G.~H.~J. and {Dallacasa}, D. and {Dettmar}, R.~J. and {Duncan}, K.~J. and {Di Gennaro}, G. and {Edler}, H.~W. and {Groeneveld}, C. and {G{\"u}rkan}, G. and {Hajduk}, M. and {Hale}, C.~L. and {Heesen}, V. and {Hoang}, D.~N. and {Hoeft}, M. and {Holties}, H. and {Horton}, M.~A. and {Iacobelli}, M. and {Jamrozy}, M. and {Jarvis}, M.~J. and {Jelic}, V. and {Kadler}, M. and {Kondapally}, R. and {Kunert-Bajraszewska}, M. and {Loose}, M. and {Magliocchetti}, M. and {Ma{\l}ek}, K. and {Manzano}, C. and {McKean}, J.~P. and {Mevius}, M. and {Mingo}, B. and {Miskolczi}, A. and {Misra}, A. and {Mold{\'o}n}, J. and {Nair}, D.~G. and {Nakoneczny}, S.~J. and {Orru}, E. and {Pashapour-Ahmadabadi}, M. and {Pasini}, T. and {Petley}, J. and {Pierce}, J.~C.~S. and {Prandoni}, I. and {Rafferty}, D. and {Rajpurohit}, K. and {Riseley}, C.~J. and {Roberts}, I.~D. and {Sethi}, S. and {Shulevski}, A. and {Stein}, M. and {Stuardi}, C. and {Sweijen}, F. and {ter Veen}, S. and {Timmerman}, R. and {Vaccari}, M. and {Wijnholds}, S.},
        title = "{The LOFAR Two-metre Sky Survey: VII. Third Data Release}",
      journal = {\aap},
         year = 2026,
        month = mar,
       volume = {707},
          eid = {A198},
        pages = {A198},
          doi = {10.1051/0004-6361/202557749},
archivePrefix = {arXiv},
       eprint = {2602.15949},
 primaryClass = {astro-ph.GA},
       adsurl = {https://ui.adsabs.harvard.edu/abs/2026A&A...707A.198S}
}

@ARTICLE{PaperI,
       author = {{Rajpurohit}, K. and {O'Sullivan}, E. and {Schellenberger}, G. and {Vrtilek}, J.~M. and {David}, L.~P. and {Giacintucci}, S. and {Appleton}, P.~N. and {Xu}, C.~K. and {Cheng}, C. and {Deb}, T.},
        title = "{A MeerKAT View of the Neutral Atomic Gas in Stephan's Quintet}",
      journal = {\apj},
         year = 2026,
        month = may,
       volume = {1002},
       number = {1},
          eid = {18},
        pages = {18},
          doi = {10.3847/1538-4357/ae58a2}
}

@ARTICLE{Rajpurohit2024,
       author = {{Rajpurohit}, K. and {O'Sullivan}, E. and {Schellenberger}, G. and {Brienza}, M. and {Vrtilek}, J.~M. and {Forman}, W. and {David}, L.~P. and {Clarke}, T. and {Botteon}, A. and {Vazza}, F. and {Giacintucci}, S. and {Jones}, C. and {Br{\"u}ggen}, M. and {Shimwell}, T.~W. and {Drabent}, A. and {Loi}, F. and {Loubser}, S.~I. and {Kolokythas}, K. and {Babyk}, I. and {R{\"o}ttgering}, H.~J.~A.},
        title = "{A Deep Dive into the NGC 741 Galaxy Group: Insights into a Spectacular Head-tail Radio Galaxy from VLA, MeerKAT, uGMRT, and LOFAR}",
      journal = {\apj},
         year = 2024,
        month = nov,
       volume = {976},
       number = {1},
          eid = {64},
        pages = {64},
          doi = {10.3847/1538-4357/ad8136},
archivePrefix = {arXiv},
       eprint = {2408.15197},
 primaryClass = {astro-ph.GA},
       adsurl = {https://ui.adsabs.harvard.edu/abs/2024ApJ...976...64R}
}

@ARTICLE{casa2022,
       author = {{CASA Team} and {Bean}, Ben and {Bhatnagar}, Sanjay and {Castro}, Sandra and {Donovan Meyer}, Jennifer and {Emonts}, Bjorn and {Garcia}, Enrique and {Garwood}, Robert and {Golap}, Kumar and {Gonzalez Villalba}, Justo and {Harris}, Pamela and {Hayashi}, Yohei and {Hoskins}, Josh and {Hsieh}, Mingyu and {Jagannathan}, Preshanth and {Kawasaki}, Wataru and {Keimpema}, Aard and {Kettenis}, Mark and {Lopez}, Jorge and {Marvil}, Joshua and {Masters}, Joseph and {McNichols}, Andrew and {Mehringer}, David and {Miel}, Renaud and {Moellenbrock}, George and {Montesino}, Federico and {Nakazato}, Takeshi and {Ott}, Juergen and {Petry}, Dirk and {Pokorny}, Martin and {Raba}, Ryan and {Rau}, Urvashi and {Schiebel}, Darrell and {Schweighart}, Neal and {Sekhar}, Srikrishna and {Shimada}, Kazuhiko and {Small}, Des and {Steeb}, Jan-Willem and {Sugimoto}, Kanako and {Suoranta}, Ville and {Tsutsumi}, Takahiro and {van Bemmel}, Ilse M. and {Verkouter}, Marjolein and {Wells}, Akeem and {Xiong}, Wei and {Szomoru}, Arpad and {Griffith}, Morgan and {Glendenning}, Brian and {Kern}, Jeff},
        title = "{CASA, the Common Astronomy Software Applications for Radio Astronomy}",
      journal = {\pasp},
         year = 2022,
        month = nov,
       volume = {134},
       number = {1041},
          eid = {114501},
        pages = {114501},
          doi = {10.1088/1538-3873/ac9642},
archivePrefix = {arXiv},
       eprint = {2210.02276},
 primaryClass = {astro-ph.IM},
       adsurl = {https://ui.adsabs.harvard.edu/abs/2022PASP..134k4501C}
}

@INPROCEEDINGS{McMullin2007,
   author = {{McMullin}, J.~P. and {Waters}, B. and {Schiebel}, D. and {Young}, W. and 
	{Golap}, K.},
    title = "{CASA Architecture and Applications}",
booktitle = {Astronomical Data Analysis Software and Systems XVI},
     year = 2007,
   series = {Astronomical Society of the Pacific Conference Series},
   volume = 376,
   editor = {{Shaw}, R.~A. and {Hill}, F. and {Bell}, D.~J.},
    month = oct,
    pages = {127},
   adsurl = {http://adsabs.harvard.edu/abs/2007ASPC..376..127M}
}

@PHDTHESIS{Briggs1995,
       author = {{Briggs}, Daniel Shenon},
        title = "{High fidelity deconvolution of moderately resolved sources}",
       school = {New Mexico Institute of Mining and Technology},
         year = 1995,
        month = jan,
       adsurl = {https://ui.adsabs.harvard.edu/abs/1995PhDT.......238B}
}

@ARTICLE{Kenyon2018,
       author = {{Kenyon}, J.~S. and {Smirnov}, O.~M. and {Grobler}, T.~L. and {Perkins}, S.~J.},
        title = "{CUBICAL - fast radio interferometric calibration suite exploiting complex optimization}",
      journal = {\mnras},
         year = 2018,
        month = aug,
       volume = {478},
       number = {2},
        pages = {2399-2415},
          doi = {10.1093/mnras/sty1221},
archivePrefix = {arXiv},
       eprint = {1805.03410},
 primaryClass = {astro-ph.IM},
       adsurl = {https://ui.adsabs.harvard.edu/abs/2018MNRAS.478.2399K}
}

@ARTICLE{Offringa2014,
       author = {{Offringa}, A.~R. and {McKinley}, B. and {Hurley-Walker}, N. and
         {Briggs}, F.~H. and {Wayth}, R.~B. and {Kaplan}, D.~L. and
         {Bell}, M.~E. and {Feng}, L. and {Neben}, A.~R. and {Hughes}, J.~D. and
         {Rhee}, J. and {Murphy}, T. and {Bhat}, N.~D.~R. and {Bernardi}, G. and
         {Bowman}, J.~D. and {Cappallo}, R.~J. and {Corey}, B.~E. and {Deshpand
        e}, A.~A. and {Emrich}, D. and {Ewall-Wice}, A. and {Gaensler}, B.~M. and
         {Goeke}, R. and {Greenhill}, L.~J. and {Hazelton}, B.~J. and
         {Hindson}, L. and {Johnston-Hollitt}, M. and {Jacobs}, D.~C. and
         {Kasper}, J.~C. and {Kratzenberg}, E. and {Lenc}, E. and
         {Lonsdale}, C.~J. and {Lynch}, M.~J. and {McWhirter}, S.~R. and
         {Mitchell}, D.~A. and {Morales}, M.~F. and {Morgan}, E. and
         {Kudryavtseva}, N. and {Oberoi}, D. and {Ord}, S.~M. and {Pindor}, B. and
         {Procopio}, P. and {Prabu}, T. and {Riding}, J. and {Roshi}, D.~A. and
         {Shankar}, N. Udaya and {Srivani}, K.~S. and {Subrahmanyan}, R. and
         {Tingay}, S.~J. and {Waterson}, M. and {Webster}, R.~L. and
         {Whitney}, A.~R. and {Williams}, A. and {Williams}, C.~L.},
        title = "{WSCLEAN: an implementation of a fast, generic wide-field imager for radio astronomy}",
      journal = {\mnras},
         year = 2014,
        month = oct,
       volume = {444},
       number = {1},
        pages = {606-619},
          doi = {10.1093/mnras/stu1368},
archivePrefix = {arXiv},
       eprint = {1407.1943},
 primaryClass = {astro-ph.IM},
       adsurl = {https://ui.adsabs.harvard.edu/abs/2014MNRAS.444..606O}
}

@ARTICLE{Astropy2013,
   author = {{Astropy Collaboration} and {Robitaille}, T.~P. and {Tollerud}, E.~J. and 
	{Greenfield}, P. and {Droettboom}, M. and {Bray}, E. and {Aldcroft}, T. and 
	{Davis}, M. and {Ginsburg}, A. and {Price-Whelan}, A.~M. and 
	{Kerzendorf}, W.~E. and {Conley}, A. and {Crighton}, N. and 
	{Barbary}, K. and {Muna}, D. and {Ferguson}, H. and {Grollier}, F. and 
	{Parikh}, M.~M. and {Nair}, P.~H. and {Unther}, H.~M. and {Deil}, C. and 
	{Woillez}, J. and {Conseil}, S. and {Kramer}, R. and {Turner}, J.~E.~H. and 
	{Singer}, L. and {Fox}, R. and {Weaver}, B.~A. and {Zabalza}, V. and 
	{Edwards}, Z.~I. and {Azalee Bostroem}, K. and {Burke}, D.~J. and 
	{Casey}, A.~R. and {Crawford}, S.~M. and {Dencheva}, N. and 
	{Ely}, J. and {Jenness}, T. and {Labrie}, K. and {Lim}, P.~L. and 
	{Pierfederici}, F. and {Pontzen}, A. and {Ptak}, A. and {Refsdal}, B. and 
	{Servillat}, M. and {Streicher}, O.},
    title = "{Astropy: A community Python package for astronomy}",
  journal = {\aap},
archivePrefix = "arXiv",
   eprint = {1307.6212},
 primaryClass = "astro-ph.IM",
     year = 2013,
    month = oct,
   volume = 558,
      eid = {A33},
    pages = {A33},
      doi = {10.1051/0004-6361/201322068},
   adsurl = {http://adsabs.harvard.edu/abs/2013A%26A...558A..33A}
}

@ARTICLE{astropy2018,
       author = {{Astropy Collaboration} and {Price-Whelan}, A.~M. and {Sip{\H{o}}cz}, B.~M. and {G{\"u}nther}, H.~M. and {Lim}, P.~L. and {Crawford}, S.~M. and {Conseil}, S. and {Shupe}, D.~L. and {Craig}, M.~W. and {Dencheva}, N. and {Ginsburg}, A. and {VanderPlas}, J.~T. and {Bradley}, L.~D. and {P{\'e}rez-Su{\'a}rez}, D. and {de Val-Borro}, M. and {Aldcroft}, T.~L. and {Cruz}, K.~L. and {Robitaille}, T.~P. and {Tollerud}, E.~J. and {Ardelean}, C. and {Babej}, T. and {Bach}, Y.~P. and {Bachetti}, M. and {Bakanov}, A.~V. and {Bamford}, S.~P. and {Barentsen}, G. and {Barmby}, P. and {Baumbach}, A. and {Berry}, K.~L. and {Biscani}, F. and {Boquien}, M. and {Bostroem}, K.~A. and {Bouma}, L.~G. and {Brammer}, G.~B. and {Bray}, E.~M. and {Breytenbach}, H. and {Buddelmeijer}, H. and {Burke}, D.~J. and {Calderone}, G. and {Cano Rodr{\'\i}guez}, J.~L. and {Cara}, M. and {Cardoso}, J.~V.~M. and {Cheedella}, S. and {Copin}, Y. and {Corrales}, L. and {Crichton}, D. and {D'Avella}, D. and {Deil}, C. and {Depagne}, {\'E}. and {Dietrich}, J.~P. and {Donath}, A. and {Droettboom}, M. and {Earl}, N. and {Erben}, T. and {Fabbro}, S. and {Ferreira}, L.~A. and {Finethy}, T. and {Fox}, R.~T. and {Garrison}, L.~H. and {Gibbons}, S.~L.~J. and {Goldstein}, D.~A. and {Gommers}, R. and {Greco}, J.~P. and {Greenfield}, P. and {Groener}, A.~M. and {Grollier}, F. and {Hagen}, A. and {Hirst}, P. and {Homeier}, D. and {Horton}, A.~J. and {Hosseinzadeh}, G. and {Hu}, L. and {Hunkeler}, J.~S. and {Ivezi{\'c}}, {\v{Z}}. and {Jain}, A. and {Jenness}, T. and {Kanarek}, G. and {Kendrew}, S. and {Kern}, N.~S. and {Kerzendorf}, W.~E. and {Khvalko}, A. and {King}, J. and {Kirkby}, D. and {Kulkarni}, A.~M. and {Kumar}, A. and {Lee}, A. and {Lenz}, D. and {Littlefair}, S.~P. and {Ma}, Z. and {Macleod}, D.~M. and {Mastropietro}, M. and {McCully}, C. and {Montagnac}, S. and {Morris}, B.~M. and {Mueller}, M. and {Mumford}, S.~J. and {Muna}, D. and {Murphy}, N.~A. and {Nelson}, S. and {Nguyen}, G.~H. and {Ninan}, J.~P. and {N{\"o}the}, M. and {Ogaz}, S. and {Oh}, S. and {Parejko}, J.~K. and {Parley}, N. and {Pascual}, S. and {Patil}, R. and {Patil}, A.~A. and {Plunkett}, A.~L. and {Prochaska}, J.~X. and {Rastogi}, T. and {Reddy Janga}, V. and {Sabater}, J. and {Sakurikar}, P. and {Seifert}, M. and {Sherbert}, L.~E. and {Sherwood-Taylor}, H. and {Shih}, A.~Y. and {Sick}, J. and {Silbiger}, M.~T. and {Singanamalla}, S. and {Singer}, L.~P. and {Sladen}, P.~H. and {Sooley}, K.~A. and {Sornarajah}, S. and {Streicher}, O. and {Teuben}, P. and {Thomas}, S.~W. and {Tremblay}, G.~R. and {Turner}, J.~E.~H. and {Terr{\'o}n}, V. and {van Kerkwijk}, M.~H. and {de la Vega}, A. and {Watkins}, L.~L. and {Weaver}, B.~A. and {Whitmore}, J.~B. and {Woillez}, J. and {Zabalza}, V. and {Astropy Contributors}},
        title = "{The Astropy Project: Building an Open-science Project and Status of the v2.0 Core Package}",
      journal = {\aj},
         year = 2018,
        month = sep,
       volume = {156},
       number = {3},
          eid = {123},
        pages = {123},
          doi = {10.3847/1538-3881/aabc4f},
archivePrefix = {arXiv},
       eprint = {1801.02634},
 primaryClass = {astro-ph.IM},
       adsurl = {https://ui.adsabs.harvard.edu/abs/2018AJ....156..123A}
}

@ARTICLE{Scaife2012,
       author = {{Scaife}, Anna M.~M. and {Heald}, George H.},
        title = "{A broad-band flux scale for low-frequency radio telescopes}",
      journal = {\mnras},
         year = 2012,
        month = jun,
       volume = {423},
       number = {1},
        pages = {L30-L34},
          doi = {10.1111/j.1745-3933.2012.01251.x},
archivePrefix = {arXiv},
       eprint = {1203.0977},
 primaryClass = {astro-ph.IM},
       adsurl = {https://ui.adsabs.harvard.edu/abs/2012MNRAS.423L..30S}
}

@ARTICLE{Brunetti1997,
       author = {{Brunetti}, G. and {Setti}, G. and {Comastri}, A.},
        title = "{Inverse Compton X-rays from strong FRII radio-galaxies.}",
      journal = {\aap},
         year = 1997,
        month = sep,
       volume = {325},
        pages = {898-910},
          doi = {10.48550/arXiv.astro-ph/9704162},
archivePrefix = {arXiv},
       eprint = {astro-ph/9704162},
 primaryClass = {astro-ph},
       adsurl = {https://ui.adsabs.harvard.edu/abs/1997A&A...325..898B}
}

@ARTICLE{Xu2003,
       author = {{Xu}, C.~K. and {Lu}, N. and {Condon}, J.~J. and {Dopita}, M. and {Tuffs}, R.~J.},
        title = "{Physical Conditions and Star Formation Activity in the Intragroup Medium of Stephan's Quintet}",
      journal = {\apj},
         year = 2003,
        month = oct,
       volume = {595},
       number = {2},
        pages = {665-684},
          doi = {10.1086/377445}}

@ARTICLE{Allen1972,
       author = {{Allen}, Ronald J. and {Hartsuiker}, Jacob W.},
        title = "{Radio Continuum Emission at 21 cm near Stephan's Quintet}",
      journal = {\nat},
         year = 1972,
        month = oct,
       volume = {239},
       number = {5371},
        pages = {324-325},
          doi = {10.1038/239324a0}}

@ARTICLE{vanderHulst1981,
       author = {{van der Hulst}, J.~M. and {Rots}, A.~H.},
        title = "{VLA observations of the radio continuum emission from Stephan's Quintet.}",
      journal = {\aj},
         year = 1981,
        month = dec,
       volume = {86},
        pages = {1775-1780},
          doi = {10.1086/113060}}

@ARTICLE{Williams2002,
       author = {{Williams}, B.~A. and {Yun}, Min S. and {Verdes-Montenegro}, L.},
        title = "{The VLA H I Observations of Stephan's Quintet (HCG 92)}",
      journal = {\aj},
         year = 2002,
        month = may,
       volume = {123},
       number = {5},
        pages = {2417-2437},
          doi = {10.1086/339839}}

@ARTICLE{Shostak1984,
       author = {{Shostak}, G.~S. and {Sullivan}, III, W.~T. and {Allen}, R.~J.},
        title = "{H I synthesis observations of the high-redshift galaxies in Stephan'sQuintet.}",
      journal = {\aap},
         year = 1984,
        month = oct,
       volume = {139},
        pages = {15-24}}

@ARTICLE{Arnaudova2024,
       author = {{Arnaudova}, M.~I. and {Das}, S. and {Smith}, D.~J.~B. and {Hardcastle}, M.~J. and {Hatch}, N. and {Trager}, S.~C. and {Smith}, R.~J. and {Drake}, A.~B. and {McGarry}, J.~C. and {Shenoy}, S. and {Stott}, J.~P. and {Knapen}, J.~H. and {Hess}, K.~M. and {Duncan}, K.~J. and {Gloudemans}, A. and {Best}, P.~N. and {Garc{\'\i}a-Benito}, R. and {Kondapally}, R. and {Balcells}, M. and {Couto}, G.~S. and {Abrams}, D.~C. and {Aguado}, D. and {Aguerri}, J.~A.~L. and {Barrena}, R. and {Benn}, C.~R. and {Bensby}, T. and {Berlanas}, S.~R. and {Bettoni}, D. and {Cano-Infantes}, D. and {Carrera}, R. and {Concepci{\'o}n}, P.~J. and {Dalton}, G.~B. and {D'Ago}, G. and {Dee}, K. and {Dom{\'\i}nguez-Palmero}, L. and {Drew}, J.~E. and {Escott}, E.~L. and {Fari{\~n}a}, C. and {Fossati}, M. and {Fumagalli}, M. and {Gafton}, E. and {Gribbin}, F.~J. and {Hughes}, S. and {Iovino}, A. and {Jin}, S. and {Lewis}, I.~J. and {Longhetti}, M. and {M{\'e}ndez-Abreu}, J. and {Mercurio}, A. and {Molaeinezhad}, A. and {Molinari}, E. and {Mongui{\'o}}, M. and {Murphy}, D.~N.~A. and {Pic{\'o}}, S. and {Pieri}, M.~M. and {Ridings}, A.~W. and {Romero-G{\'o}mez}, M. and {Schallig}, E. and {Shimwell}, T.~W. and {Skvar{\v{c}}}, J. and {Stuik}, R. and {Vallenari}, A. and {van der Hulst}, J.~M. and {Walton}, N.~A. and {Worley}, C.~C.},
        title = "{WEAVE First Light Observations: Origin and Dynamics of the Shock Front in Stephan's Quintet}",
      journal = {\mnras},
         year = 2024,
        month = dec,
       volume = {535},
       number = {3},
        pages = {2269-2290},
          doi = {10.1093/mnras/stae2235}}

@ARTICLE{Aoki1999,
       author = {{Aoki}, Kentaro and {Kosugi}, George and {Wilson}, Andrew S. and {Yoshida}, Michitoshi},
        title = "{The Radio Emission of the Seyfert Galaxy NGC 7319}",
      journal = {\apj},
         year = 1999,
        month = aug,
       volume = {521},
       number = {2},
        pages = {565-571},
          doi = {10.1086/307559}}

@ARTICLE{Xanthopoulos2004,
       author = {{Xanthopoulos}, E. and {Muxlow}, T.~W.~B. and {Thomasson}, P. and {Garrington}, S.~T.},
        title = "{MERLIN observations of Stephan's Quintet}",
      journal = {\mnras},
         year = 2004,
        month = oct,
       volume = {353},
       number = {4},
        pages = {1117-1125},
          doi = {10.1111/j.1365-2966.2004.08133.x}}

@ARTICLE{NikielWroczynski2013,
       author = {{Nikiel-Wroczy{\'n}ski}, B. and {Soida}, M. and {Urbanik}, M. and {Beck}, R. and {Bomans}, D.~J.},
        title = "{Intergalactic magnetic fields in Stephan's Quintet}",
      journal = {\mnras},
         year = 2013,
        month = oct,
       volume = {435},
       number = {1},
        pages = {149-157},
          doi = {10.1093/mnras/stt1263}}

@ARTICLE{NikielWroczynski2020,
       author = {{Nikiel-Wroczy{\'n}ski}, B{\l}a{\.z}ej and {Soida}, Marian and {Heald}, George and {Urbanik}, Marek},
        title = "{A Large-scale, Regular Intergalactic Magnetic Field Associated with Stephan's Quintet?}",
      journal = {\apj},
         year = 2020,
        month = aug,
       volume = {898},
       number = {2},
          eid = {110},
        pages = {110},
          doi = {10.3847/1538-4357/ab9d89}}

@ARTICLE{Konstantopoulos2010,
   author = {{Konstantopoulos}, I.~S. and {Gallagher}, S.~C. and {Fedotov}, K. and {Durrell}, P.~R. and {Heiderman}, A. and {Elmegreen}, D.~M. and {Charlton}, J.~C. and {Hibbard}, J.~E. and {Tzanavaris}, P. and {Chandar}, R. and {Johnson}, K.~E. and {et al.}},
    title = "{Galaxy Evolution in a Complex Environment: A Multi-wavelength Study of HCG 7}",
  journal = {\apj},
     year = 2010,
    month = nov,
   volume = 723,
    pages = {197},
      doi = {10.1088/0004-637X/723/1/197}}

@ARTICLE{OSullivan2009,
   author = {{O'Sullivan}, E. and {Giacintucci}, S. and {Vrtilek}, J.~M. and {Raychaudhury}, S. and {David}, L.~P.},
    title = "{A Chandra X-ray View of Stephan's Quintet: Shocks and Star Formation}",
  journal = {\apj},
archivePrefix = "arXiv",
   eprint = {0812.0383},
     year = 2009,
    month = aug,
   volume = 701,
    pages = {1560},
      doi = {10.1088/0004-637X/701/2/1560}}

@ARTICLE{Sulentic2001,
   author = {{Sulentic}, J.~W. and {Rosado}, M. and {Dultzin-Hacyan}, D. and {Verdes-Montenegro}, L. and {Trinchieri}, G. and {Xu}, C. and {Pietsch}, W.},
    title = "{A Multiwavelength Study of Stephan's Quintet}",
  journal = {\aj},
   eprint = {arXiv:astro-ph/0111155},
     year = 2001,
    month = dec,
   volume = 122,
    pages = {2993},
      doi = {10.1086/324455}}

@ARTICLE{Appleton2013,
       author = {{Appleton}, P.~N. and {Guillard}, P. and {Boulanger}, F. and {Cluver}, M.~E. and {Ogle}, P. and {Falgarone}, E. and {Pineau des For{\^e}ts}, G. and {O'Sullivan}, E. and {Duc}, P. -A. and {Gallagher}, S. and {Gao}, Y. and {Jarrett}, T. and {Konstantopoulos}, I. and {Lisenfeld}, U. and {Lord}, S. and {Lu}, N. and {Peterson}, B.~W. and {Struck}, C. and {Sturm}, E. and {Tuffs}, R. and {Valchanov}, I. and {van der Werf}, P. and {Xu}, K.~C.},
        title = "{Shock-enhanced C$^{+}$ Emission and the Detection of H$_{2}$O from the Stephan's Quintet Group-wide Shock Using Herschel}",
      journal = {\apj},
         year = 2013,
        month = nov,
       volume = {777},
       number = {1},
          eid = {66},
        pages = {66},
          doi = {10.1088/0004-637X/777/1/66}}

@ARTICLE{Appleton2017,
       author = {{Appleton}, P.~N. and {Guillard}, P. and {Togi}, A. and {Alatalo}, K. and {Boulanger}, F. and {Cluver}, M. and {Pineau des For{\^e}ts}, G. and {Lisenfeld}, U. and {Ogle}, P. and {Xu}, C.~K.},
        title = "{Powerful H$_{2}$ Line Cooling in Stephan{\textquoteright}s Quintet. II. Group-wide Gas and Shock Modeling of the Warm H$_{2}$ and a Comparison with [C II] 157.7 {\ensuremath{\mu}}m Emission and Kinematics}",
      journal = {\apj},
         year = 2017,
        month = feb,
       volume = {836},
       number = {1},
          eid = {76},
        pages = {76},
          doi = {10.3847/1538-4357/836/1/76}}

@ARTICLE{DuartePuertas2019,
       author = {{Duarte Puertas}, S. and {Iglesias-P{\'a}ramo}, J. and {Vilchez}, J.~M. and {Drissen}, L. and {Kehrig}, C. and {Martin}, T.},
        title = "{Searching for intergalactic star forming regions in Stephan's Quintet with SITELLE. I. Ionised gas structures and kinematics}",
      journal = {\aap},
         year = 2019,
        month = sep,
       volume = {629},
          eid = {A102},
        pages = {A102},
          doi = {10.1051/0004-6361/201935686}}

@ARTICLE{Emonts2025,
       author = {{Emonts}, B.~H.~C. and {Appleton}, P.~N. and {Lisenfeld}, U. and {Guillard}, P. and {Xu}, C.~K. and {Reach}, W.~T. and {Barcos-Mu{\~n}oz}, L. and {Labiano}, A. and {Ogle}, P.~M. and {O'Sullivan}, E. and {Togi}, A. and {Gallagher}, S.~C. and {Aromal}, P. and {Duc}, P. -A. and {Alatalo}, K. and {Boulanger}, F. and {D{\'\i}az-Santos}, T. and {Helou}, G.},
        title = "{Bird's-eye View of Molecular Gas across Stephan's Quintet Galaxy Group and Intragroup Medium}",
      journal = {\apj},
         year = 2025,
        month = jan,
       volume = {978},
       number = {1},
          eid = {111},
        pages = {111},
          doi = {10.3847/1538-4357/ad957c}}

@ARTICLE{Appleton2023,
       author = {{Appleton}, P.~N. and {Guillard}, P. and {Emonts}, Bjorn and {Boulanger}, Francois and {Togi}, Aditya and {Reach}, William T. and {Alatalo}, Katherine and {Cluver}, M. and {Diaz Santos}, T. and {Duc}, P. -A. and {Gallagher}, S. and {Ogle}, P. and {O'Sullivan}, E. and {Voggel}, K. and {Xu}, C.~K.},
        title = "{Multiphase Gas Interactions on Subarcsec Scales in the Shocked Intergalactic Medium of Stephan's Quintet with JWST and ALMA}",
      journal = {\apj},
         year = 2023,
        month = jul,
       volume = {951},
       number = {2},
          eid = {104},
        pages = {104},
          doi = {10.3847/1538-4357/accc2a}}

@ARTICLE{Guillard2009,
       author = {{Guillard}, P. and {Boulanger}, F. and {Pineau Des For{\^e}ts}, G. and {Appleton}, P.~N.},
        title = "{H$_{2}$ formation and excitation in the Stephan's Quintet galaxy-wide collision}",
      journal = {\aap},
         year = 2009,
        month = aug,
       volume = {502},
       number = {2},
        pages = {515-528},
          doi = {10.1051/0004-6361/200811263}}

@ARTICLE{Guillard2012,
       author = {{Guillard}, P. and {Boulanger}, F. and {Pineau des For{\^e}ts}, G. and {Falgarone}, E. and {Gusdorf}, A. and {Cluver}, M.~E. and {Appleton}, P.~N. and {Lisenfeld}, U. and {Duc}, P. -A. and {Ogle}, P.~M. and {Xu}, C.~K.},
        title = "{Turbulent Molecular Gas and Star Formation in the Shocked Intergalactic Medium of Stephan's Quintet}",
      journal = {\apj},
         year = 2012,
        month = apr,
       volume = {749},
       number = {2},
          eid = {158},
        pages = {158},
          doi = {10.1088/0004-637X/749/2/158}}

@ARTICLE{Xu2022,
       author = {{Xu}, C.~K. and {Cheng}, C. and {Appleton}, P.~N. and {Duc}, P. -A. and {Gao}, Y. and {Tang}, N. -Y. and {Yun}, M. and {Dai}, Y.~S. and {Huang}, J. -S. and {Lisenfeld}, U. and {Renaud}, F.},
        title = "{A 0.6 Mpc H I structure associated with Stephan's Quintet}",
      journal = {\nat},
         year = 2022,
        month = oct,
       volume = {610},
       number = {7932},
        pages = {461-466},
          doi = {10.1038/s41586-022-05206-x}
}

@ARTICLE{Cheng2023,
       author = {{Cheng}, Cheng and {Xu}, Cong Kevin and {Appleton}, P.~N. and {Duc}, P. -A. and {Tang}, N. -Y. and {Dai}, Y. -S. and {Huang}, J. -S. and {Lisenfeld}, U. and {Renaud}, F. and {He}, Chuan and {Feng}, Hai-Cheng},
        title = "{Deep H I Mapping of Stephan's Quintet and Its Neighborhood}",
      journal = {\apj},
         year = 2023,
        month = sep,
       volume = {954},
       number = {1},
          eid = {74},
        pages = {74},
          doi = {10.3847/1538-4357/ace03e}
}

@ARTICLE{Borthakur2010,
       author = {{Borthakur}, Sanchayeeta and {Yun}, Min Su and {Verdes-Montenegro}, Lourdes},
        title = "{Detection of Diffuse Neutral Intragroup Medium in Hickson Compact Groups}",
      journal = {\apj},
         year = 2010,
        month = feb,
       volume = {710},
       number = {1},
        pages = {385-407},
          doi = {10.1088/0004-637X/710/1/385}
}

@article{matplotlib,
  author    = {Hunter, J. D.},
  title     = {Matplotlib: A 2D graphics environment},
  journal   = {Computing in Science \& Engineering},
  volume    = {9},
  number    = {3},
  pages     = {90--95},
  publisher = {IEEE COMPUTER SOC},
  doi       = {10.1109/MCSE.2007.55},
  year      = 2007
}

@misc{aplpy,
  author        = {{Robitaille}, Thomas and {Bressert}, Eli},
  title         = {{APLpy: Astronomical Plotting Library in Python}},
  year          = 2012,
  month         = aug,
  eid           = {ascl:1208.017},
  pages         = {ascl:1208.017},
  archiveprefix = {ascl},
  eprint        = {1208.017},
  adsurl        = {https://ui.adsabs.harvard.edu/abs/2012ascl.soft08017R}
}

@ARTICLE{Offringa2010,
       author = {{Offringa}, A.~R. and {de Bruyn}, A.~G. and {Biehl}, M. and
         {Zaroubi}, S. and {Bernardi}, G. and {Pandey}, V.~N.},
        title = "{Post-correlation radio frequency interference classification methods}",
      journal = {\mnras},
         year = "2010",
        month = "Jun",
       volume = {405},
       number = {1},
        pages = {155-167},
          doi = {10.1111/j.1365-2966.2010.16471.x},
archivePrefix = {arXiv},
       eprint = {1002.1957},
 primaryClass = {astro-ph.IM},
       adsurl = {https://ui.adsabs.harvard.edu/abs/2010MNRAS.405..155O}
}

@MISC{caracal2020,
       author = {{J{\'o}zsa}, Gyula I.~G. and {White}, Sarah V. and {Thorat}, Kshitij and {Smirnov}, Oleg M. and {Serra}, Paolo and {Ramatsoku}, Mpati and {Ramaila}, Athanaseus J.~T. and {Perkins}, Simon J. and {Moln{\'a}r}, D{\'a}niel Cs. and {Makhathini}, Sphesihle and {Maccagni}, Filippo M. and {Kleiner}, Dane and {Kamphuis}, Peter and {Hugo}, Benjamin V. and {de Blok}, W.~J.~G. and {Andati}, Lexy A.~L.},
        title = "{CARACal: Containerized Automated Radio Astronomy Calibration pipeline}",
 howpublished = {Astrophysics Source Code Library, record ascl:2006.014},
         year = 2020,
        month = jun,
          eid = {ascl:2006.014},
        pages = {ascl:2006.014},
archivePrefix = {ascl},
       eprint = {2006.014},
       adsurl = {https://ui.adsabs.harvard.edu/abs/2020ascl.soft06014J}
}

@ARTICLE{Shimwell2022,
       author = {{Shimwell}, T.~W. and {Hardcastle}, M.~J. and {Tasse}, C. and {Best}, P.~N. and {R{\"o}ttgering}, H.~J.~A. and {Williams}, W.~L. and {Botteon}, A. and {Drabent}, A. and {Mechev}, A. and {Shulevski}, A. and {van Weeren}, R.~J. and {Bester}, L. and {Br{\"u}ggen}, M. and {Brunetti}, G. and {Callingham}, J.~R. and {Chy{\.z}y}, K.~T. and {Conway}, J.~E. and {Dijkema}, T.~J. and {Duncan}, K. and {de Gasperin}, F. and {Hale}, C.~L. and {Haverkorn}, M. and {Hugo}, B. and {Jackson}, N. and {Mevius}, M. and {Miley}, G.~K. and {Morabito}, L.~K. and {Morganti}, R. and {Offringa}, A. and {Oonk}, J.~B.~R. and {Rafferty}, D. and {Sabater}, J. and {Smith}, D.~J.~B. and {Schwarz}, D.~J. and {Smirnov}, O. and {O'Sullivan}, S.~P. and {Vedantham}, H. and {White}, G.~J. and {Albert}, J.~G. and {Alegre}, L. and {Asabere}, B. and {Bacon}, D.~J. and {Bonafede}, A. and {Bonnassieux}, E. and {Brienza}, M. and {Bilicki}, M. and {Bonato}, M. and {Calistro Rivera}, G. and {Cassano}, R. and {Cochrane}, R. and {Croston}, J.~H. and {Cuciti}, V. and {Dallacasa}, D. and {Danezi}, A. and {Dettmar}, R.~J. and {Di Gennaro}, G. and {Edler}, H.~W. and {En{\ss}lin}, T.~A. and {Emig}, K.~L. and {Franzen}, T.~M.~O. and {Garc{\'\i}a-Vergara}, C. and {Grange}, Y.~G. and {G{\"u}rkan}, G. and {Hajduk}, M. and {Heald}, G. and {Heesen}, V. and {Hoang}, D.~N. and {Hoeft}, M. and {Horellou}, C. and {Iacobelli}, M. and {Jamrozy}, M. and {Jeli{\'c}}, V. and {Kondapally}, R. and {Kukreti}, P. and {Kunert-Bajraszewska}, M. and {Magliocchetti}, M. and {Mahatma}, V. and {Ma{\l}ek}, K. and {Mandal}, S. and {Massaro}, F. and {Meyer-Zhao}, Z. and {Mingo}, B. and {Mostert}, R.~I.~J. and {Nair}, D.~G. and {Nakoneczny}, S.~J. and {Nikiel-Wroczy{\'n}ski}, B. and {Orr{\'u}}, E. and {Pajdosz-{\'S}mierciak}, U. and {Pasini}, T. and {Prandoni}, I. and {van Piggelen}, H.~E. and {Rajpurohit}, K. and {Retana-Montenegro}, E. and {Riseley}, C.~J. and {Rowlinson}, A. and {Saxena}, A. and {Schrijvers}, C. and {Sweijen}, F. and {Siewert}, T.~M. and {Timmerman}, R. and {Vaccari}, M. and {Vink}, J. and {West}, J.~L. and {Wo{\l}owska}, A. and {Zhang}, X. and {Zheng}, J.},
        title = "{The LOFAR Two-metre Sky Survey. V. Second data release}",
      journal = {\aap},
         year = 2022,
        month = mar,
       volume = {659},
          eid = {A1},
        pages = {A1},
          doi = {10.1051/0004-6361/202142484},
archivePrefix = {arXiv},
       eprint = {2202.11733},
 primaryClass = {astro-ph.GA},
       adsurl = {https://ui.adsabs.harvard.edu/abs/2022A&A...659A...1S}
}

@ARTICLE{Rajpurohit2022b,
       author = {{Rajpurohit}, K. and {van Weeren}, R.~J. and {Hoeft}, M. and {Vazza}, F. and {Brienza}, M. and {Forman}, W. and {Wittor}, D. and {Dom{\'\i}nguez-Fern{\'a}ndez}, P. and {Rajpurohit}, S. and {Riseley}, C.~J. and {Botteon}, A. and {Osinga}, E. and {Brunetti}, G. and {Bonnassieux}, E. and {Bonafede}, A. and {Rajpurohit}, A.~S. and {Stuardi}, C. and {Drabent}, A. and {Br{\"u}ggen}, M. and {Dallacasa}, D. and {Shimwell}, T.~W. and {R{\"o}ttgering}, H.~J.~A. and {Gasperin}, F. de and {Miley}, G.~K. and {Rossetti}, M.},
        title = "{Deep Low-frequency Radio Observations of A2256. I. The Filamentary Radio Relic}",
      journal = {\apj},
         year = 2022,
        month = mar,
       volume = {927},
       number = {1},
          eid = {80},
        pages = {80},
          doi = {10.3847/1538-4357/ac4708},
archivePrefix = {arXiv},
       eprint = {2111.04449},
 primaryClass = {astro-ph.CO},
       adsurl = {https://ui.adsabs.harvard.edu/abs/2022ApJ...927...80R}
}

@ARTICLE{Wittor2021,
       author = {{Wittor}, Denis and {Ettori}, Stefano and {Vazza}, Franco and {Rajpurohit}, Kamlesh and {Hoeft}, Matthias and {Dom{\'\i}nguez-Fern{\'a}ndez}, Paola},
        title = "{Exploring the spectral properties of radio relics I: Integrated spectral index and Mach number}",
      journal = {arXiv e-prints},
         year = 2021,
        month = jun,
          eid = {arXiv:2106.08351},
        pages = {arXiv:2106.08351},
archivePrefix = {arXiv},
       eprint = {2106.08351},
 primaryClass = {astro-ph.CO},
       adsurl = {https://ui.adsabs.harvard.edu/abs/2021arXiv210608351W}
}

@ARTICLE{Intema2009,
       author = {{Intema}, H.~T. and {van der Tol}, S. and {Cotton}, W.~D. and {Cohen}, A.~S. and {van Bemmel}, I.~M. and {R{\"o}ttgering}, H.~J.~A.},
        title = "{Ionospheric calibration of low frequency radio interferometric observations using the peeling scheme. I. Method description and first results}",
      journal = {\aap},
         year = 2009,
        month = jul,
       volume = {501},
       number = {3},
        pages = {1185-1205},
          doi = {10.1051/0004-6361/200811094},
archivePrefix = {arXiv},
       eprint = {0904.3975},
 primaryClass = {astro-ph.IM},
       adsurl = {https://ui.adsabs.harvard.edu/abs/2009A&A...501.1185I}
}

@ARTICLE{Rajpurohit2021a,
       author = {{Rajpurohit}, K. and {Wittor}, D. and {van Weeren}, R.~J. and {Vazza}, F. and {Hoeft}, M. and {Rudnick}, L. and {Locatelli}, N. and {Eilek}, J. and {Forman}, W.~R. and {Bonafede}, A. and {Bonnassieux}, E. and {Riseley}, C.~J. and {Brienza}, M. and {Brunetti}, G. and {Br{\"u}ggen}, M. and {Loi}, F. and {Rajpurohit}, A.~S. and {R{\"o}ttgering}, H.~J.~A. and {Botteon}, A. and {Clarke}, T.~E. and {Drabent}, A. and {Dom{\'\i}nguez-Fern{\'a}ndez}, P. and {Di Gennaro}, G. and {Gastaldello}, F.},
        title = "{Understanding the radio relic emission in the galaxy cluster MACS J0717.5+3745: Spectral analysis}",
      journal = {\aap},
         year = 2021,
        month = feb,
       volume = {646},
          eid = {A56},
        pages = {A56},
          doi = {10.1051/0004-6361/202039428},
archivePrefix = {arXiv},
       eprint = {2011.14436},
 primaryClass = {astro-ph.GA},
       adsurl = {https://ui.adsabs.harvard.edu/abs/2021A&A...646A..56R}
}

@ARTICLE{Shimwell2019,
       author = {{Shimwell}, T.~W. and {Tasse}, C. and {Hardcastle}, M.~J. and
         {Mechev}, A.~P. and {Williams}, W.~L. and {Best}, P.~N. and
         {R{\"o}ttgering}, H.~J.~A. and {Callingham}, J.~R. and
         {Dijkema}, T.~J. and {de Gasperin}, F. and {Hoang}, D.~N. and
         {Hugo}, B. and {Mirmont}, M. and {Oonk}, J.~B.~R. and {Prandoni}, I. and
         {Rafferty}, D. and {Sabater}, J. and {Smirnov}, O. and
         {van Weeren}, R.~J. and {White}, G.~J. and {Atemkeng}, M. and
         {Bester}, L. and {Bonnassieux}, E. and {Br{\"u}ggen}, M. and
         {Brunetti}, G. and {Chy{\.z}y}, K.~T. and {Cochrane}, R. and
         {Conway}, J.~E. and {Croston}, J.~H. and {Danezi}, A. and {Duncan}, K. and
         {Haverkorn}, M. and {Heald}, G.~H. and {Iacobelli}, M. and
         {Intema}, H.~T. and {Jackson}, N. and {Jamrozy}, M. and
         {Jarvis}, M.~J. and {Lakhoo}, R. and {Mevius}, M. and {Miley}, G.~K. and
         {Morabito}, L. and {Morganti}, R. and {Nisbet}, D. and {Orr{\'u}}, E. and
         {Perkins}, S. and {Pizzo}, R.~F. and {Schrijvers}, C. and
         {Smith}, D.~J.~B. and {Vermeulen}, R. and {Wise}, M.~W. and
         {Alegre}, L. and {Bacon}, D.~J. and {van Bemmel}, I.~M. and
         {Beswick}, R.~J. and {Bonafede}, A. and {Botteon}, A. and {Bourke}, S. and
         {Brienza}, M. and {Calistro Rivera}, G. and {Cassano}, R. and
         {Clarke}, A.~O. and {Conselice}, C.~J. and {Dettmar}, R.~J. and
         {Drabent}, A. and {Dumba}, C. and {Emig}, K.~L. and
         {En{\ss}lin}, T.~A. and {Ferrari}, C. and {Garrett}, M.~A. and
         {G{\'e}nova-Santos}, R.~T. and {Goyal}, A. and {G{\"u}rkan}, G. and
         {Hale}, C. and {Harwood}, J.~J. and {Heesen}, V. and {Hoeft}, M. and
         {Horellou}, C. and {Jackson}, C. and {Kokotanekov}, G. and
         {Kondapally}, R. and {Kunert-Bajraszewska}, M. and {Mahatma}, V. and
         {Mahony}, E.~K. and {Mandal}, S. and {McKean}, J.~P. and {Merloni}, A. and
         {Mingo}, B. and {Miskolczi}, A. and {Mooney}, S. and
         {Nikiel-Wroczy{\'n}ski}, B. and {O'Sullivan}, S.~P. and {Quinn}, J. and
         {Reich}, W. and {Roskowi{\'n}ski}, C. and {Rowlinson}, A. and
         {Savini}, F. and {Saxena}, A. and {Schwarz}, D.~J. and {Shulevski}, A. and
         {Sridhar}, S.~S. and {Stacey}, H.~R. and {Urquhart}, S. and
         {van der Wiel}, M.~H.~D. and {Varenius}, E. and {Webster}, B. and
         {Wilber}, A.},
        title = "{The LOFAR Two-metre Sky Survey. II. First data release}",
      journal = {\aap},
         year = 2019,
        month = feb,
       volume = {622},
          eid = {A1},
        pages = {A1},
          doi = {10.1051/0004-6361/201833559},
archivePrefix = {arXiv},
       eprint = {1811.07926},
 primaryClass = {astro-ph.GA},
       adsurl = {https://ui.adsabs.harvard.edu/abs/2019A&A...622A...1S}
}

@ARTICLE{vanWeeren2009,
       author = {{van Weeren}, R.~J. and {R{\"o}ttgering}, H.~J.~A. and
         {Br{\"u}ggen}, M. and {Cohen}, A.},
        title = "{Diffuse radio emission in the merging cluster MACS J0717.5+3745: the discovery of the most powerful radio halo}",
      journal = {\aap},
         year = "2009",
        month = "Oct",
       volume = {505},
       number = {3},
        pages = {991-997},
          doi = {10.1051/0004-6361/200912528},
archivePrefix = {arXiv},
       eprint = {0905.3650},
 primaryClass = {astro-ph.CO},
       adsurl = {https://ui.adsabs.harvard.edu/abs/2009A&A...505..991V}
}

@BOOK{Pacholczyk1970,
   author = {{Pacholczyk}, A.~G.},
    title = "{Radio astrophysics. Nonthermal processes in galactic and extragalactic sources}",
booktitle = {Series of Books in Astronomy and Astrophysics, San Francisco},
publisher = {Freeman},
     year = 1970,
   adsurl = {https://ui.adsabs.harvard.edu/abs/1970ranp.book.....P}
}

@ARTICLE{Kardashev1962,
   author = {{Kardashev}, N.~S.},
    title = "{Nonstationarity of Spectra of Young Sources of Nonthermal Radio Emission}",
  journal = {\sovast},
     year = 1962,
    month = dec,
   volume = 6,
    pages = {317},
   adsurl = {https://ui.adsabs.harvard.edu/abs/1962SvA.....6..317K}
}

@ARTICLE{Gennaro2018,
       author = {{Di Gennaro}, G. and {van Weeren}, R.~J. and {Hoeft}, M. and {Kang}, H. and
         {Ryu}, D. and {Rudnick}, L. and {Forman}, W. and
         {R{\"o}ttgering}, H.~J.~A. and {Br{\"u}ggen}, M. and {Dawson}, W.~A. and
         {Golovich}, N. and {Hoang}, D.~N. and {Intema}, H.~T. and {Jones}, C. and
         {Kraft}, R.~P. and {Shimwell}, T.~W. and {Stroe}, A.},
        title = "{Deep Very Large Array Observations of the Merging Cluster CIZA J2242.8+5301: Continuum and Spectral Imaging}",
      journal = {\apj},
         year = "2018",
        month = "Sep",
       volume = {865},
       number = {1},
          eid = {24},
        pages = {24},
          doi = {10.3847/1538-4357/aad738},
archivePrefix = {arXiv},
       eprint = {1808.02447},
 primaryClass = {astro-ph.GA},
       adsurl = {https://ui.adsabs.harvard.edu/abs/2018ApJ...865...24D}
}

@ARTICLE{Rudnick1994,
       author = {{Rudnick}, Lawrence and {Katz-Stone}, Debora M. and
         {Anderson}, Martha C.},
        title = "{Do Relativistic Electrons either Gain or Lose Energy, outside of Extragalactic Nuclei?}",
      journal = {\apjs},
         year = "1994",
        month = "Feb",
       volume = {90},
        pages = {955},
          doi = {10.1086/191931},
       adsurl = {https://ui.adsabs.harvard.edu/abs/1994ApJS...90..955R}
}

@ARTICLE{Rajpurohit2020a,
       author = {{Rajpurohit}, K. and {Hoeft}, M. and {Vazza}, F. and {Rudnick}, L. and {van Weeren}, R.~J. and {Wittor}, D. and {Drabent}, A. and {Brienza}, M. and {Bonnassieux}, E. and {Locatelli}, N. and {Kale}, R. and {Dumba}, C.},
        title = "{New mysteries and challenges from the Toothbrush relic: wideband observations from 550 MHz to 8 GHz}",
      journal = {\aap},
         year = 2020,
        month = apr,
       volume = {636},
          eid = {A30},
        pages = {A30},
          doi = {10.1051/0004-6361/201937139},
archivePrefix = {arXiv},
       eprint = {1911.08904},
 primaryClass = {astro-ph.HE},
       adsurl = {https://ui.adsabs.harvard.edu/abs/2020A&A...636A..30R}
}

@ARTICLE{Rajpurohit2018,
	author = {{Rajpurohit}, K. and {Hoeft}, M. and {van Weeren}, R.~J. and 
	{Rudnick}, L. and {R{\"o}ttgering}, H.~J.~A. and {Forman}, W.~R. and 
	{Br{\"u}ggen}, M. and {Croston}, J.~H. and {Andrade-Santos}, F. and 
	{Dawson}, W.~A. and {Intema}, H.~T. and {Kraft}, R.~P. and {Jones}, C. and 
	{Jee}, M.~J.},
	title = "{Deep VLA Observations of the Cluster 1RXS J0603.3+4214 in the Frequency Range of 1-2 GHz}",
	journal = {\apj},
	archivePrefix = "arXiv",
	eprint = {1712.01327},
	year = 2018,
	month = jan,
	volume = 852,
	eid = {65},
	pages = {65},
	doi = {10.3847/1538-4357/aa9f13},
	adsurl = {http://adsabs.harvard.edu/abs/2018ApJ...852...65R}
}

@ARTICLE{Katz1993,
   author = {{Katz-Stone}, D.~M. and {Rudnick}, L. and {Anderson}, M.~C.},
    title = "{Determining the shape of spectra in extended radio sources}",
  journal = {\apj},
     year = 1993,
    month = apr,
   volume = 407,
    pages = {549-555},
      doi = {10.1086/172536},
   adsurl = {http://adsabs.harvard.edu/abs/1993ApJ...407..549K}
}

@ARTICLE{Perley2013,
   author = {{Perley}, R.~A. and {Butler}, B.~J.},
    title = "{An Accurate Flux Density Scale from 1 to 50 GHz}",
  journal = {\apjs},
archivePrefix = "arXiv",
   eprint = {1211.1300},
 primaryClass = "astro-ph.IM",
     year = 2013,
    month = feb,
   volume = 204,
      eid = {19},
    pages = {19},
      doi = {10.1088/0067-0049/204/2/19},
   adsurl = {http://adsabs.harvard.edu/abs/2013ApJS..204...19P}
}

@ARTICLE{vanWeeren2012a,
   author = {{van Weeren}, R.~J. and {R{\"o}ttgering}, H.~J.~A. and {Intema}, H.~T. and 
	{Rudnick}, L. and {Br{\"u}ggen}, M. and {Hoeft}, M. and {Oonk}, J.~B.~R.
	},
    title = "{The ''toothbrush-relic'': evidence for a coherent linear 2-Mpc scale shock wave in a massive merging galaxy cluster?}",
  journal = {\aap},
archivePrefix = "arXiv",
   eprint = {1209.2196},
     year = 2012,
    month = oct,
   volume = 546,
      eid = {A124},
    pages = {A124},
      doi = {10.1051/0004-6361/201219000},
   adsurl = {http://adsabs.harvard.edu/abs/2012A%26A...546A.124V}
}

@ARTICLE{Drury1983,
   author = {{Drury}, L.~O.},
    title = "{An introduction to the theory of diffusive shock acceleration of energetic particles in tenuous plasmas}",
  journal = {Reports on Progress in Physics},
     year = 1983,
    month = aug,
   volume = 46,
    pages = {973-1027},
      doi = {10.1088/0034-4885/46/8/002},
   adsurl = {http://adsabs.harvard.edu/abs/1983RPPh...46..973D}
}

@ARTICLE{EnsslinGopalKrishna2001,
       author = {{En{\ss}lin}, T.~A. and {Gopal-Krishna}},
        title = "{Reviving fossil radio plasma in clusters of galaxies by adiabatic compression in environmental shock waves}",
      journal = {\aap},
         year = 2001,
        month = jan,
       volume = {366},
        pages = {26-34},
          doi = {10.1051/0004-6361:20000198}
          }

@ARTICLE{tribble1991,
   author = {{Tribble}, P.~C.},
    title = "{Depolarization of extended radio sources by a foreground Faraday screen}",
  journal = {\mnras},
     year = 1991,
    month = jun,
   volume = 250,
    pages = {726-736},
      doi = {10.1093/mnras/250.4.726},
   adsurl = {http://adsabs.harvard.edu/abs/1991MNRAS.250..726T}
}

@ARTICLE{Hoeft2007,
   author = {{Hoeft}, M. and {Br{\"u}ggen}, M.},
    title = "{Radio signature of cosmological structure formation shocks}",
  journal = {\mnras},
   eprint = {astro-ph/0609831},
     year = 2007,
    month = feb,
   volume = 375,
    pages = {77-91},
      doi = {10.1111/j.1365-2966.2006.11111.x},
   adsurl = {http://adsabs.harvard.edu/abs/2007MNRAS.375...77H}
}

@ARTICLE{Verdes-Montenegro2001,
       author = {{Verdes-Montenegro}, L. and {Yun}, M.~S. and {Williams}, B.~A. and {Huchtmeier}, W.~K. and {Del Olmo}, A. and {Perea}, J.},
        title = "{Where is the neutral atomic gas in Hickson groups?}",
      journal = {\aap},
         year = 2001,
        month = oct,
       volume = {377},
        pages = {812-826},
          doi = {10.1051/0004-6361:20011127},
archivePrefix = {arXiv},
       eprint = {astro-ph/0108223},
 primaryClass = {astro-ph},
       adsurl = {https://ui.adsabs.harvard.edu/abs/2001A&A...377..812V}
}

@ARTICLE{Jaffe1973,
	author = {{Jaffe}, W.~J. and {Perola}, G.~C.},
	title = "{Dynamical Models of Tailed Radio Sources in Clusters of Galaxies}",
	journal = {\aap},
	year = 1973,
	month = aug,
	volume = 26,
	pages = {423},
	adsurl = {http://adsabs.harvard.edu/abs/1973A%26A....26..423J}
}

@ARTICLE{Haarlem2013,
   author = {{van Haarlem}, M.~P. and {Wise}, M.~W. and {Gunst}, A.~W. and 
	{Heald}, G. and {McKean}, J.~P. and {Hessels}, J.~W.~T. and 
	{de Bruyn}, A.~G. and {Nijboer}, R. and {Swinbank}, J. and {Fallows}, R. and 
	{Brentjens}, M. and {Nelles}, A. and {Beck}, R. and {Falcke}, H. and 
	{Fender}, R. and {H{\"o}randel}, J. and {Koopmans}, L.~V.~E. and 
	{Mann}, G. and {Miley}, G. and {R{\"o}ttgering}, H. and {Stappers}, B.~W. and 
	{Wijers}, R.~A.~M.~J. and {Zaroubi}, S. and {van den Akker}, M. and 
	{Alexov}, A. and {Anderson}, J. and {Anderson}, K. and {van Ardenne}, A. and 
	{Arts}, M. and {Asgekar}, A. and {Avruch}, I.~M. and {Batejat}, F. and 
	{B{\"a}hren}, L. and {Bell}, M.~E. and {Bell}, M.~R. and {van Bemmel}, I. and 
	{Bennema}, P. and {Bentum}, M.~J. and {Bernardi}, G. and {Best}, P. and 
	{B{\^i}rzan}, L. and {Bonafede}, A. and {Boonstra}, A.-J. and 
	{Braun}, R. and {Bregman}, J. and {Breitling}, F. and {van de Brink}, R.~H. and 
	{Broderick}, J. and {Broekema}, P.~C. and {Brouw}, W.~N. and 
	{Br{\"u}ggen}, M. and {Butcher}, H.~R. and {van Cappellen}, W. and 
	{Ciardi}, B. and {Coenen}, T. and {Conway}, J. and {Coolen}, A. and 
	{Corstanje}, A. and {Damstra}, S. and {Davies}, O. and {Deller}, A.~T. and 
	{Dettmar}, R.-J. and {van Diepen}, G. and {Dijkstra}, K. and 
	{Donker}, P. and {Doorduin}, A. and {Dromer}, J. and {Drost}, M. and 
	{van Duin}, A. and {Eisl{\"o}ffel}, J. and {van Enst}, J. and 
	{Ferrari}, C. and {Frieswijk}, W. and {Gankema}, H. and {Garrett}, M.~A. and 
	{de Gasperin}, F. and {Gerbers}, M. and {de Geus}, E. and {Grie{\ss}meier}, J.-M. and {Grit}, T. and {Gruppen}, P. and {Hamaker}, J.~P. and {Hassall}, T. and 
	{Hoeft}, M. and {Holties}, H.~A. and {Horneffer}, A. and {van der Horst}, A. and 
	{van Houwelingen}, A. and {Huijgen}, A. and {Iacobelli}, M. and 
	{Intema}, H. and {Jackson}, N. and {Jelic}, V. and {de Jong}, A. and 
	{Juette}, E. and {Kant}, D. and {Karastergiou}, A. and {Koers}, A. and 
	{Kollen}, H. and {Kondratiev}, V.~I. and {Kooistra}, E. and 
	{Koopman}, Y. and {Koster}, A. and {Kuniyoshi}, M. and {Kramer}, M. and 
	{Kuper}, G. and {Lambropoulos}, P. and {Law}, C. and {van Leeuwen}, J. and 
	{Lemaitre}, J. and {Loose}, M. and {Maat}, P. and {Macario}, G. and 
	{Markoff}, S. and {Masters}, J. and {McFadden}, R.~A. and {McKay-Bukowski}, D. and 
	{Meijering}, H. and {Meulman}, H. and {Mevius}, M. and {Middelberg}, E. and 
	{Millenaar}, R. and {Miller-Jones}, J.~C.~A. and {Mohan}, R.~N. and 
	{Mol}, J.~D. and {Morawietz}, J. and {Morganti}, R. and {Mulcahy}, D.~D. and 
	{Mulder}, E. and {Munk}, H. and {Nieuwenhuis}, L. and {van Nieuwpoort}, R. and 
	{Noordam}, J.~E. and {Norden}, M. and {Noutsos}, A. and {Offringa}, A.~R. and 
	{Olofsson}, H. and {Omar}, A. and {Orr{\'u}}, E. and {Overeem}, R. and 
	{Paas}, H. and {Pandey-Pommier}, M. and {Pandey}, V.~N. and 
	{Pizzo}, R. and {Polatidis}, A. and {Rafferty}, D. and {Rawlings}, S. and 
	{Reich}, W. and {de Reijer}, J.-P. and {Reitsma}, J. and {Renting}, G.~A. and 
	{Riemers}, P. and {Rol}, E. and {Romein}, J.~W. and {Roosjen}, J. and 
	{Ruiter}, M. and {Scaife}, A. and {van der Schaaf}, K. and {Scheers}, B. and 
	{Schellart}, P. and {Schoenmakers}, A. and {Schoonderbeek}, G. and 
	{Serylak}, M. and {Shulevski}, A. and {Sluman}, J. and {Smirnov}, O. and 
	{Sobey}, C. and {Spreeuw}, H. and {Steinmetz}, M. and {Sterks}, C.~G.~M. and 
	{Stiepel}, H.-J. and {Stuurwold}, K. and {Tagger}, M. and {Tang}, Y. and 
	{Tasse}, C. and {Thomas}, I. and {Thoudam}, S. and {Toribio}, M.~C. and 
	{van der Tol}, B. and {Usov}, O. and {van Veelen}, M. and {van der Veen}, A.-J. and 
	{ter Veen}, S. and {Verbiest}, J.~P.~W. and {Vermeulen}, R. and 
	{Vermaas}, N. and {Vocks}, C. and {Vogt}, C. and {de Vos}, M. and 
	{van der Wal}, E. and {van Weeren}, R. and {Weggemans}, H. and 
	{Weltevrede}, P. and {White}, S. and {Wijnholds}, S.~J. and 
	{Wilhelmsson}, T. and {Wucknitz}, O. and {Yatawatta}, S. and 
	{Zarka}, P. and {Zensus}, A. and {van Zwieten}, J.},
    title = "{LOFAR: The LOw-Frequency ARray}",
  journal = {\aap},
archivePrefix = "arXiv",
   eprint = {1305.3550},
 primaryClass = "astro-ph.IM",
     year = 2013,
    month = aug,
   volume = 556,
      eid = {A2},
    pages = {A2},
      doi = {10.1051/0004-6361/201220873},
   adsurl = {http://adsabs.harvard.edu/abs/2013A%26A...556A...2V}
}

@ARTICLE{Trinchieri2003,
       author = {{Trinchieri}, G. and {Sulentic}, J. and {Breitschwerdt}, D. and {Pietsch}, W.},
        title = "{Stephan's Quintet: The X-ray anatomy of a multiple galaxy collision}",
      journal = {\aap},
         year = 2003,
        month = apr,
       volume = {401},
        pages = {173-183},
          doi = {10.1051/0004-6361:20030108}}

@ARTICLE{Trinchierietal05,
   author = {{Trinchieri}, G. and {Sulentic}, J. and {Pietsch}, W. and {Breitschwerdt}, D.},
    title = "{Stephan's Quintet with XMM-Newton}",
  journal = {\aap},
   eprint = {arXiv:astro-ph/0506761},
     year = 2005,
    month = dec,
   volume = 444,
    pages = {697},
      doi = {10.1051/0004-6361:20052910}}

@ARTICLE{Stephan1877,
   author = {{Stephan}, M.},
    title = "{Nebul{\ae} (new) discovered and observed at the observatory of Marseilles, 1876 and 1877, M. Stephan}",
  journal = {\mnras},
     year = 1877,
    month = apr,
   volume = 37,
    pages = {334}}

@article{Hickson82,
        author = "Hickson, P.",
        title = "Systematic properties of compact groups of galaxies",
        journal = {\apj},
        year = "1982",
        volume = "255",
        pages = "382"}

@ARTICLE{Xuetal25,
       author = {{Xu}, C.~K. and {Cheng}, C. and {Yun}, M.~S. and {Appleton}, P.~N. and {Emonts}, B.~H.~C. and {Braine}, J. and {Gallagher}, S.~C. and {Guillard}, P. and {Lisenfeld}, U. and {O'Sullivan}, E. and {Renaud}, F. and {Aromal}, P. and {Duc}, P.-A. and {Labiano}, A. and {Togi}, A.},
        title = "{SQ-A: A Collision-triggered Starburst in the Intragroup Medium of Stephan's Quintet}",
      journal = {\apj},
         year = 2025,
        month = oct,
       volume = {991},
       number = {2},
          eid = {197},
        pages = {197},
          doi = {10.3847/1538-4357/adfbf6}}

@ARTICLE{Makarovetal14,
       author = {{Makarov}, Dmitry and {Prugniel}, Philippe and {Terekhova}, Nataliya and {Courtois}, H{\'e}l{\`e}ne and {Vauglin}, Isabelle},
        title = "{HyperLEDA. III. The catalogue of extragalactic distances}",
      journal = {\aap},
         year = 2014,
        month = oct,
       volume = {570},
          eid = {A13},
        pages = {A13},
          doi = {10.1051/0004-6361/201423496}}

@ARTICLE{Xuetal99,
       author = {{Xu}, Cong and {Sulentic}, Jack W. and {Tuffs}, Richard},
        title = "{Starburst in the Intragroup Medium of Stephan's Quintet}",
      journal = {\apj},
         year = 1999,
        month = feb,
       volume = {512},
       number = {1},
        pages = {178-183},
          doi = {10.1086/306771}}

@ARTICLE{Lisenfeldetal04,
       author = {{Lisenfeld}, U. and {Braine}, J. and {Duc}, P.-A. and {Brinks}, E. and {Charmandaris}, V. and {Leon}, S.},
        title = "{Molecular and ionized gas in the tidal tail in Stephan's Quintet}",
      journal = {\aap},
         year = 2004,
        month = nov,
       volume = {426},
        pages = {471-479},
          doi = {10.1051/0004-6361:20041330}}

@ARTICLE{Maedaetal2025,
       author = {{Maeda}, Fumiya and {Komugi}, Shinya and {Muraoka}, Kazuyuki and {Yamamoto}, Misaki and {Egusa}, Fumi and {Ohta}, Kouji and {Asada}, Yoshihisa and {Habe}, Asao and {Hatsukade}, Bunyo and {Kaneko}, Hiroyuki and {Kobayashi}, Masato I.~N. and {Kohno}, Kotaro and {Konishi}, Ayu and {Matsusaka}, Ren and {Morokuma-Matsui}, Kana and {Nakanishi}, Kouichiro and {Tosaki}, Tomoka and {Tsujita}, Akiyoshi},
        title = "{Spatially and Dynamically Extended Molecular Gas in Stephan's Quintet Revealed by ALMA CO(1{\textendash}0) Total Power Mapping}",
      journal = {\apj},
         year = 2025,
        month = sep,
       volume = {990},
       number = {2},
          eid = {221},
        pages = {221},
          doi = {10.3847/1538-4357/adfc56}}

@ARTICLE{Guillardetal22,
       author = {{Guillard}, P. and {Appleton}, P.~N. and {Boulanger}, F. and {Shull}, J.~M. and {Lehnert}, M.~D. and {Pineau des Forets}, G. and {Falgarone}, E. and {Cluver}, M.~E. and {Xu}, C.~K. and {Gallagher}, S.~C. and {Duc}, P.~A.},
        title = "{Extremely Broad Ly{\ensuremath{\alpha}} Line Emission from the Molecular Intragroup Medium in Stephan's Quintet: Evidence for a Turbulent Cascade in a Highly Clumpy Multiphase Medium?}",
      journal = {\apj},
         year = 2022,
        month = jan,
       volume = {925},
       number = {1},
          eid = {63},
        pages = {63},
          doi = {10.3847/1538-4357/ac313f}
}

@ARTICLE{Cluveretal10,
   author = {{Cluver}, M.~E. and {Appleton}, P.~N. and {Boulanger}, F. and {Guillard}, P. and {Ogle}, P. and {Duc}, P.-A. and {Lu}, N. and {Rasmussen}, J. and {Reach}, W.~T. and {Smith}, J.~D. and {Tuffs}, R. and {Xu}, C.~K. and {Yun}, M.~S.},
    title = "{Powerful H$_{2}$ Line Cooling in Stephan's Quintet. I. Mapping the Significant Cooling Pathways in Group-wide Shocks}",
  journal = {\apj},
     year = 2010,
    month = feb,
   volume = 710,
    pages = {248},
      doi = {10.1088/0004-637X/710/1/248}
}

@ARTICLE{Aromaletal25,
       author = {{Aromal}, P. and {Gallagher}, S.~C. and {Fedotov}, K. and {Bastian}, N. and {Lisenfeld}, U. and {Charlton}, J.~C. and {Appleton}, P.~N. and {Braine}, J. and {Johnson}, K.~E. and {Tzanavaris}, P. and {Emonts}, B.~H.~C. and {Togi}, A. and {Xu}, C.~K. and {Guillard}, P. and {Barcos-Mu{\~n}oz}, L. and {Smith}, L.~J. and {Konstantopoulos}, I.~S.},
        title = "{Characterizing the Star Cluster Populations in Stephan's Quintet Using HST and JWST Observations}",
      journal = {\apj},
         year = 2025,
        month = nov,
       volume = {994},
       number = {1},
          eid = {90},
        pages = {90},
          doi = {10.3847/1538-4357/ae0cb4}}

@ARTICLE{Xuetal05,
       author = {{Xu}, C. Kevin and {Iglesias-P{\'a}ramo}, Jorge and {Burgarella}, Denis and {Rich}, R. Michael and {Neff}, Susan G. and {Lauger}, Sebastien and {Barlow}, Tom A. and {Bianchi}, Luciana and {Byun}, Yong-Ik and {Forster}, Karl and {Friedman}, Peter G. and {Heckman}, Timothy M. and {Jelinsky}, Patrick N. and {Lee}, Young-Wook and {Madore}, Barry F. and {Malina}, Roger F. and {Martin}, D. Christopher and {Milliard}, Bruno and {Morrissey}, Patrick and {Schiminovich}, David and {Siegmund}, Oswald H.~W. and {Small}, Todd and {Szalay}, Alex S. and {Welsh}, Barry Y. and {Wyder}, Ted K.},
        title = "{Ultraviolet Emission and Star Formation in Stephan's Quintet}",
      journal = {\apjl},
         year = 2005,
        month = jan,
       volume = {619},
       number = {1},
        pages = {L95-L98},
          doi = {10.1086/425130}
}

@ARTICLE{Edleretal24,
       author = {{Edler}, H.~W. and {Roberts}, I.~D. and {Boselli}, A. and {de Gasperin}, F. and {Heesen}, V. and {Br{\"u}ggen}, M. and {Ignesti}, A. and {Gajovi{\'c}}, L.},
        title = "{ViCTORIA project: The LOFAR view of environmental effects in Virgo cluster star-forming galaxies}",
      journal = {\aap},
         year = 2024,
        month = mar,
       volume = {683},
          eid = {A149},
        pages = {A149}
}

@ARTICLE{Heesenetal22,
       author = {{Heesen}, V. and {Staffehl}, M. and {Basu}, A. and {Beck}, R. and {Stein}, M. and {Tabatabaei}, F.~S. and {Hardcastle}, M.~J. and {Chy{\.z}y}, K.~T. and {Shimwell}, T.~W. and {Adebahr}, B. and {Beswick}, R. and {Bomans}, D.~J. and {Botteon}, A. and {Brinks}, E. and {Br{\"u}ggen}, M. and {Dettmar}, R.-J. and {Drabent}, A. and {de Gasperin}, F. and {G{\"u}rkan}, G. and {Heald}, G.~H. and {Horellou}, C. and {Nikiel-Wroczynski}, B. and {Paladino}, R. and {Piotrowska}, J. and {R{\"o}ttgering}, H.~J.~A. and {Smith}, D.~J.~B. and {Tasse}, C.},
        title = "{Nearby galaxies in the LOFAR Two-metre Sky Survey. I. Insights into the non-linearity of the radio-SFR relation}",
      journal = {\aap},
         year = 2022,
        month = aug,
       volume = {664},
          eid = {A83},
        pages = {A83},
          doi = {10.1051/0004-6361/202142878}
          }

@ARTICLE{LisenfeldVolk10,
       author = {{Lisenfeld}, U. and {V{\"o}lk}, H.~J.},
        title = "{Shock acceleration of relativistic particles in galaxy-galaxy collisions}",
      journal = {\aap},
         year = 2010,
        month = dec,
       volume = {524},
          eid = {A27},
        pages = {A27},
          doi = {10.1051/0004-6361/201015083}
}

@ARTICLE{Reynolds10,
       author = {{Reynolds}, S.~P.},
        title = "{Particle acceleration in supernova-remnant shocks}",
      journal = {\apss},
         year = 2011,
        month = nov,
       volume = {336},
       number = {1},
        pages = {257-262},
          doi = {10.1007/s10509-010-0559-8}
}

@ARTICLE{OSullivanetal19,
       author = {{O'Sullivan}, Ewan and {Schellenberger}, Gerrit and {Burke}, D.~J. and {Sun}, Ming and {Vrtilek}, Jan M. and {David}, Laurence P. and {Sarazin}, Craig},
        title = "{Building a cluster: shocks, cavities, and cooling filaments in the group-group merger NGC 6338}",
      journal = {\mnras},
         year = 2019,
        month = sep,
       volume = {488},
       number = {2},
        pages = {2925-2946},
          doi = {10.1093/mnras/stz1711}
}

@INPROCEEDINGS{ShullDraine87,
       author = {{Shull}, J. Michael and {Draine}, Bruce T.},
        title = "{The Physics of Interstellar Shock Waves}",
    booktitle = {Interstellar Processes},
         year = 1987,
       editor = {{Hollenbach}, David J. and {Thronson}, Harley A., Jr.},
    publisher = "Dordrecht: Reidel",
       volume = {134},
        month = jan,
        pages = {283},
          doi = {10.1007/978-94-009-3861-8_13}}

@ARTICLE{Jones_relics2023,
       author = {{Jones}, A. and {de Gasperin}, F. and {Cuciti}, V. and {Botteon}, A. and {Zhang}, X. and {Gastaldello}, F. and {Shimwell}, T. and {Simionescu}, A. and {Rossetti}, M. and {Cassano}, R. and {Akamatsu}, H. and {Bonafede}, A. and {Br{\"u}ggen}, M. and {Brunetti}, G. and {Camillini}, L. and {Di Gennaro}, G. and {Drabent}, A. and {Hoang}, D.~N. and {Rajpurohit}, K. and {Natale}, R. and {Tasse}, C. and {van Weeren}, R.~J.},
        title = "{The Planck clusters in the LOFAR sky. VI. LoTSS-DR2: Properties of radio relics}",
      journal = {\aap},
         year = 2023,
        month = dec,
       volume = {680},
          eid = {A31},
        pages = {A31},
          doi = {10.1051/0004-6361/202245102}
}

@ARTICLE{Colafrancesco2017,
       author = {{Colafrancesco}, S. and {Marchegiani}, P. and {Paulo}, C.~M.},
        title = "{The correlation between radio power and Mach number for radio relics in galaxy clusters}",
      journal = {\mnras},
         year = 2017,
        month = nov,
       volume = {471},
       number = {4},
        pages = {4747-4759},
          doi = {10.1093/mnras/stx1806}
}

@ARTICLE{Appletonetal15,
       author = {{Appleton}, P.~N. and {Lanz}, L. and {Bitsakis}, T. and {Wang}, J. and {Peterson}, B.~W. and {Lisenfeld}, U. and {Alatalo}, K. and {Guillard}, P. and {Boulanger}, F. and {Cluver}, M. and {Gao}, Y. and {Helou}, G. and {Ogle}, P. and {Struck}, C.},
        title = "{X-Ray Emission from the Taffy (VV254) Galaxies and Bridge}",
      journal = {\apj},
         year = 2015,
        month = oct,
       volume = {812},
       number = {2},
          eid = {118},
        pages = {118},
          doi = {10.1088/0004-637X/812/2/118}
}

@ARTICLE{Joshietal19,
       author = {{Joshi}, Bhavin A. and {Appleton}, Philip N. and {Blanc}, Guillermo A. and {Guillard}, Pierre and {Rich}, Jeffrey and {Struck}, Curtis and {Freeland}, Emily E. and {Peterson}, Bradley W. and {Helou}, George and {Alatalo}, Katherine},
        title = "{Evidence for Shock-heated Gas in the Taffy Galaxies and Bridge from Optical Emission-line IFU Spectroscopy}",
      journal = {\apj},
         year = 2019,
        month = jun,
       volume = {878},
       number = {2},
        pages = {161},
          doi = {10.3847/1538-4357/ab2124}}

@ARTICLE{OSullivanetal25,
       author = {{O'Sullivan}, Ewan and {Appleton}, P.~N. and {Joshi}, B.~A. and {Lanz}, Lauranne and {Alatalo}, Katherine and {Vrtilek}, Jan M. and {Zezas}, Andreas and {David}, Laurence P.},
        title = "{HCG 57: Evidence for Shock-heated Intergalactic Gas from X-Rays and Optical Emission Line Spectroscopy}",
      journal = {\apj},
         year = 2025,
        month = feb,
       volume = {979},
       number = {2},
          eid = {240},
        pages = {240},
          doi = {10.3847/1538-4357/ada14b}
}

@ARTICLE{Rasmussenetal08,
   author = {{Rasmussen}, J. and {Ponman}, T.~J. and {Verdes-Montenegro}, L. and {Yun}, M.~S. and {Borthakur}, S.},
    title = "{Galaxy evolution in Hickson compact groups: the role of ram-pressure stripping and strangulation}",
  journal = {\mnras},
     year = 2008,
    month = aug,
   volume = 388,
    pages = {1245},
      doi = {10.1111/j.1365-2966.2008.13451.x}
}

@ARTICLE{Rasmussenetal06a,
   author = {{Rasmussen}, J. and {Ponman}, T.~J. and {Mulchaey}, J.~S.},
    title = "{Gas stripping in galaxy groups - the case of the starburst spiral NGC 2276}",
  journal = {\mnras},
     year = 2006,
    month = jun,
   volume = 370,
    pages = {453},
      doi = {10.1111/j.1365-2966.2006.10492.x}
}

@article{OSullivanetal14b,
   author = {{O'Sullivan}, E. and {Zezas}, A. and {Vrtilek}, J.~M. and {Giacintucci}, S. and {Trevisan}, M. and {David}, L.~P. and {Ponman}, T.~J. and {Mamon}, G.~A. and {Raychaudhury}, S.},
    title = "{Deep Chandra Observations of HCG 16. I. Active Nuclei, Star Formation, and Galactic Winds}",
  journal = {\apj},
     year = 2014,
    month = oct,
   volume = 793,
    pages = {73},
      doi = {10.1088/0004-637X/793/2/73}
}

@article{OSullivanetal14c,
   author = {{O'Sullivan}, E. and {Vrtilek}, J.~M. and {David}, L.~P. and {Giacintucci}, S. and {Zezas}, A. and {Ponman}, T.~J. and {Mamon}, G.~A. and {Nulsen}, P. and {Raychaudhury}, S.},
    title = "{Deep Chandra Observations of HCG 16. II. The Development of the Intra-group Medium in a Spiral-rich Group}",
  journal = {\apj},
     year = 2014,
    month = oct,
   volume = 793,
    pages = {74},
      doi = {10.1088/0004-637X/793/2/74}
}

@ARTICLE{Ianjamasimananaetal25,
       author = {{Ianjamasimanana}, R. and {Verdes-Montenegro}, L. and {Sorgho}, A. and {Hess}, K.~M. and {Jones}, M.~G. and {Cannon}, J.~M. and {Solanes}, J.~M. and {Cluver}, M.~E. and {Mold{\'o}n}, J. and {Namumba}, B. and {Rom{\'a}n}, J. and {Labadie-Garc{\'\i}a}, I. and {de la Casa}, C.~C. and {Borthakur}, S. and {Wang}, J. and {Garc{\'\i}a-Benito}, R. and {del Olmo}, A. and {Perea}, J. and {Wiegert}, T. and {Yun}, M. and {Garrido}, J. and {Sanchez-Exp{\'o}sito}, S. and {Bosma}, A. and {Athanassoula}, E. and {J{\'o}zsa}, G.~I.~G. and {Jarrett}, T.~H. and {Xu}, C.~K. and {Smirnov}, O.~M.},
        title = "{MeerKAT view of Hickson Compact Groups: I. Data description and release}",
      journal = {\aap},
         year = 2025,
        month = apr,
       volume = {696},
          eid = {A176},
        pages = {A176},
          doi = {10.1051/0004-6361/202453005}
}

@ARTICLE{Fedotovetal11,
       author = {{Fedotov}, K. and {Gallagher}, S.~C. and {Konstantopoulos}, I.~S. and {Chandar}, R. and {Bastian}, N. and {Charlton}, J.~C. and {Whitmore}, B. and {Trancho}, G.},
        title = "{Star Clusters as Tracers of Interactions in Stephan's Quintet (Hickson Compact Group 92)}",
      journal = {\aj},
         year = 2011,
        month = aug,
       volume = {142},
       number = {2},
          eid = {42},
        pages = {42},
          doi = {10.1088/0004-6256/142/2/42}
}

\appendix
\restartappendixnumbering 

\renewcommand{\thefigure}{\thesection.\arabic{figure}}

\section{Optical and Radio continuum overlays: SQ-R, SQA, Ridge and Bridge}
\label{app:SQR}
Northeast of the shock ridge lies SQ-R, a multi-component radio source that is believed to be unrelated to the group and is probably at considerably higher redshift \citep{Xu2003,NikielWroczynski2013,Xanthopoulos2004}. In our new images, SQ-R appears to comprise at least four distinct sources surrounded by diffuse emission (see Figure\,\ref{fig::optical-zoom} left panel) in agreement with the findings of \citet{Xanthopoulos2004}. As reported by \cite{Xu2003, NikielWroczynski2013}, the brightest component of SQ-R is SQ-Ra. Our images reveal that this component is extended in the east-west direction. Additionally, at lower frequencies ($\leq 400~\text{MHz}$), another compact component (SQ-Rb) is visible to the east of SQ-Ra (see Figure\,\ref{fig::optical-zoom}). We do not find any optical counterpart for either of these components. Just below SQ-Ra, we observe another weaker component (SQ-Rc) which is connected to SQ-Ra. \jwst\ imaging reveals a faint source coincident with SQ-Rc in the 7.7, 10 and 15~$\mu$m bands. To the south of SQ-Rc is the fourth component, SQ-Rd, not connected to SQ-Ra, which is also noted by \cite{Xu2003} and \citet{Xanthopoulos2004}. We do not find evidence of any direct morphological connection between the diffuse emission and either SQ-A or NGC 7318B. We find that SQ-R as a whole has a spectral index of $\alpha_{\rm 144~MHz}^{\rm 6~GHz}$=-0.93$\pm$0.03, in agreement with previous studies. 

An overlay of the \jwst/MIRI 15~$\mu$m image with the VLA C-band (8\arcsec\ resolution) and CS-band combined images (3.7\arcs$\times$2.8\arcs) contours is shown in the middle panel of Figure~\ref{fig::optical-zoom}, and highlights the correlations and offsets between the radio and near-IR emission. The right panel shows a wider view of the \jwst\ image with MeerKAT L-band contours overlaid, highlighting the radio bridge connecting NGC~7319 and the shock ridge, and its close correlation with the infrared (IR) emission. 

\begin{figure*}[!thbp]
    \centering
    \includegraphics[width=0.98\textwidth]{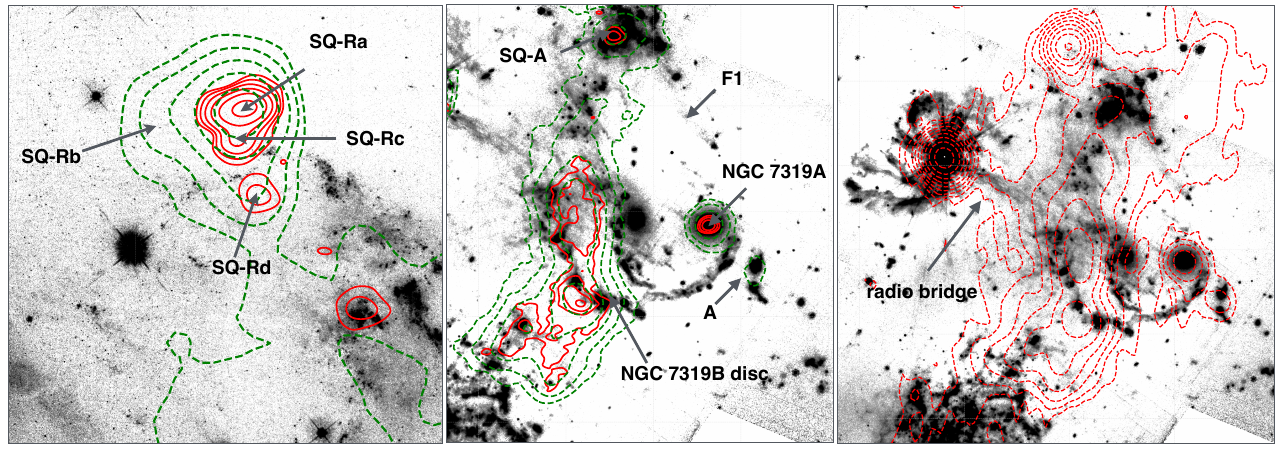}
 \caption{Zoom-in view of SQ-R and NGC~7218B overlaid with radio contours. \textit{Left}: HST WFC3 cut out of the region around SQ-R overlaid with the VLA CS-band  (solid red) and LOFAR contours (dashed green). The radio images reveal that SQ-R consists of four sources.  The VLA high resolution ($3.7\arcsec\times2.8\arcsec$)  contour levels are drawn at  $[1, 2, 4, 8 ...]\times 4.5\sigma_{\rm rms}$  and the LOFAR one at  $[1, 2, 4, 8 ...]\times 15\sigma_{\rm rms}$.  \textit{Middle}: \jwst\ 15~$\mu$m image superimposed with the VLA CS-band (solid red) and C-band contours (dashed green). The radio contour levels of the VLA CS-band image (similar to the left panel image) and the VLA C-band 8\arcsec image contour levels are drawn at  $[1, 2, 4, 8 ...]\times 3.5\sigma_{\rm rms}$. \textit{Right}: \jwst\ 15~$\mu$m image overlaid with the MeerKAT L-band 8\arcsec image radio contours. The image shows that the radio bridge traces an optical filament. Contour levels are drawn at  $[1, 2, 4, 8 ...]\times 4\sigma_{\rm rms}$.}
      \label{fig::optical-zoom}
      \vspace{0.5cm}
\end{figure*} 

\section{Radio spectral age}
\label{app:agemap2}
\setcounter{figure}{0}    
\renewcommand{\thefigure}{\thesection.\arabic{figure}}

Figure\,\ref{fig::age_NWextension} shows the spectral age map at 8\arcsec~ resolution derived using {\tt BRATS} from the 144~MHz, 400~MHz, 1.28~GHz, and 3~GHz maps. The injection index and magnetic field strength were fixed to the values adopted in Section\,\ref{sec::age}. The C-band data were excluded to investigate the spatial age trends across the NW extension. The northern part of the ridge and the NW extension show similar ages. A clear age gradient is also visible from south to north.

\begin{figure*}[!thbp]
    \centering
     \includegraphics[width=0.45\textwidth]{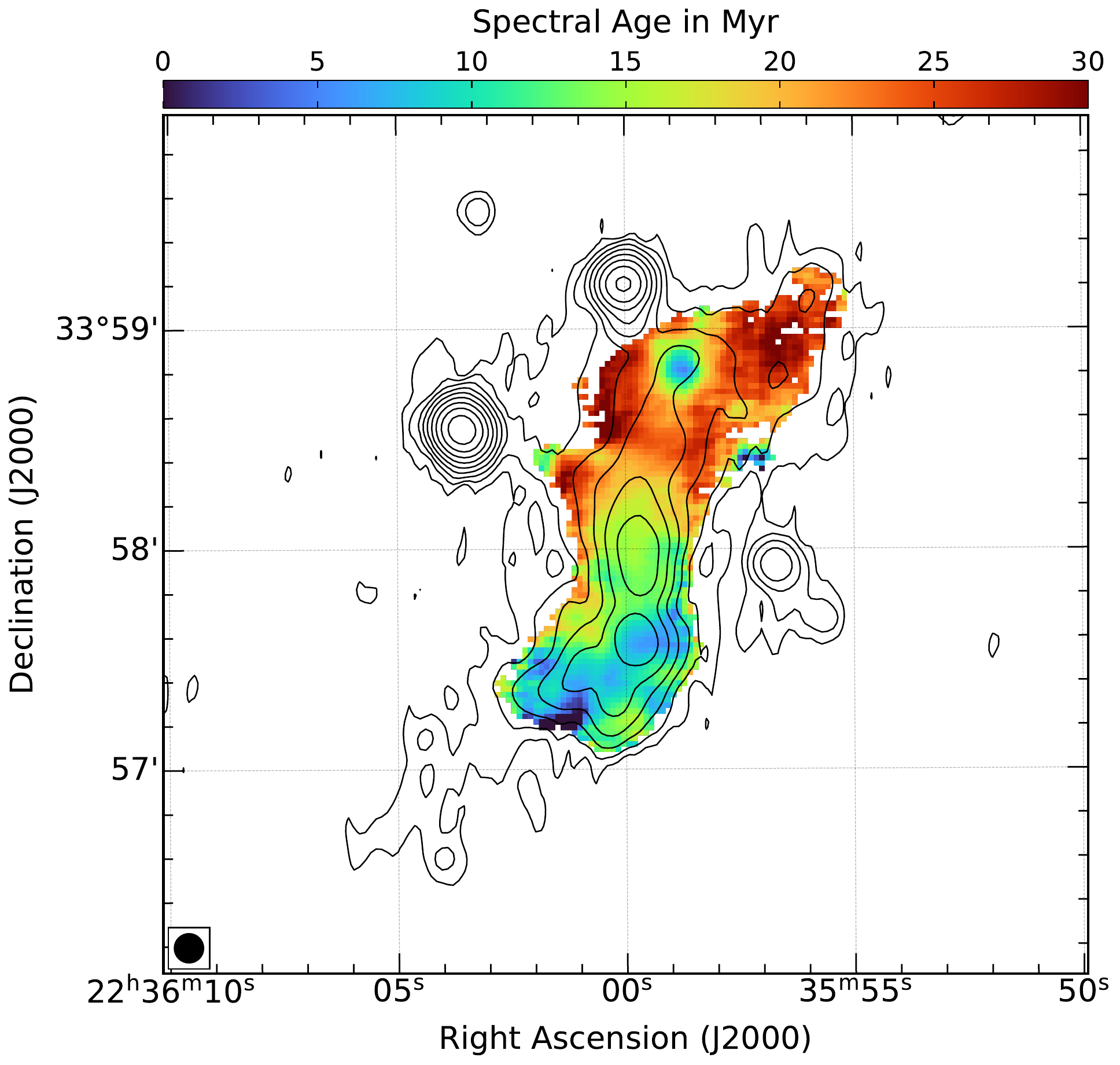}        
 \caption{The spectral age map obtained using 144~MHz, 400~MHz, 1.28~GHz, and 3~GHz maps. We adopted the JP model, an injection index of $-0.70$ and a magnetic field of $\rm 10\mu G$. The radio contour levels are drawn at  $[1, 2, 4, 8 ...]\times 3.5\sigma_{\rm rms}$ and are from the MeerKAT L-band. }
      \label{fig::age_NWextension}
\end{figure*}

\end{document}